\documentclass[12pt,a4paper]{article}
\usepackage{epsfig}
\usepackage[round]{natbib}
\newcommand\dg{${}^\circ$}

\newlength{\figwidth}
\ifx\stupidformat\undefined
\else

\usepackage{a4wide}
\usepackage{doublespace}
\begin{spacing}{1.66}
\fi
\ifx\colourpictures\undefined
\newcommand\coloureps{}
\else
\newcommand\coloureps{c}
\fi
\begin{document}
%
%  #] preamble:
%  #[ titlepage:
%
\title{Tracking down the ENSO delayed oscillator with an adjoint OGCM}
\author{%
Geert Jan van Oldenborgh, 
Gerrit Burgers\\
{\it KNMI, De Bilt, The Netherlands}\\
Stephan Venzke,
Christian Eckert\\
{\it Max-Planck-Institut f\"{u}r Meteorologie, Hamburg, Germany}\\
Ralf Giering\\
{\it Massachusetts Institute of Technology, Cambridge, USA}
}
\date{August 1997\\
revised March 1998, June 1998}
\maketitle
\thispagestyle{empty}
\begin{abstract}
According to the delayed-oscillator picture of ENSO, a positive SST
anomaly in the eastern tropical Pacific will cause westerly wind
anomalies more close to the dateline that first give a positive
feedback, and later, via planetary wave reflection at the western
boundary, a negative feedback.  The aim of this study is to follow a
chain of sensitivities that lead to a delayed-oscillator mechanism in
a general circulation model.  To this end, the adjoint of such an
ocean model is used for studying sensitivities of ENSO indices.

The ocean model used in this study is the HOPE ocean general
circulation model.  Its adjoint has been constructed using the Adjoint
Model Compiler.  Applied to a scalar function computed with a
forward model run, an adjoint run goes back in time and calculates
sensitivities as the derivatives of this function to forcing fields or
ocean state variables at earlier times.

Results from six adjoint runs are reported, tracing the sensitivities
of the NINO3 and NINO3.4 indices in October 1987, December 1987 and
December 1988, as simulated by a Pacfic-only version of HOPE forced by
ECHAM-3 fluxes.

The sensitivities to sea level can be followed back in time for more
than a year.  They are non-local: patterns propagate back in time that
are identified as adjoint Kelvin and $n=1,2$ and 3 Rossby waves, with
speeds compatible with those obtained from model density profiles.
Both the first and the second baroclinic modes seem to play a role.
In contrast, the model sensitivities to heat flux, zonal surface
currents and SST are local and decay in about a month.

The sensitivities to the wind stress agree with the wave
interpretation of the sea-level sensitivities, but only the $n=1$
Rossby wave is visible.  Going back in time, the sensitivity to
westerly anomalies along the equator changes sign, in agreement with
the delayed-oscillator picture.  

Finally, we use a statistical atmosphere model to convert
sensitivities to wind stress at a given time to sensitivities to SST
through the atmosphere at that time.  Focussing on the sensitivities
to the ENSO index region itself at an earlier time then closes the
circle.  These sensitivities have a natural interpretation as
delayed-oscillator coefficients and show the expected behaviour of a
positive sensitivity in the recent past changing to a negative
sensitivity at longer lags.  However, the strength of these feedbacks,
and hence the relevance of this mechanism in ENSO simulated in HOPE,
cannot be determined accurately.
\end{abstract}
\centerline{\it submitted to Monthly Weather Review}
\vspace*{\fill}
\begin{flushleft}
\footnotesize
Corresponding author:\\
\quad G. J. van Oldenborgh, KNMI, P.O. Box 201,\\
\quad NL-3730 AE De Bilt, Netherlands,\\
\quad oldenbor@knmi.nl
\end{flushleft}
\begin{flushright}
KNMI PR 97-23\\
physics/9708037
\end{flushright}
\thispagestyle{empty}
\setcounter{page}{0}
\clearpage

%  #] titlepage:
%  #[ introduction:

\section{Introduction}

% standard BS on ENSO ``It was a dark and stormy night.''
%The interannual ocean and atmosphere variability in the tropical
%Pacific denoted by El-Ni\~{n}o -- Southern-Oscillation (ENSO) is the
%largest climate anomaly on this time scale in the world.  Locally, it
%changes the weather in the Pacific from Peru to Indonesia and eastern
%Australia, and correlations have been found to almost all regions of
%the world.  Naturally, a large effort has been made over the last
%century to describe, understand and predict ENSO.

The first descriptions of the Southern Oscillation 
%the warm cycle of ENSO, El Ni\~{n}o, 
were given by \citet{Hildebrandsson1897,WalkerCirculation}.  
%These
%include anomalously warm surface water off the Peruvian coast,
%combined with relatively low air pressure and abundant precipitation
%in this normally arid region.  The usual trade winds also relax over
%much of the Pacific, or even reverse into westerlies.  
\citet{BjerknesENSO} recognized that the periodic warming of the
eastern tropical Pacific, El Ni\~no, was part of the same coupled
system, now called ENSO.  He also gave a physical mechanism that
identified a positive feedback loop that explains the large
amplitude of the oscillations.  The next goal is the
identification of the mechanisms behind the changes:
the onset of a warm (or cold) episode and the negative feedback that
limits the duration of an episode.  This knowledge can be used, if 
possible, to predict ENSO 3--12 months ahead of time.
%added 
The last decade yielded much improved data from the TOGA-TAO
experiment and remote sensing, and modelling
capabilities have increased steadily.

These models span the range from a stochastically
forced harmonic oscillator \citep{BurgersOsc} and canonical correlation
analyses \citep{BarnstonCCA} to coupled shallow-water models 
\citep[e.g.,][]{ZebiakCaneModel,SchopfSuarezVacillations,KleemanSubsurface}
and general circulation models \citep{NCEPNinoModel,Barnett93}. 
Several hypotheses for the mechanisms that are important in the observed
ENSO signal have been proposed, and identified in relatively simple
ocean-atmosphere models that give rise to ENSO-like oscillations but
can still be analysed analytically.  

A mechanism that operates in many models is the 
delayed oscillator \citep{SuarezSchopfDelayed,BattistiHirstDelayed}.
Consider SST anomalies 
in the eastern equatorial Pacific.  On short time scales a positive
anomaly will weaken the temperature difference with the western
Pacific, and hence weaken the trade winds, which in turn diminishes
the amount of upwelling. The subsiding trade winds also generate
Kelvin waves that propagate to the east and deepen the thermocline
there, levelling off the wind-generated slope.  Both effects lead 
to even higher temperatures.  This classical
positive feedback mechanism \citep{BjerknesENSO} is
kept in check by non-linear effects and by the effect of an
accompanying equatorial Rossby wave, which reflects off the western
coasts as a negative anomaly Kelvin wave and hence gives a negative
feedback some time $\delta$ later.  The delayed oscillator contains
these three effects as
\begin{equation}
	\frac{dN}{dt} = aN - bN^3 - c N(t-\delta)
\;,
\label{eq:delayed}
\end{equation}
with $N$ some index of the SST in the eastern Pacific, and
$a,b,c,\delta>0$.  The $N^3$ term does not influence the period very
much, but is the simplest way to represent non-linear effects that keep
the oscillations bounded.

This relation has been diagnosed in many models of the tropical
Pacific \citep{McCrearyAnderson1991}, 
and gives rise to oscillations with time scales comparable
with the observed ENSO record.
%(Of course, a plethora of mechanisms
%has been proposed that have this feature.)  
Evidence that this mechanism is present in nature has been presented
by \citet{Kessler9193}, although only the ending of the studied warm
event is attributed to western boundary reflection.

However, many other possible modes have been proposed to be active in 
ENSO oscillations.  The coupling between ocean and atmosphere may give
rise to a `slow SST mode' \citep{NeelinSlow}, whereas other autors
stress the importance of the `recharge' mechanism
\citep{Wyrtki75,JinRecharge}.  Also, atmospheric events (westerly
wind-burst, Madden-Julian oscillations) may be responsible for driving
ENSO oscillations \citep{KesslerForcing}.  A full investigation of
these mechanisms falls outside the scope of this article, which focuses 
on the role of the ocean.

In this study we try to identify the delayed oscillator mechanism operating 
in an ocean general circulation model.  The approach we take is based on 
the observation that most indicators for ENSO are highly correlated.
% (which, incidentally, frustrates the search for a second independent
% variable to set up an oscillator).
Many effects can be related to a single
function, for example, the NINO3 index of sea surface temperature in
the eastern Pacific.  We rephrase the question to the cause of ENSO as
the computation of the sensitivity of this index to changes in the
forcing fields or state of the ocean at earlier times.  For small
deviations this question can be addressed with an adjoint model, which
can trace \emph{all} influences on \emph{one} scalar function back in 
time.

We describe the HOPE OGCM in section~\ref{sec:model},
together with a statistical atmosphere model that is used in some
experiments.  This ocean model describes ENSO reasonably well
(section~\ref{sec:enso}).  In section~\ref{sec:adjoint} we describe
the adjoint ocean model.  Using this we search for
the sensitivity of a NINO index to forcing fields (wind stress, heat flux) 
and state variables (sea level height, zonal currents) in section
\ref{sec:sensitivity}.  With the help of the adjoint statistical
atmosphere model we transform the sensitivities to fluxes into
sensitivities to SST anomalies, which are then simplified into a
description resembling the delayed oscillator in 
section~\ref{sec:delayed}.  Our
conclusions are presented in section \ref{sec:conclusions}.  Two
appendices contain technical details and validation runs of the
adjoint model.

%  #] introduction:
%  #[ model description:

\section{Model Description}
\label{sec:model}

%  #[ ocean model:

\subsection{The ocean model}
\label{sec:ocean}

The ocean model used in this study is HOPE \citep[The Hamburg Ocean
Primitive Equation Model,][]{Frey97,HOPE97}.  As we are mainly interested in
ENSO we use a domain that is limited to the Pacific Ocean (120\dg
E--70\dg W, 55\dg S--67\dg N).  The model has a 2.8\dg\ resolution,
with the meridional resolution gradually increased to 0.5\dg\ within
the region 10\dg N to 10\dg S{}.  There are 20 vertical levels which
are irregularly spaced in such a fashion that ten levels are placed in
the upper 300 meters.  The model contains a realistic bottom
topography.  The numerical scheme uses the Arakawa E-grid
\citep{Arakawa77} and a two-hour time step.  We use a 360-day year
subdivided in 12 equal-length months of 4 weeks each.  There is no
diurnal cycle.

A Newtonian relaxation is employed to
restore the surface salinity within the whole domain to the
climatology of \citet{Levitus82} using a time constant
of 40 days.  Otherwise, no explicit fresh water flux is provided.
Prognostic variables are the three-dimensional fields of horizontal 
velocity, temperature and salinity, and the sea surface elevation.
Physical parametrizations in HOPE include solar penetration below the
surface.  About 14\% of the solar radiation incident on the ocean
surface is allowed to penetrate beneath the 20 m surface layer of the
model \citep{Paulson77,Schneider95}.  The vertical mixing in
HOPE is based on a Richardson-number dependent formulation and a
simple mixed layer scheme to represent the effects of wind stirring
(see \cite{Latif94a} for details).  Further details about the model
physics as well as the numerical scheme are given in the HOPE
documentation \citep{HOPE97}.

%  #] ocean model:
%  #[ atmosphere model:

\subsection{The atmosphere model}
\label{sec:atmosphere}

For most experiments in this paper we used prescribed heat flux and
wind stress fields.  These were generated with the ECHAM-3 atmosphere 
model from
observed SST fields and stored as daily values.  ECHAM-3 is the
Hamburg version of the European Centre operational weather forecasting
model.  It is described in detail in two reports
\citep{Roeckner92,DKRZ92}.  The SSTs were taken from a data set of the
British Meteorological Office (UKMO), which is referred to as the
`GISST' data set \citep{GISST}.  Additionally, we imposed a relaxation
to observed SSTs of $40\:\mathrm{Wm^{-2}/K}$, using a Newtonian
formulation.  This is required to compensate for model drift, which
produces systematic differences of the simulated SSTs compared to the
observed ones that were used to generate the atmospheric forcing with
the ECHAM-3 model.
%\footnote{Some quality
%control proved necessary {\em do any MPI'ers object against this?}.}

A statistical atmosphere model similar to the one used by
\citet{Barnett93} is used to convert the sensitivities of the ocean
model to these fluxes into sensitivities to SST.  This simple
atmosphere model is essentially a regression relationship: 
wind stress anomalies and surface heat flux anomalies are related to sea
surface temperature anomalies.  This dependence is almost instantaneous and
non-local.  It reproduces the large-scale response of the atmosphere
to large-scale SST anomalies.  In particular, the Bjerknes positive
feedback of reduced trade winds resulting from a reduced east-west SST
gradient is well-reproduced in this type of model.  On the other hand,
smaller-scale atmospheric responses are not caught.

The regression matrix is derived from an integration of the
atmospheric general circulation model ECHAM-3 with prescribed observed
global SSTs during the period 1949 to 1990.  The linear regression is
done in an Empirical Orthogonal Function space, 
retaining the first 10 EOFs of the anomalies, which
account for 63\% and 49\% of the field variance in the SST and the
atmospheric fields, respectively.  The dominant modes of interannual
variability are found to be well captured by the EOF truncation.  To
compensate for systematic deficiencies of the OGCM, the AGCM's simulated SST
anomaly is corrected using another regression matrix based on the leading 10
EOFs of observed and simulated SST anomaly.

Let the observed and simulated SST anomalies be denoted by
$T_{\mathrm{obs}}(x,t)$ and $T_{\mathrm{sim}}(x,t)$, respectively.
The zonal and meridional wind stress anomalies were pooled together
with the surface heat flux anomalies in one data set of atmospheric
flux anomalies denoted by $\mathcal{F}_\mathrm{echam}(x,t)$.  
The EOF expansions of the observed and simulated fields are
\begin{eqnarray}
 T_{\mathrm{obs}}(x,t) & = & \sum_n \alpha_n (t) \, e_n (x) \\
 \mathcal{F}_{\mathrm{echam}}(x,t) & = & \sum_m \beta_m (t) \, f_m (x) \\
 T_{\mathrm{sim}}(x,t) & = & \sum_l \gamma_l (t) \, g_l (x) 
\;.
\end{eqnarray}
The linear regressions yield matrices $C^{(1)}$ and $C^{(2)}$
whose entries are given by
\begin{eqnarray}
C^{(1)}_{mn} & = & {\langle \beta_m \alpha_n \rangle \over 
                    \langle \alpha_n^2 \rangle } \\
C^{(2)}_{nl} & = & {\langle \alpha_n \gamma_l \rangle \over 
                    \langle \gamma_l^2 \rangle } 
\;.
\end{eqnarray}
Angle brackets signify time averages. The estimated wind stress anomaly, 
$\mathcal{F}_\mathrm{est}(x,t)$, may
now be obtained by the simulated SST anomaly via
\begin{equation}
\mathcal{F}_{\mathrm{est}}(x,t) = \sum_m \beta_{\mathrm{est},m} (t) f_m (x) 
\;,
\end{equation}
with
\begin{equation}
\beta_{\mathrm{est},m} (t) = \sum_{n,l} C^{(1)}_{mn\phantom{l}} C^{(2)}_{nl}
                         \,\gamma_l (t) 
\;.
\end{equation}

% Stephan:
SST anomalies simulated by the HOPE model when forced with 
the ECHAM-3 data mentioned above are not identical to those
observed (Fig.~\ref{fig:EOFs}). To compensate for these systematic
differences, the SST anomalies produced by the HOPE model
are corrected before they are used to infer the atmospheric
fields. For this correction another regression matrix based
on the leading 10 EOFs of simulated and observed SST anomalies
is used.  A similar approach was taken by \citet{Barnett93}.
The additional surface heat flux correction is given by
\begin{eqnarray}
  Q_{\mathrm{cor}}(x,t) & = & \lambda \bigl( T_\mathrm{sim}(x,t) - 
	T_\mathrm{obs,est}(x,t) \bigr) \\
  T_\mathrm{obs,est}(x,t) & = &
	\sum_n \sum_{l} C^{(2)}_{nl\phantom{l}}\gamma_l(t)
        \, e_n (x)
\;,
\end{eqnarray}
%where $\lambda$ is the relaxation time constant.  We found that the
%combination of HOPE and this statistical atmosphere 
%showed most realistic oscillations when $\lambda$
%was chosen to be $-10\:\mathrm{Wm^{-2}/K}$.  
with the relaxation constant $\lambda = -10\:\mathrm{Wm^{-2}/K}$.  As
the estimates for the wind stress and surface heat flux anomalies are
always made for the next month to be integrated, the linear
regressions are computed with the SSTs leading the atmospheric fields
by one month.

%new - comparison with fluxes used.
\ifx\stupidformat\undefined
\begin{figure}[htbp]
\else
\begin{figure}[b]
\fi
\setlength{\figwidth}{110mm}
\begin{center}
\Large a \raisebox{1.5ex}{%
\psfig{file=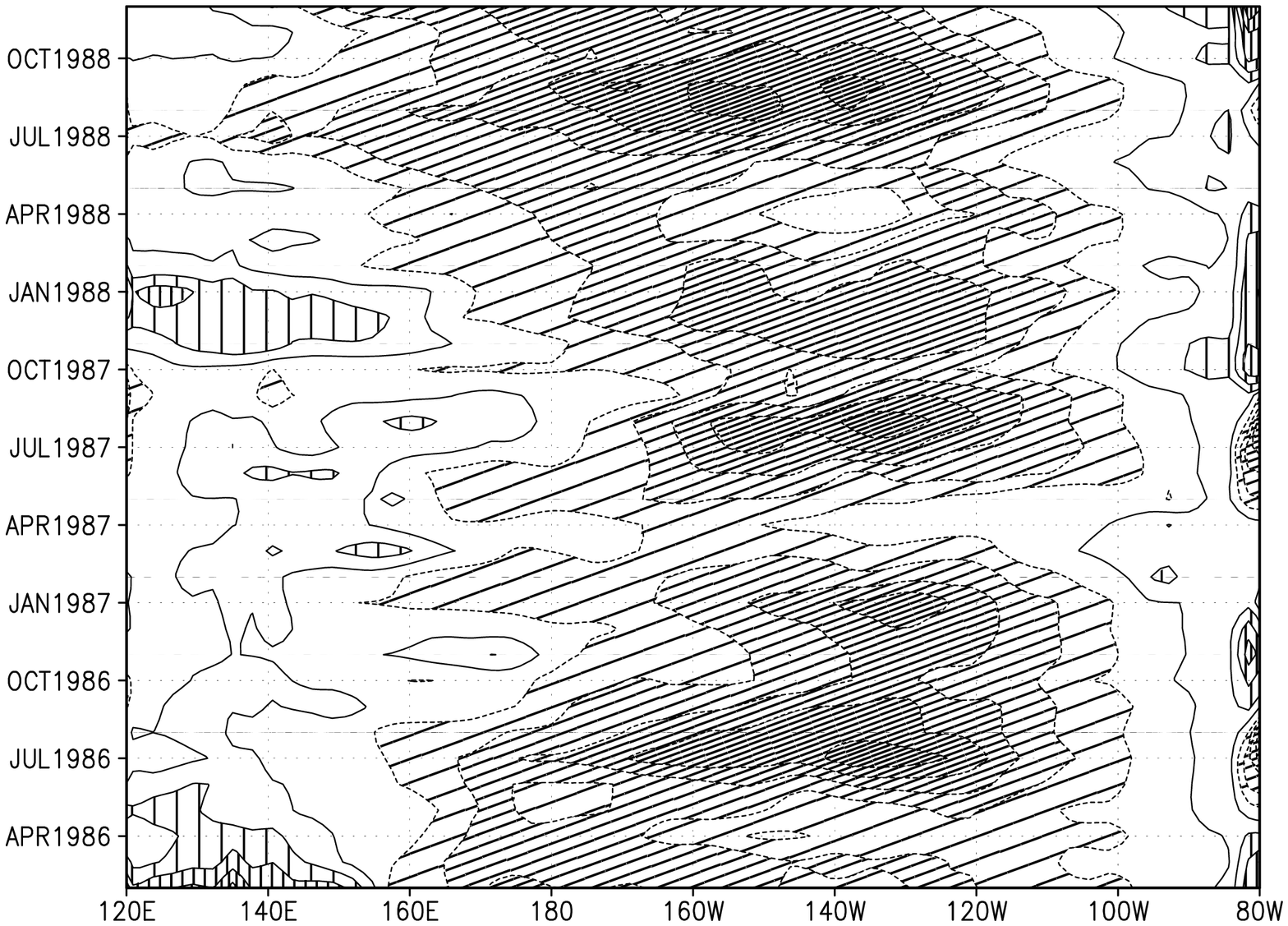,angle=-90,width=\figwidth}}\\
\Large b \raisebox{1.5ex}{%
\psfig{file=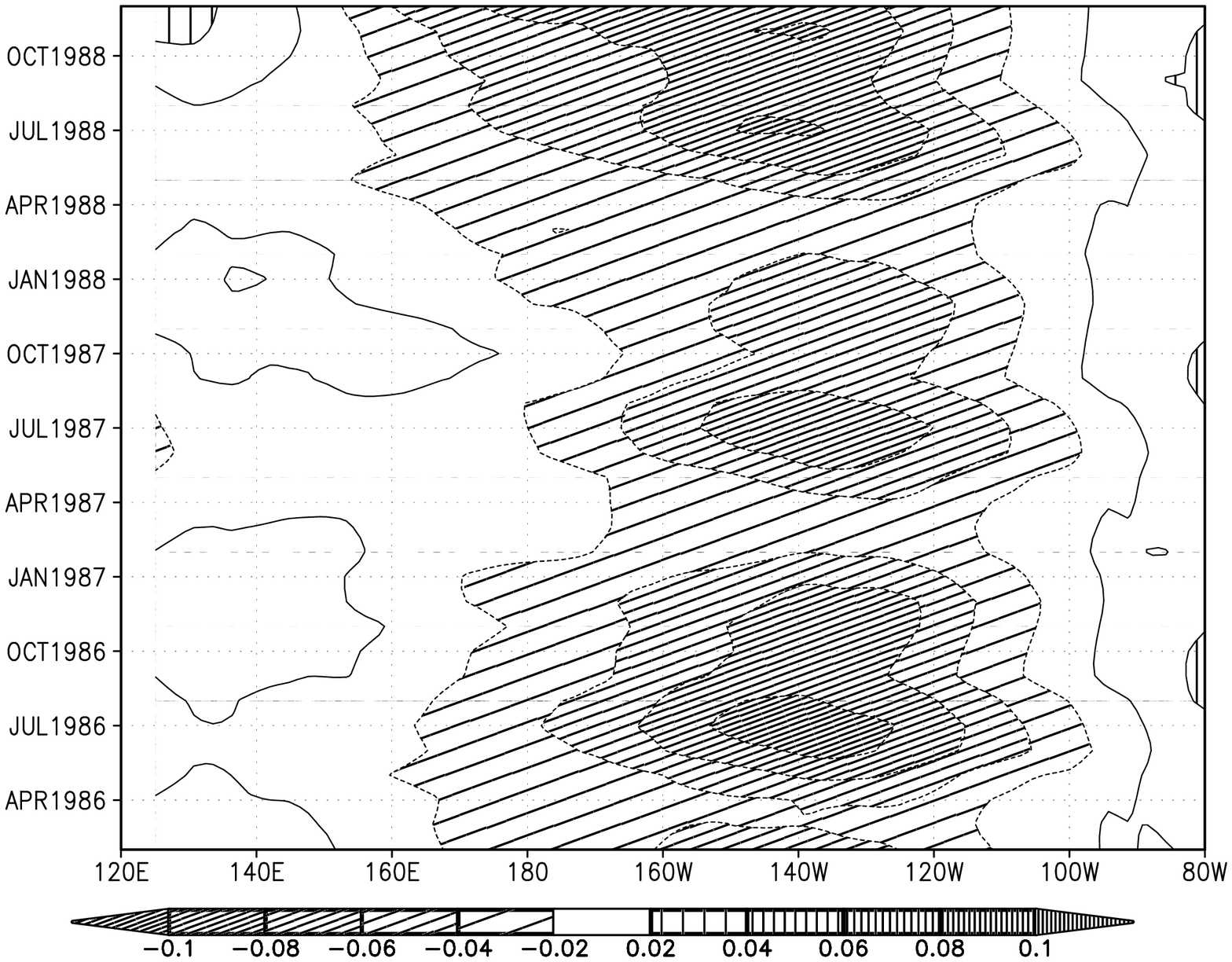,angle=-90,width=\figwidth}}\\
\end{center}
\caption{The zonal wind stress $\tau_x$ in $\mathrm{Nm^{-2}}$ averaged
from 5\dg S to 5\dg N generated from observed SST by the ECHAM-3 model
(a) and from simulated SST by the statistical atmosphere model (b).}
\label{fig:txlat}
\end{figure}

In Fig.~\ref{fig:txlat} we compare the zonal wind stress along the
equator generated with the statistical atmosphere model with the flux
generated by the ECHAM-3 model, which was used as a forcing field.
The large-scale features are seen to correspond quite well.

%not used
%\begin{figure}
%\figwidth\textwidth
%\advance\figwidth by -2cm
%\begin{center}
%\unitlength 1cm
%\begin{picture}(10,5)(0,0)
%\put(0,0){\framebox(10,1)[c]{\Large ocean model}}
%\put(0,4){\framebox(10,1)[c]{\Large statistical atmosphere model}}
%\put(0,2){\framebox(3,1)[c]{\large\shortstack{temperature\\interface}}}
%\put(7,2){\framebox(3,1)[c]{\large\shortstack{heat flux\\interface}}}
%\put(1,1.1){\vector(0,1){0.8}}
%\put(1.2,1.5){\makebox(0,0)[l]{\large$\mathrm{SST'_O}$}}
%\put(8,1.1){\vector(0,1){0.8}}
%\put(8.2,1.5){\makebox(0,0)[l]{\large$\mathrm{SST'_O}$}}
%\put(1,3.1){\vector(0,1){0.8}}
%\put(1.2,3.5){\makebox(0,0)[l]{\large$\mathrm{SST'_A}$}}
%\put(4,3.9){\vector(0,-1){2.8}}
%\put(4.2,3.5){\makebox(0,0)[l]{\Large$\vec{\tau}'_\mathrm{est}$}}
%\put(5.5,3.9){\vector(0,-1){0.9}}
%\put(5.7,3.5){\makebox(0,0)[l]{\Large$Q'_\mathrm{est}$}}
%\put(5.5,2.5){\circle{0.8}}
%\put(5.5,2.5){\makebox(0,0)[c]{\Large$+$}}
%\put(5.5,2.0){\vector(0,-1){0.9}}
%\put(6.9,2.5){\vector(-1,0){0.9}}
%\put(6.5,2.3){\makebox(0,0)[t]{\Large$Q'_c$}}
%\end{picture}
%\end{center}
%\caption[]{A schematic overview of the hybrid coupled model.}
%\label{fig:hcm}
%\end{figure}
%
%A schematic picture of the Hybrid Coupled Model consisting of the HOPE
%OGCM and the statistical atmosphere model is given in
%Fig.~\ref{fig:hcm}.  

%  #] atmosphere model:
%  #[ ENSO simulated by HOPE:

\subsection{ENSO simulated by HOPE}
\label{sec:enso}

Next we establish whether the model used is in fact able to reproduce
the observed behaviour of ENSO{}.  The ocean model was spun up from the
Levitus climatology for 30 years with climatological forcing.  It was
then integrated for 42 years starting in January 1949, using wind
stress and surface heat fluxes from the ECHAM-3 atmosphere model
integration described in section \ref{sec:atmosphere} as upper
boundary conditions, including the $40\:\mathrm{Wm^{-2}}$ relaxation
to observed SST.

\setlength{\figwidth}{12cm}
\ifx\stupidformat\undefined
\begin{figure}[htbp]
\else
\begin{figure}[b]
\fi
\begin{center}
\Large a\raisebox{2ex}%
{\psfig{file=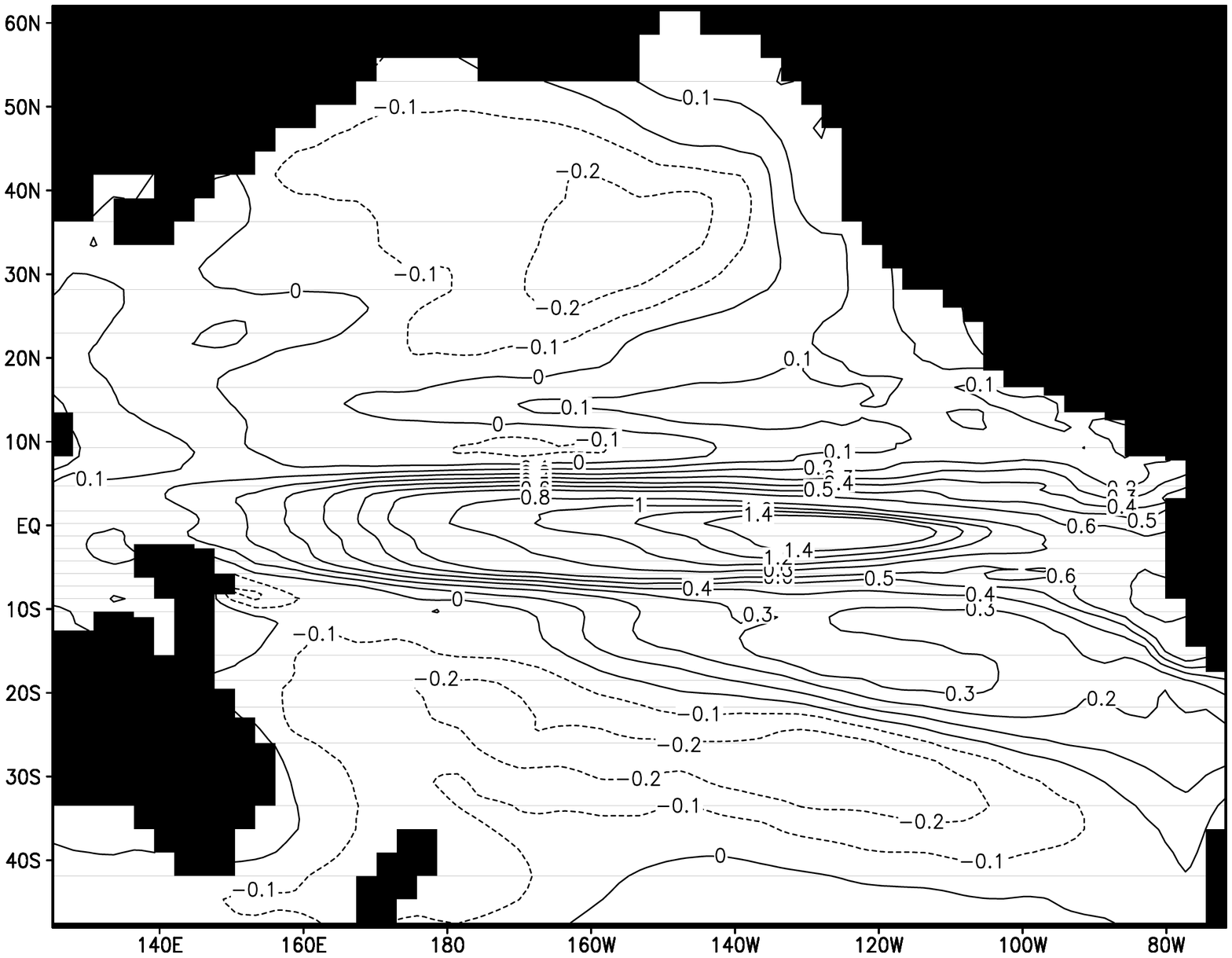,angle=-90,width=\figwidth}}
\vspace{3mm}
\Large b\raisebox{2ex}%
{\psfig{file=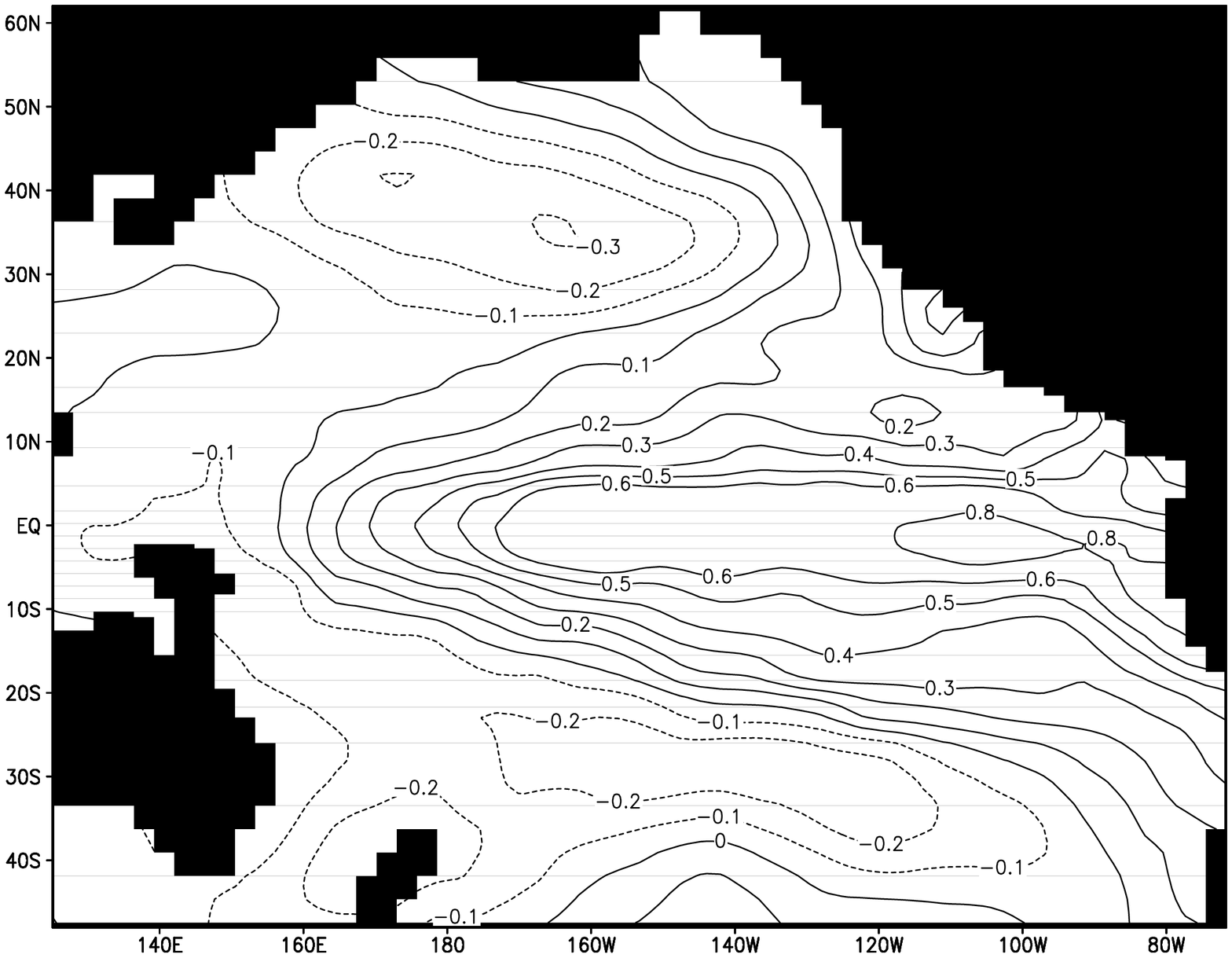,angle=-90,width=\figwidth}}\\
\end{center}
\caption{Maps of the first EOF of monthly mean SST anomalies 
from 1949 to 1990 as
derived from (a) the HOPE model simulation and (b) observations (GISST
data set) explaining 26\% and 29\% of the total variance,
respectively.}
\label{fig:EOFs}
\end{figure}
\begin{figure}[htbp]
\begin{center}
\psfig{file=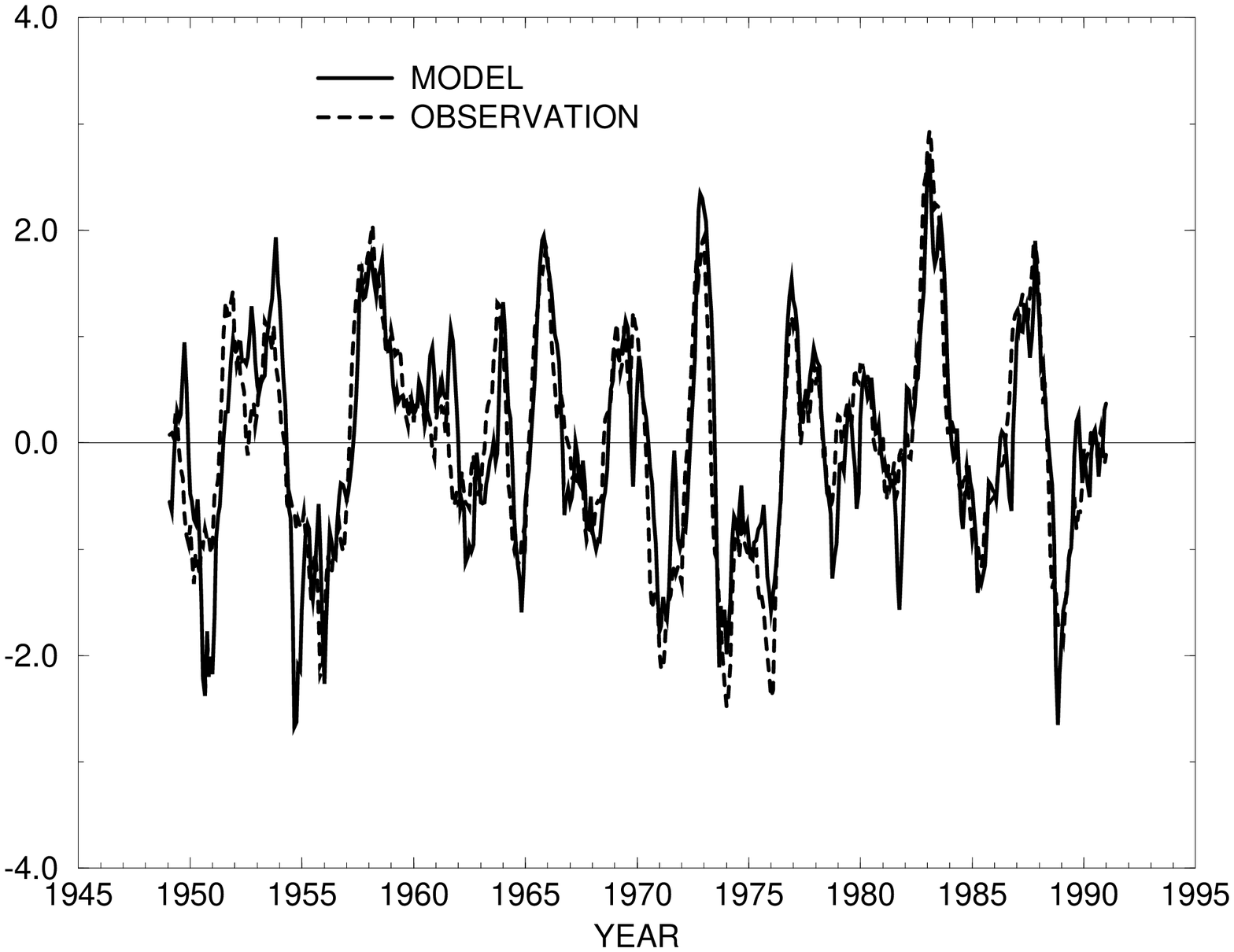,angle=-90,width=\figwidth}
\end{center}
\caption{Time series of the principal components.}
\label{fig:ENSO}
\end{figure}

The spatial structure as well as the temporal evolution of observed
and simulated interannual SST anomalies in the Pacific Ocean can be
inferred from the pattern of the first EOF and timeseries of the 
associated principal component.  Overall, the anomaly structure, shown 
in Fig.~\ref{fig:EOFs}, is simulated reasonably well by the HOPE model.
There are, however, some differences to the observations.  The
simulated SST anomalies are too much equatorially confined, and they
extend too far into the western Pacific.  The latter deficiency is
related to the too strong cold tongue which extends too far into the
western Pacific.  Hence, in addition to the average SST in the NINO3
region (5\dg S--5\dg N, 90\dg W--150\dg W) we will use a slightly
further western region (NINO3.4: 5\dg S--5\dg N, 120\dg W--170\dg W)
as an ENSO index in our sensitivity studies.  The comparison of the
principal components, shown in Fig.~\ref{fig:ENSO}, reveals 
satisfactory agreement ($r=0.88$) between
the temporal evolution of the simulated and observed SST anomalies.

The HOPE model can be coupled to the statistical atmosphere described in
the previous section instead of being forced by prescribed upper boundary
conditions.  Such a combination of an OGCM and a statistical
atmosphere model is generally referred to as a Hybrid Coupled Model (HCM){}.
It shows a regular oscillation of the SST anomalies with a period of 2.5
years. The spatial structure and amplitude of the SST anomalies are
comparable to those obtained with prescribed forcing (Fig.~\ref{fig:EOFs}a).

%  #] ENSO simulated by HOPE:

%  #] model description:
%  #[ adjoint model:

%  #[ intro:

\section{The adjoint model}
\label{sec:adjoint}

In this section we define what we mean by the term `adjoint
model' and present some aspects of the adjoint of the OGCM we
constructed.

%  #] intro:
%  #[ definition:

\subsection{Definition}

% Algemeen concept
The system being studied is described by a state vector $x$ of
prognostic variables.  Let the model $\mathcal{M}$ be some function,
in general non-linear, that computes from an initial state $x_0$ and a
set of forcing fields $y_i$ at times $t_i \;(i=1,\ldots,n)$ a state
vector $x_n$ at time $t_n$
\begin{equation}
	x_n = \mathcal{M}(x_0,y_i)
\;.
\end{equation}
%  Afgeleide, gevoeligheid, infinitesimaal

The tangent model $M(x_0,y_i)$ is the linear mapping that gives the
effect on $x_n$ of infinitesimal changes in the initial state $\delta
x_0$ and in the set of forcing fields $\delta y_i$
\begin{equation}
	\delta x_n = M(x_0,y_i) \; \left( 
		\begin{array}{@{}c@{}}\delta x_0\\\delta y_i\end{array} 
		\right)
	= \lim_{\epsilon\to0} 
		\frac{\mathcal{M}(x_0+\epsilon\,\delta x_0,
				  y_i+\epsilon\,\delta y_i)
			- \mathcal{M}(x_0,y_i)}{\epsilon}
\;.
\label{eq:limit}
\end{equation}
Let $f(x_n)$ be some scalar function of the final state, for instance
an averaged field, index or cost function.
We can now compute the change in this function of the final state as a
result of a change in the initial state and forcing fields as
\begin{eqnarray}
\label{eq:derivatives}
	\delta f(x_n) 
	& \!=\! & \Biggl(\frac{\partial f(x_n)}{\partial x_n}\Biggr)^\mathrm{T} 
			\, \delta x_n 
	        = \Biggl(\frac{\partial f(x_n)}{\partial x_n}\Biggr)^\mathrm{T} 
         		\, M(x_0,y_i) \; \left( 
		\begin{array}{@{}c@{}}\delta x_0\\\delta y_i\end{array} 
					\right)
\\
	& \!=\! & \left(\delta x_0^\mathrm{T},\delta y_i^\mathrm{T}\right) \; 
		M^\dagger(x_0,y_i) \, \frac{\partial f(x_n)}{\partial x_n}
	= \left(\delta x_0^\mathrm{T},\delta y_i^\mathrm{T}\right) 
		\left( \begin{array}{@{}c@{}}
			\partial f(x_n)/\partial x_0 \\
			\partial f(x_n)/\partial y_i
		\end{array} \right)
\;,\nonumber
\end{eqnarray}
with the mapping $M^\dagger(x_0,y_i)$ the adjoint model.  It
propagates the sensitivities of the scalar $f$ to the final state back
to the initial state, and can be seen as a backward derivative of the
model.

In practice, the model $\mathcal{M}$ consists of a series of discrete 
time steps $x_i 
\to x_{i+1}(x_i,y_i)$ for $i=0,\ldots,n-1$, so 
\begin{eqnarray}
	x_n & = & x_n(x_{n-1},y_{n-1}) 
\nonumber\\
	    & = & x_n(x_{n-1}(x_{n-2},y_{n-2}),y_{n-1}) 
\nonumber\\
	    & = & \ldots
\end{eqnarray}
The adjoint model is then evaluated by time-stepping
the discrete model back from $t_n$ to $t_o$, at each step applying
the chain rule of differentiation with respect to the model variables
\begin{equation}
\label{eq:state}
	\frac{\partial f(x_n)}{\partial x_i} = 
		\frac{\partial f(x_n)}{\partial x_{i+1}}
		\frac{\partial x_{i+1}}{\partial x_i}
\end{equation}
and the forcing fields
\begin{equation}
\label{eq:force}
	\frac{\partial f(x_n)}{\partial y_i} = 
		\frac{\partial f(x_n)}{\partial x_{i+1}}
		\frac{\partial x_{i+1}}{\partial y_i}
\;.
\end{equation}
Note that the evolution of the adjoint fields $\partial f/\partial x_i$ 
does not depend on the function $f$, hence except for the 
initialization of $\partial f/\partial x_n$ the adjoint model is 
independent 
of the function $f$.
Physically, the adjoint field represents the sensitivity of $f$ to
infinitesimal perturbations of the prior state $x_i$ and forcing field $y_i$.

For a non-linear model $\mathcal{M}$, the response to a finite 
perturbation may or may not be similar.  An example of the latter case 
is convection, which is often represented by a step function of the 
difference in density of two layers.  The derivative of this function
to the density is zero almost everywhere, yet a finite perturbation
can induce or inhibit convection.  The effect of non-linear terms in a 
particular problem can be gauged to some extent by comparing the effect of 
a finite perurbation $\delta x_0, \delta y_i$ on the scalar function 
$f(x_n)$ to the effect deduced from the derivatives, Eqs~(\ref{eq:limit}, 
\ref{eq:derivatives}).

%  Twee soorten afgeleides: toestand en invoer
The two types of adjoint fields in Eqs~(\ref{eq:state}) and
(\ref{eq:force}) are conceptually different.
Derivatives to forcing fields $\partial f/\partial y_i$ (for instance
externally prescribed fluxes) are not used at times $t < t_i$.  These
adjoint fields thus represent quantities that can be varied
independently.  On the other hand, derivatives to internal model 
state fields $\partial f/\partial x_i$ are used in the previous time
step to compute $\partial f/\partial x_{i-1}$, and hence cannot
explicitly be varied without invalidating the adjoint model for
$t<t_i$.

%  Normalisatie
As defined above the adjoint field is an extensive variable: the
influence of a single grid box is proportional to its size.  Also, the
influence is proportional to the time step used.  We therefore
normalize the adjoint field to the grid box size and a time interval
of one day and give the derivative to forcing fields (fluxes) in units
$[f]/([x]\:\mathrm{day}\:\mathrm{sr})$.

%  #] definition:
%  #[ adjoint HOPE:

\subsection{Adjoint HOPE and atmosphere models}
\label{sec:adjointHOPE}

We constructed adjoint models (as defined above)
of the ocean and atmosphere models in the following way.

%  (T)AMC
In HOPE, a time step consists of a cycle of separate tasks.  
For each of these subroutines (or easily definable
subdivisions) the adjoint code was generated using the Adjoint Model
Compiler \citep{GieringRecipes,AMC}.  This tool gives computer code
which is the adjoint of the original algorithm.  These adjoint routines 
were hand-optimized in some cases and integrated into an adjoint
model.  Further details can be found in Appendix~\ref{technical}.

This construction gives the exact
adjoint of the complete (discrete) model.  We realize that there
are many partial derivatives that can probably be safely neglected,
like the dependence of the mixing parameters on the temperature and
the detailed intricacies of the equation of state.  However, it is a
nontrivial task to verify that omissions are in fact harmless, and we
did not (yet) perform this analysis.

%  Instabiliteit doodgeslagen.
There is one point where it was necessary to deviate from the exact
derivatives to include some non-linear effects.  In our model the
vertical eddy viscosity $A_V$ and diffusivity $D_V$ depend on the
Richardson number $R$ as
\begin{eqnarray}
\label{eq:AV}
	A_V & = & \frac{A_{V,0}}{(1+C_A R')^2} + \mbox{other terms} \\
\label{eq:DV}
	D_V & = & \frac{D_{V,0}}{(1+C_D R')^3} + \mbox{other terms} \\
	R   & = & \frac{-g\partial\rho/\partial z}
	     {(\partial u/\partial z)^2 + (\partial v/\partial z)^2}\\
	R'  & = & \max(0,R)
\end{eqnarray}
with $C_A=C_D=5$.  For marginally stable layers ($0 < R <
R_\mathrm{cut} = 0.3$) the large derivatives $\partial A_V/\partial
R$, $\partial D_V/\partial R$ are not very good 
estimates for the effect of a finite perturbation, as the functions have
very sharp kinks at $R=0$.  In practice this manifests itself as
dipoles of very strong positive and negative influence one vertical
grid point apart in the adjoint temperature and salinity fields.  This
instability can be suppressed by smoothing the derivatives of 
Eqs~(\ref{eq:AV}--\ref{eq:DV}).  In the evaluation of the derivative, we
therefore use $R' = R_\mathrm{cut}^2/R$ for $0 < R < R_\mathrm{cut}$.

% mention convection again
In contrast, the non-linear terms associated with convection were
found not to influence the validity of adjoint runs with targets the
NINO3 and NINO3.4 indices.  This was verified by using different-sign
perturbations in the finite difference experiments of
App.~\ref{sec:kicks}, which show effects of less than 35\%.  The
physical reason is that the signal region, in the cold tongue, is
usually stably stratified, so that convection does not play a role
there.  Signals to the target region often move in the thermocline,
where convection also is irrelevant.

% finally, Tziperman article.
A recent article \citep{SirkesTziperman1997} cautions against using the
adjoint of the finite differences model to trace sensitivities in
time, as a discretization that is stable in the forward runs may be
unstable in the adjoint model.  The specific example shown, a
leap-frog scheme, is not used in HOPE, and we did not find this kind
of instabilities in our adjoint runs in the variables shown.  There is
a slight two-point instability in $\partial N_3/\partial\tau_y$ in the
equatorial western Pacific that may be caused by similar mechanisms.

% statistical atmosphere.
The adjoint of the statistical atmosphere model was constructed
in a similar fashion.  As this is a linear model, it does not depend on the
forward fields, which simplifies the code.  
The time lag between the SST and fluxes that the atmosphere model 
introduces has been taken into account.  

%  #] adjoint HOPE:

%  #] adjoint model:
%  #[ sensitivity:

\section{Sensitivity experiments}
\label{sec:sensitivity}

%  #[ intro:

% Verschil met kick-experimenten
To investigate the causes of temperature changes in the eastern
tropical Pacific in the HOPE model we performed a few experiments with
the adjoint HOPE model described in section \ref{sec:adjointHOPE} using
prescribed fluxes.  These experiments are the exact opposite of
perturbation experiments that search for all effects of one given
perturbation, usually a wind burst \citep[e.g.,][]{GieseResponse}.  We
search backwards in time to find all sensitivities of one target
function, which we take to be the NINO3 (or NINO3.4) index of SST in
the eastern Pacific.  A sensitivity field gives the change in the
target function if we would have added a perturbation at that time
and place.

% Invoeren Equatoriale golven.
{}From perturbation studies and lag-correlation analyses 
\citep{GillBook,PhilanderBook} it has been
found that many perturbations propagate as equatorial Kelvin or Rossby
waves.  The baroclinic Kelvin waves travel eastwards within a few
degrees of the equator, with a speed of 2--$3\:\mathrm{ms^{-1}}$.  The
\mbox{$n$-th} mode Rossby waves travel to the west at $1/(2n+1)$ times this
speed, with a maximum perturbation to the sea level for the first three 
modes 
at about 5\dg, 8--10\dg\ and 12--15\dg\ off the equator (in theory, these 
are 
$\sqrt{3/2}$, $\sqrt{3}$ and $\sqrt{(5+\sqrt{18})/2}$ Rossby radii).  
This propagation implies that sensitivities
propagate backwards along the same trajectories, which we refer to as
adjoint Kelvin or Rossby waves.  Reflections of these waves at the
coast are therefore the backward analogon of reflection of the usual
waves: at the western boundary, an adjoint Kelvin wave coming from the
west will reflect into an opposite sign adjoint Rossby wave coming
from the east.  We will verify this interpretation by
checking that the propagation speed is consistent with predictions from
linear theory.

% Beschrijving experimenten.
As background for the sensitivity experiments we used the years 1987
and 1988.  In 1987 there was an El-Ni\~{n}o episode (the observed NINO3
index was $+1.5\:\mathrm{K}$ to $+2.0\:\mathrm{K}$ for much of the
year, but had returned to $1.0\:\mathrm{K}$ by the end of December),
whereas 1988 ended on a strong cold anomaly ($-1.5\:\mathrm{K}$){}.
As target dates we took October 1987 (warm), December 1987
(transition) and December 1988 (cold).  These three dates give an
impression of the dependance of the sensitivities to the background
state.  For a linear model the three runs should give identical
results.  Our resources were not sufficient to do a systematic study
of these non-linear effects.  We studied the sensitivity of
the NINO3 index $N_3$ (NINO3.4 index $N_{3.4}$) at these times to
various fields at earlier times.  The interesting ones are the input
fluxes (wind stress $\tau_{x,y}$ and the heat flux $Q$) and state
variable sea level height $\zeta$.  The former have been averaged over
$7.5$ day `weeks', the latter is sampled at the beginning of these
periods%
\footnote{{\sc Mpeg} movies and space-time diagrams of the experiments
are available at 
{\tt http://www.knmi.nl/$\sim$oldenbor/Sense/pictures.html}.}.
Unless otherwise noticed
the illustrations are from the run to the NINO3 index at the end of
December 1987.

%  #] intro:
%  #[ sea level:

\subsection{Sea level}
\label{sec:sealevel}

%\setlength{\figwidth}{\textwidth}
%\addtolength{\figwidth}{-4.2cm}
\setlength{\figwidth}{95mm}
\begin{figure}[htbp]
\begin{center}
\Large a \raisebox{1.5ex}{%
\makebox[0mm][l]%
{\psfig{file=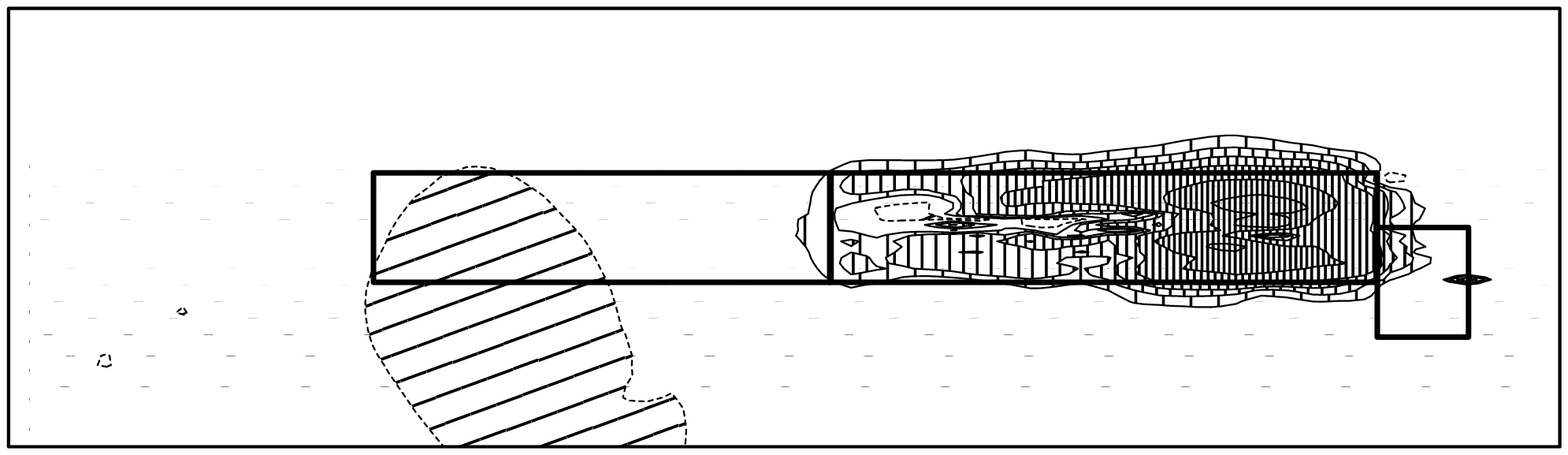,angle=-90,width=\figwidth}}%
\psfig{file=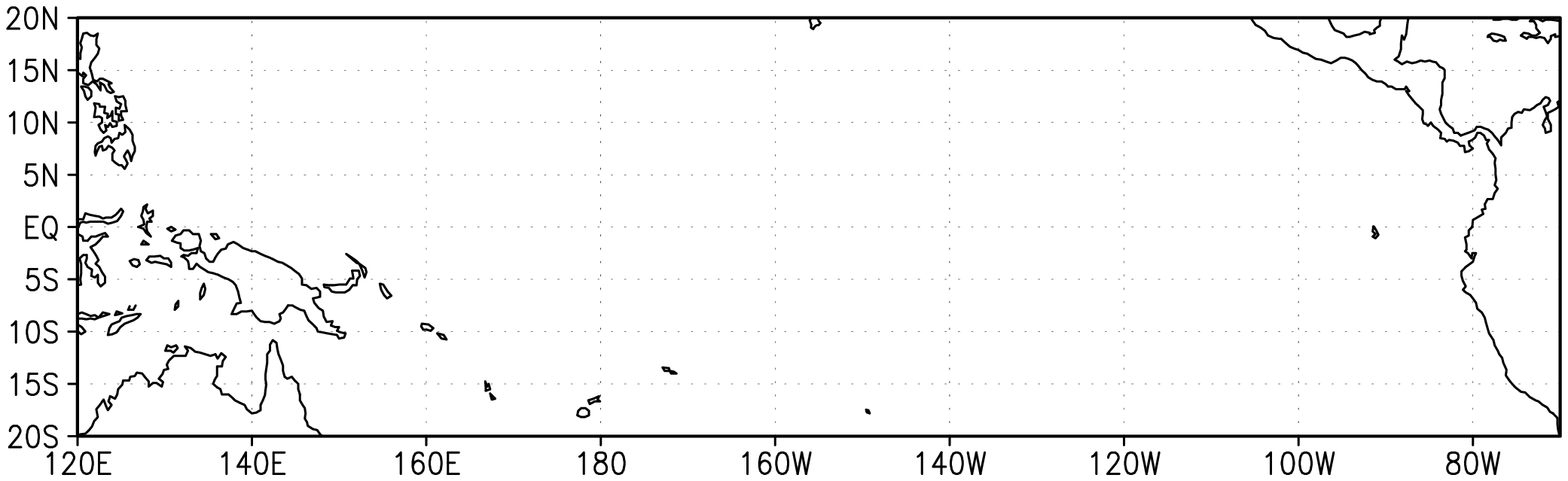,angle=-90,width=\figwidth}}\\
\Large b \raisebox{1.5ex}{%
\makebox[0mm][l]%
{\psfig{file=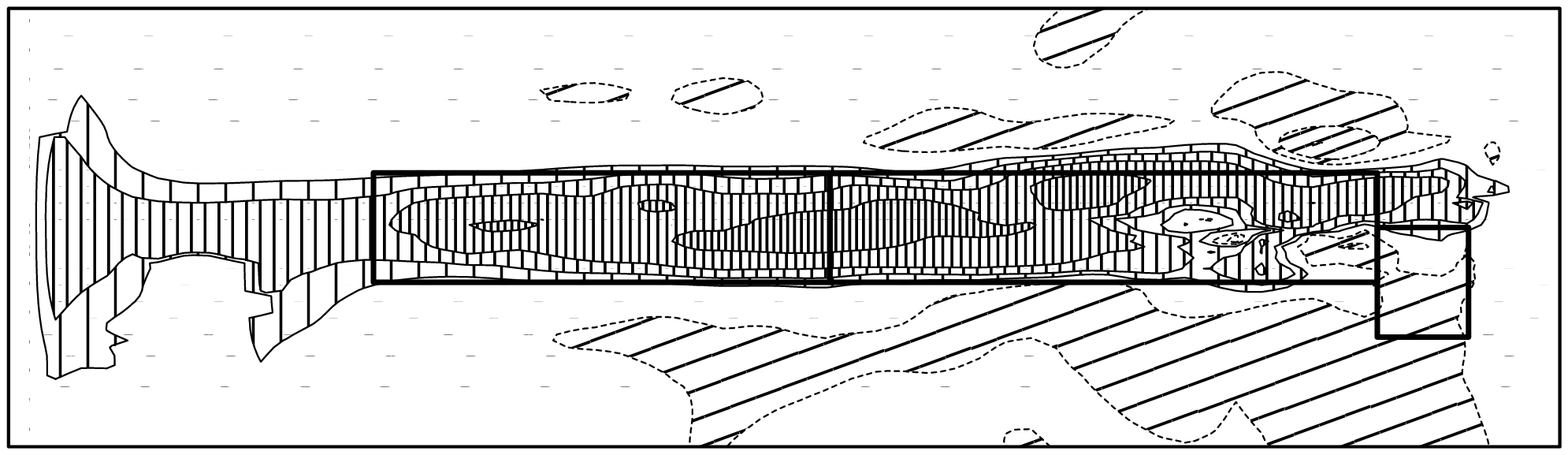,angle=-90,width=\figwidth}}%
\psfig{file=frame.eps,angle=-90,width=\figwidth}}\\
\Large c \raisebox{1.5ex}{%
\makebox[0mm][l]%
{\psfig{file=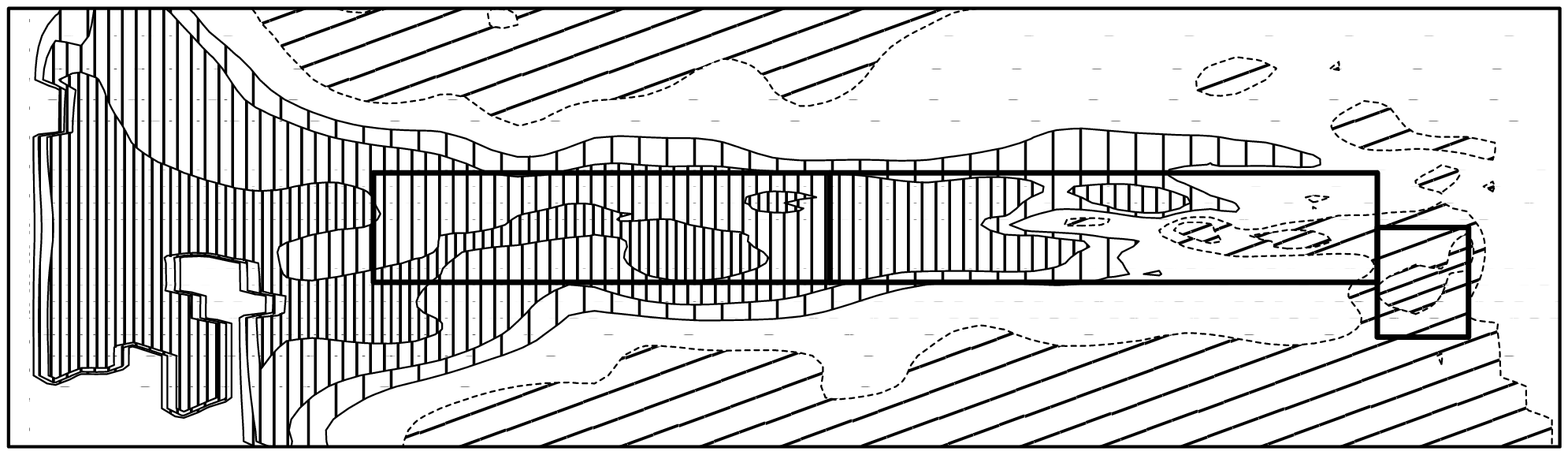,angle=-90,width=\figwidth}}%
\psfig{file=frame.eps,angle=-90,width=\figwidth}}\\
\Large d \raisebox{1.5ex}{%
\makebox[0mm][l]%
{\psfig{file=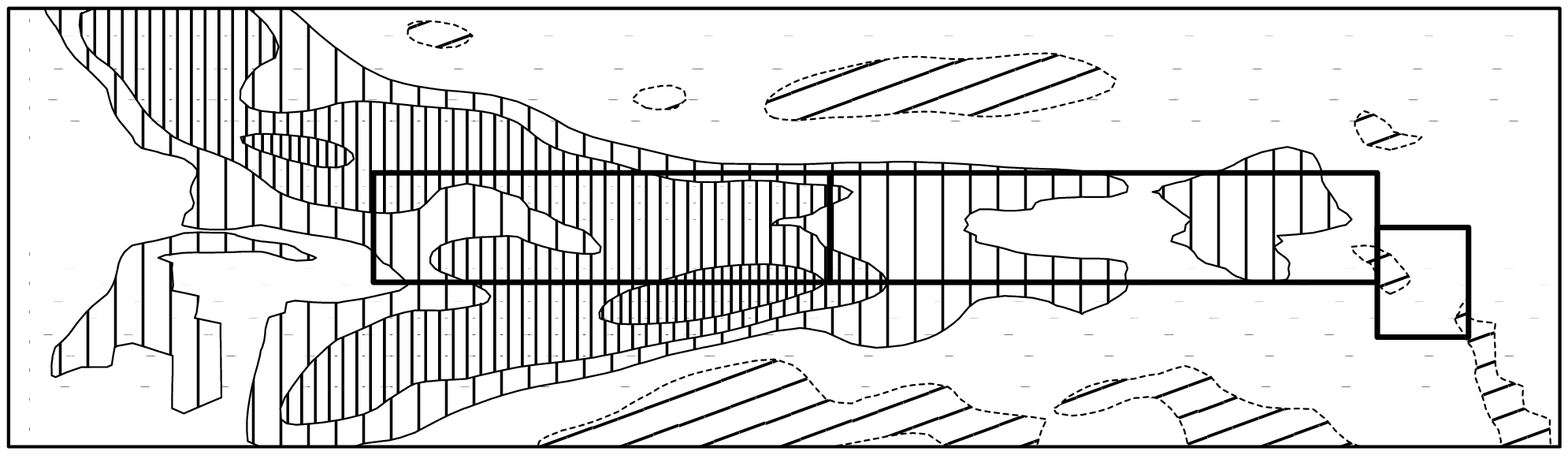,angle=-90,width=\figwidth}}%
\psfig{file=frame.eps,angle=-90,width=\figwidth}}\\
\Large e \raisebox{1.5ex}{%
\makebox[0mm][l]%
{\psfig{file=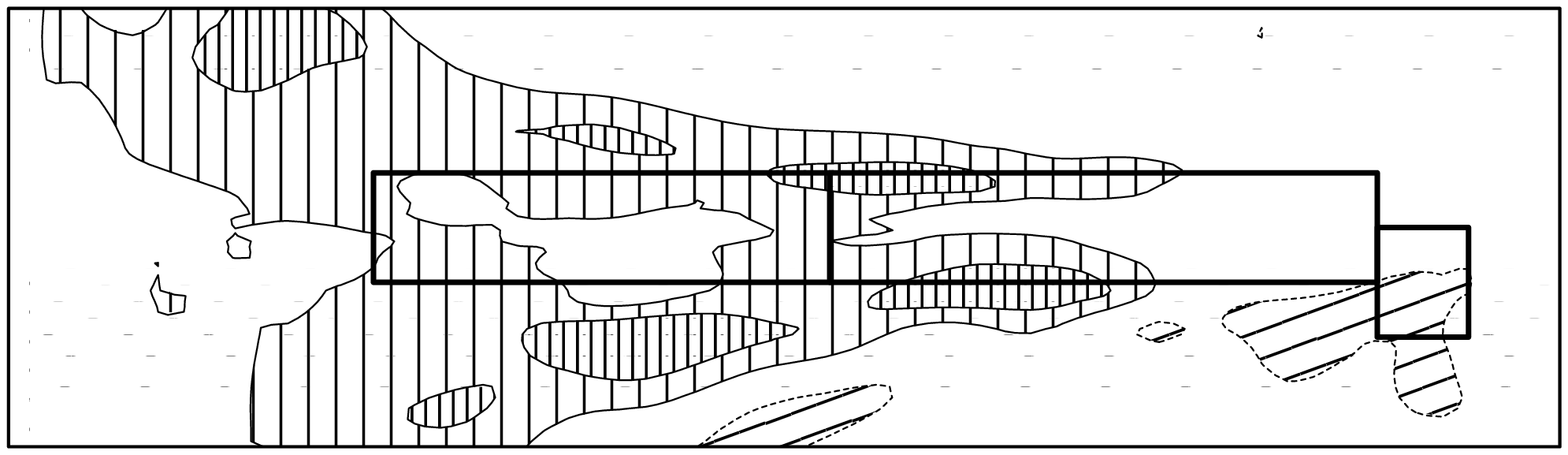,angle=-90,width=\figwidth}}%
\psfig{file=frame.eps,angle=-90,width=\figwidth}}\\
\Large f \raisebox{1.5ex}{%
\makebox[0mm][l]%
{\psfig{file=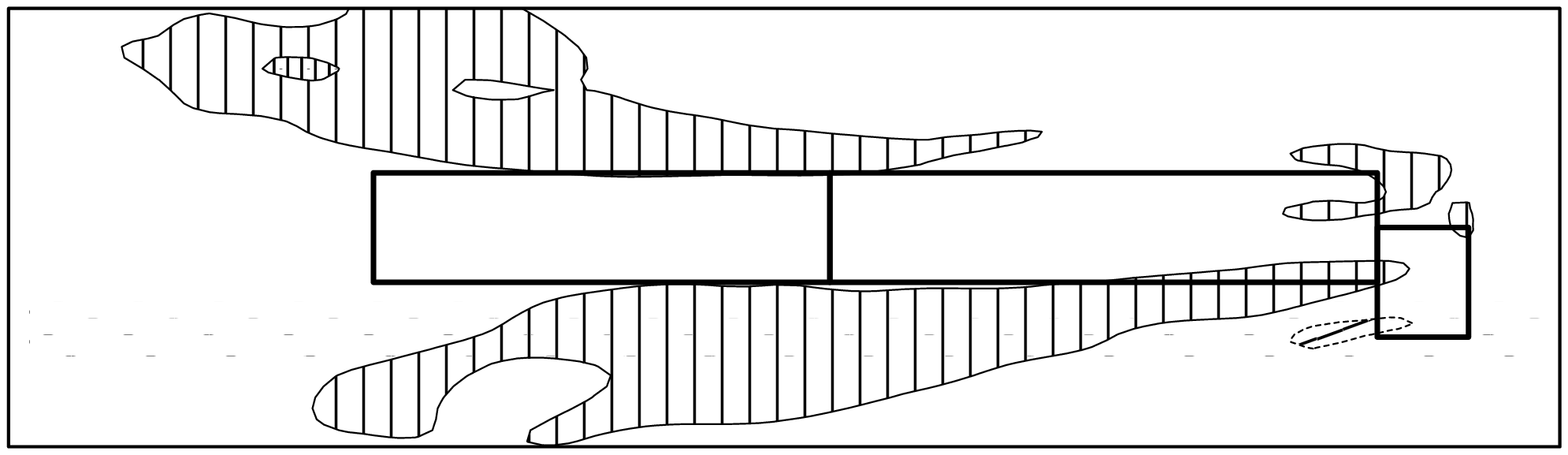,angle=-90,width=\figwidth}}%
\psfig{file=frame.eps,angle=-90,width=\figwidth}}\\
\psfig{file=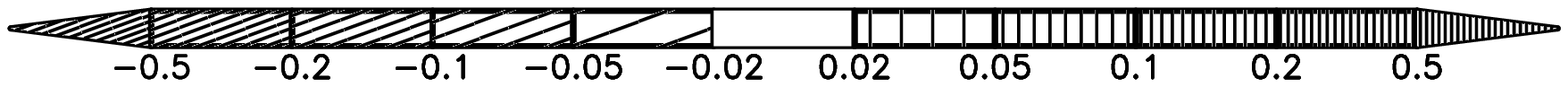,angle=-90,width=\figwidth}
\end{center}
\caption{The sensitivity of the NINO3 index to sea level changes 
$\partial
N_3/\partial\zeta$ in $\mathrm{K}/\mathrm{m}/\mathrm{sr}$ at the end
of 1987 in the last week of December (a), the first weeks of November
(b), July (c), April (d) and January 1987 (e), and July 1986 (f).  The 
boxes denote, from east to west, the NINO12, NINO3 and NINO4 index regions.}
\label{fig:dz87}
\end{figure}

The clearest signals can be seen in the sensitivity to the sea level.
This is due to the fact that, as in the forward case, the adjoint
Kelvin and Rossby waves are clearly separated by latitude.  The
interpretation of the adjoint field is however complicated, as there
is no physical way to change the value of this state variable directly.  A
balanced change in the sea level requires changes to the subsurface
structure as well, and the effect of these are not included in the
partial derivatives shown here.

The sensitivity of the SST to the sea level on short time scales is
caused by the following chain of influences.  An increase in sea level
will depress the thermocline slightly\footnote{Most extra water will
just flow away though \citep{FischerLatif95}.}.  This deeper
thermocline in an upwelling region in turn causes the sea surface to
be less cold.  It is therefore no surprise to see that at the end of
December 1987 the NINO3 index would have been higher if the sea level
would have been higher one week earlier in the same area.  The NINO3
index is the average temperature in the middle box shown in
Fig.~\ref{fig:dz87}a.  The sensitivity is thus positive and confined
to this region, with an emphasis on the upwelling region.  (The
slightly negative field in the western Pacific is a weak barotropic
mode.)

Going back in time,
at the beginning of November (Fig.~\ref{fig:dz87}b), one can see an
adjoint Kelvin wave coming from the western Pacific, and some adjoint
$n=1$ Rossby waves coming from the eastern boundary.  The adjoint Kelvin 
wave should be
interpreted as follows: {\em if\/} the sea level would have been
(infinitesimally) higher in the vertically hatched area in November 
1987, then there would have been a
baroclinic Kelvin wave added to the actual state at
the position indicated, and the temperature would have been
(infinitesimally) higher in the NINO3 area 2 months later.  The
Kelvin wave would have depressed the thermocline in that region.
(Note that this wave is not actually added to the state: the
linearization is performed around the actual background, not
around climatology.)

In the beginning of July (Fig.~\ref{fig:dz87}c), all the adjoint
Rossby waves going to the measurement region from the eastern boundary
have reached the coast.  Comparing with the adjoint run of a linear
shallow water model (not shown), it seems that these are partially
reflected into an adjoint Kelvin wave, which merges with the tail of
the original one.  The resolution does not permit coastal Kelvin waves
to travel north and south.  On the other side, the adjoint equatorial
Kelvin wave has reflected at the western coasts into a set of adjoint
Rossby waves of opposite sign.  The first blob at 165\dg W corresponds
to the $n=1$ reflection off New Guinea (cf.\ Fig.~\ref{fig:dz87st}),
the second one (at 150\dg E) has been reflected off the Philippines.
(The HOPE land-sea mask has approximately two north-south walls at
these locations.)  At the same longitude the $n=2$ reflection off New
Guinea is forming at 8\dg S, and further west the beginnings of
higher-order adjoint Rossby waves can be discerned.  The presence of
the parity-odd $n=2$ mode is due to the fact that the western coast
has some structure in our model.

Three months earlier (Fig.~\ref{fig:dz87}d) the first adjoint Rossby
wave has almost dissipated, the strongest signal is now at 160\dg W.
The southern $n=2$ signal is now at the dateline, whereas the northern
reflection has reached 150\dg E at 8\dg N, clearly separated from the
$n=3$ reflection at 15\dg--20\dg N.  The situation at the beginning of
January is not much different (Fig.~\ref{fig:dz87}e).

At the beginning of July 1986, 18 months before the NINO3 index is
evaluated (Fig.~\ref{fig:dz87}f), the only influence remaining is a
pattern of higher order Rossby waves that slowly traverses the basin.
This sensitivity on the sea level, and equivalently the thermocline
depth, of the off-equatorial warm pool is in agreement with the
original recharge hypothesis \citep{Wyrtki75}.  However, a sensitivity
of the NINO indices to the zonally-averaged thermocline depth along
the equator at this time scale \citep{JinRecharge} is not seen.

\setlength{\figwidth}{0.5\textwidth}
\addtolength{\figwidth}{-3mm}
\begin{figure}[p]
%{\psfig{file=adzjr87lat5N.eps,angle=-90,width=\figwidth}}\\
%{\psfig{file=adzjr87lat5S.eps,angle=-90,width=\figwidth}}
\begin{center}
\mbox{%
\makebox[0pt][r]{\Large a }%
\raisebox{3ex}%
{\psfig{file=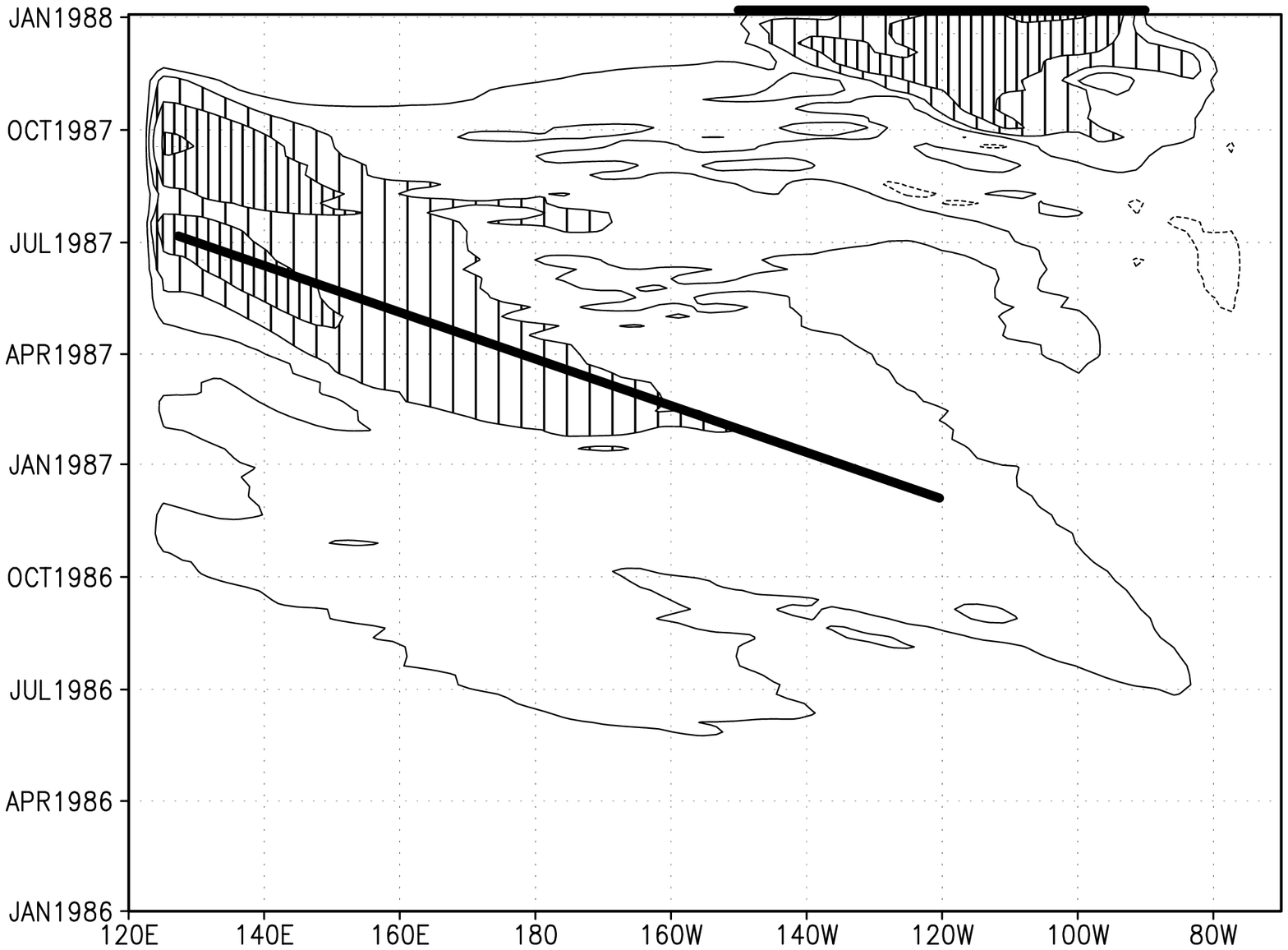,angle=-90,width=\figwidth}}%
\hspace{0.5mm}
\raisebox{3ex}%
{\psfig{file=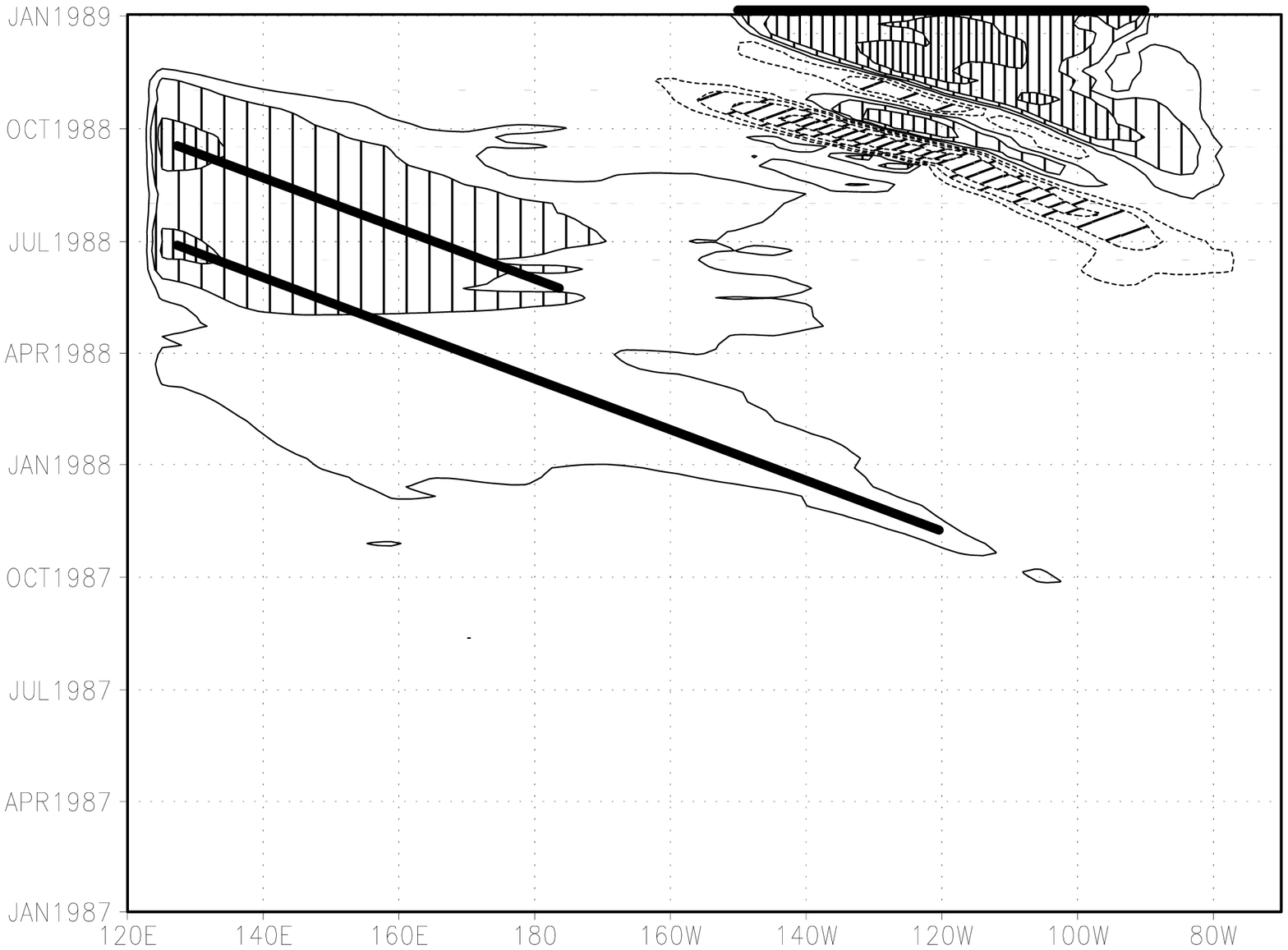,angle=-90,width=\figwidth}}
\makebox[0pt][l]{\Large\ d}%
}\\
\mbox{%
\makebox[0pt][r]{\Large b }%
\raisebox{3ex}%
{\psfig{file=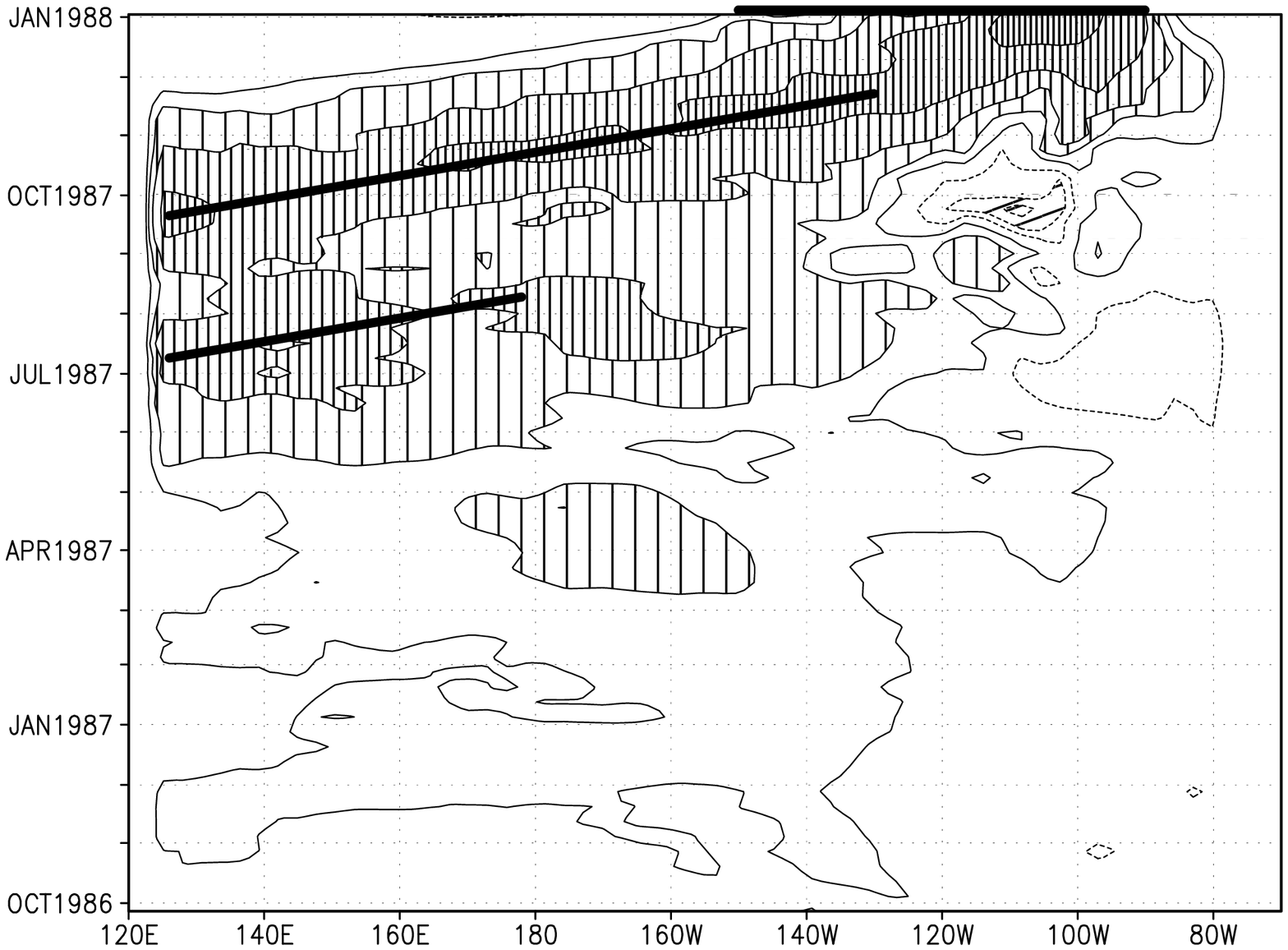,angle=-90,width=\figwidth}}%
\hspace{1mm}
\raisebox{3ex}%
{\psfig{file=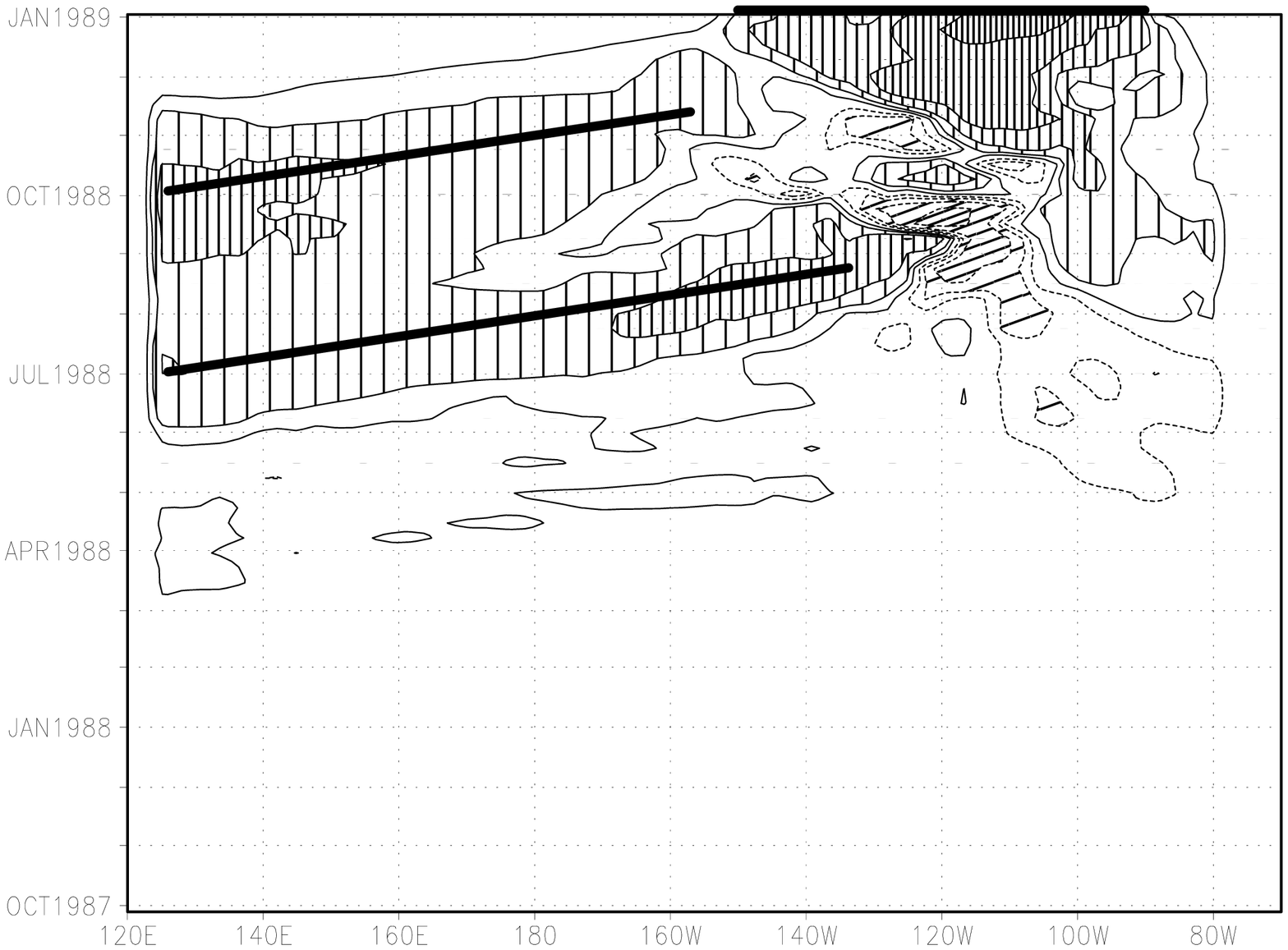,angle=-90,width=\figwidth}}%
\makebox[0pt][l]{\Large\ e}%
}\\
\mbox{%
\makebox[0pt][r]{\Large c }%
\raisebox{3ex}%
{\psfig{file=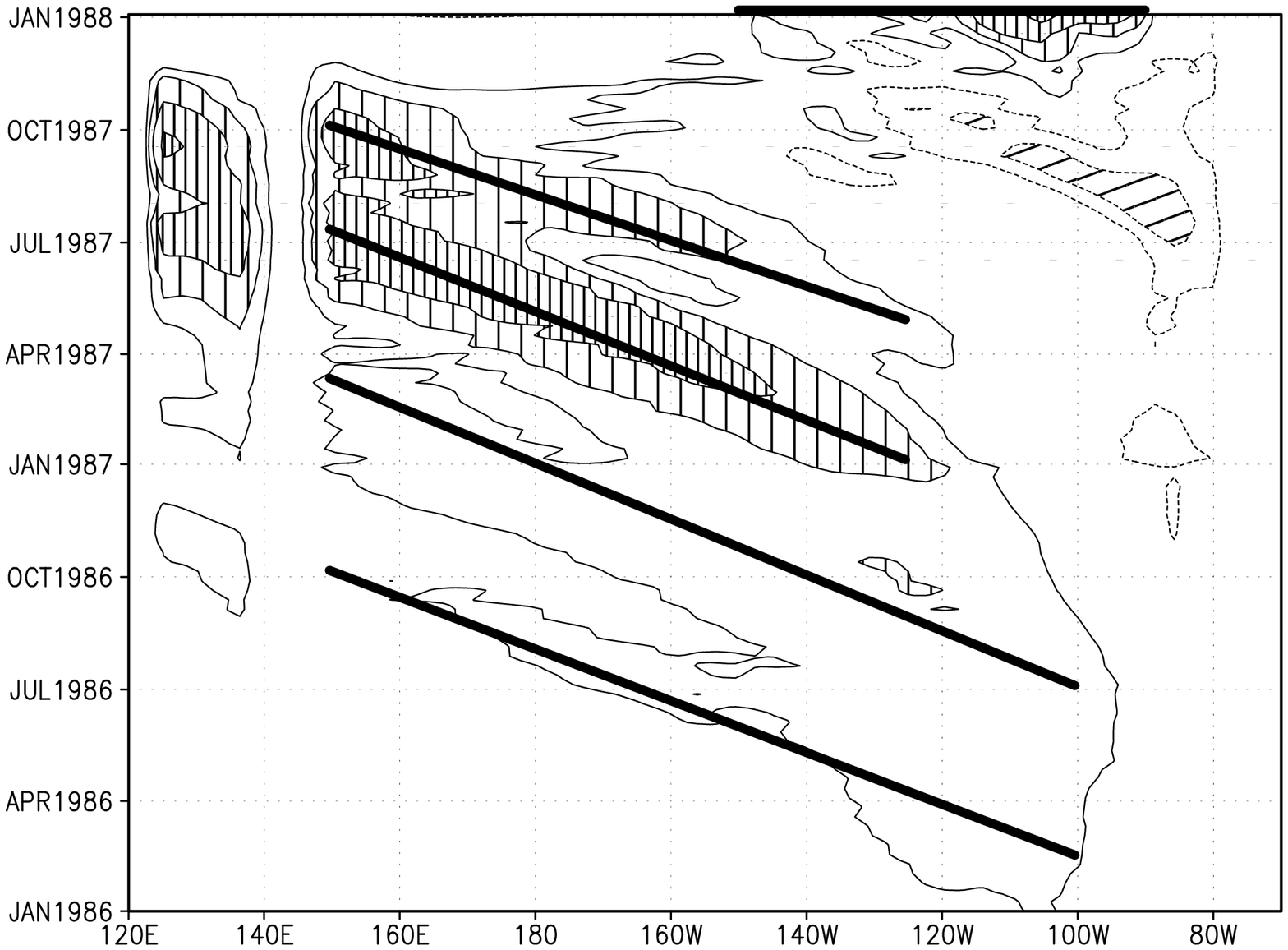,angle=-90,width=\figwidth}}%
\hspace{0.5mm}
\raisebox{3ex}%
{\psfig{file=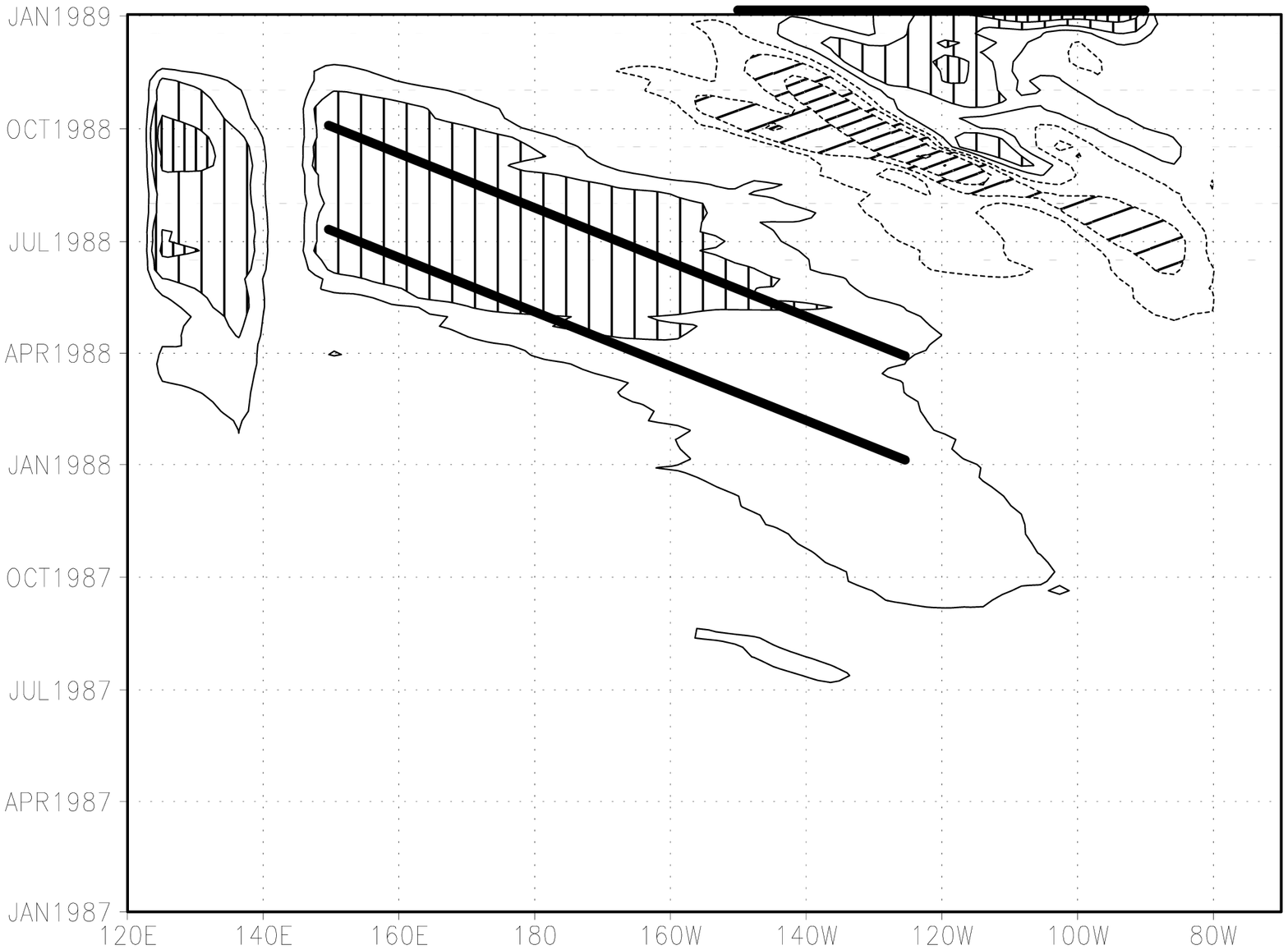,angle=-90,width=\figwidth}}%
\makebox[0pt][l]{\Large\ f}%
}\\
\psfig{file=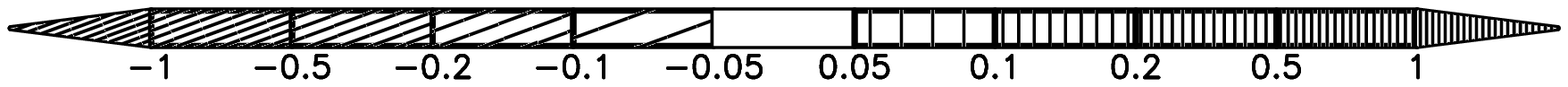,angle=-90,width=\figwidth}
\end{center}
\caption{The sensitivity of the NINO3 index to sea level changes 
in December 1987 
$\partial N_3/\partial\zeta$ in $\mathrm{K}/\mathrm{m}/\mathrm{sr}$ at
5\dg N (a), the equator (b) and 5\dg S (c), averaged over 5\dg\
latitude bands; the same one year later, to December 1988 (d,e,f).}
\label{fig:dz87st}
\end{figure}

The reflection pattern off the western coasts is clearly visible in
the space-time diagrams of 5\dg\ latitude bands at 5\dg N, the equator
and 5\dg S in Fig.~\ref{fig:dz87st}.  In 1987 the adjoint Kelvin wave,
travelling at about $2\:\mathrm{ms^{-1}}$ (solid line in Fig.\
\ref{fig:dz87st}b) reaches the coast of New Guinea in the first week
of October, generating a Rossby wave at 5\dg S that picks up a
northern component some time later.  The main reflection at 120\dg E
occurs at the end of September.  In 1988 the adjoint Kelvin wave is
both faster and weaker, arriving at the eastern boundary one week
earlier.  East of 150\dg W there are also the adjoint Rossby waves
generated by the edges of the NINO3 region, and adjoint Rossby waves
with opposite sign caused by the adjoint Kelvin wave leaving the
NINO3 signal region.  They reach the eastern coast within a few
months.  These waves are much stronger in the cold conditions of 1988.
In HOPE the NINO3 region is characterized by intense currents and
changes in thermocline depth at this time, so that linear theory is
not applicable.

A second adjoint Kelvin wave is present in the western and central
Pacific in August and July.  It seems to be generated from three
sources: partial reflections at the eastern boundary (especially in
the 1988 experiment with strong activity in the eastern Pacific), a
partial reflection of the first adjoint Rossby wave at the edge of the
warm pool (mainly in 1987), and finally a reinforcement near the
western boundary by a reflection off the western coast of New Guinea
by the Rossby wave reflected off the Philippines.  The arrival of this
second Kelvin wave generates a third adjoint Rossby wave during July.
Similar effects seem to produce two more adjoint Kelvin and Rossby
waves in February and November the year before respectively.

These reflected Rossby waves merge into a structure with an effective
group velocity that is significantly lower than the phase velocity of
the individual $n=1$ Rossby waves, indicated by the dashed lines in
Fig.~\ref{fig:dz87st}a,c with speed $0.6\:\mathrm{ms^{-1}}$.  This
behaviour is very similar to the properties of the observed Kelvin and
Rossby waves that constitute ENSO oscillations
\citep{ChaoPhilanderStructure,KesslerForcing}.

These reflection patterns are compatible with observations from sea
level measurements near the eastern and western boundaries
\citep{BoulangerReflection}.  Some of the discrepancies at the western
coast reported in this paper may be explained by the multiple
reflections observed in our model.

%  #] sea level:
%  #[ wave speeds:

\subsection{Wave speeds}
\label{sec:speeds}

\setlength{\figwidth}{0.5\textwidth}
\addtolength{\figwidth}{-7mm}
\begin{figure}[tbp]
\begin{center}
\mbox{\small
\shortstack{\psfig{file=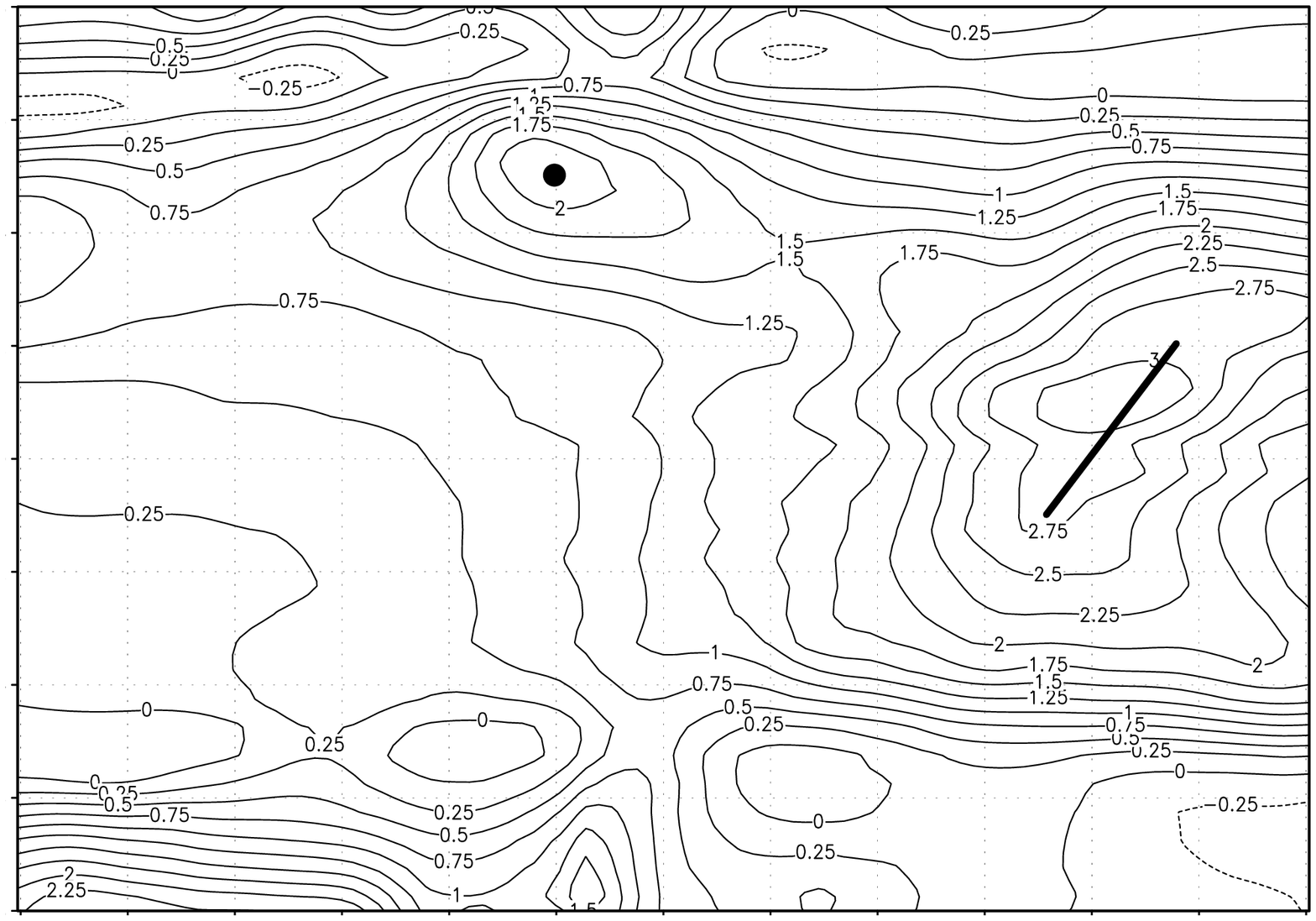,angle=-90,width=\figwidth}\\
\unitlength\figwidth
\begin{picture}(0,0)(-.01,0)
%\put(-.5,0.045){\makebox(0,0)[r]{$-8^\circ$}}
%\put(-.5,0.208){\makebox(0,0)[r]{$-4^\circ$}}
\put(-.5,0.045){\makebox(0,0)[r]{8\dg S}}
\put(-.5,0.208){\makebox(0,0)[r]{4\dg S}}
\put(-.5,0.370){\makebox(0,0)[r]{EQ}}
\put(-.5,0.538){\makebox(0,0)[r]{4\dg N}}
\put(-.5,0.705){\makebox(0,0)[r]{8\dg N}}
%\put(-.5,0.538){\makebox(0,0)[r]{$+4^\circ$}}
%\put(-.5,0.705){\makebox(0,0)[r]{$+8^\circ$}}
\put(0.0 ,0.0){\makebox(0,0){$0$}}
\put(0.16,0.0){\makebox(0,0){$1$}}
\put(0.32,0.0){\makebox(0,0){$2$}}
\put(0.48,0.0){\makebox(0,0){$3$}}
\put(-.16,0.0){\makebox(0,0){$-1$}}
\put(-.32,0.0){\makebox(0,0){$-2$}}
\put(-.48,0.0){\makebox(0,0){$-3$}}
\put(0.48,-.06){\makebox(0,0)[r]{c [m/s]}}
\end{picture}\\[5pt]
\Large\vphantom{b}a\quad}
%\hspace{2mm}
\shortstack{\psfig{file=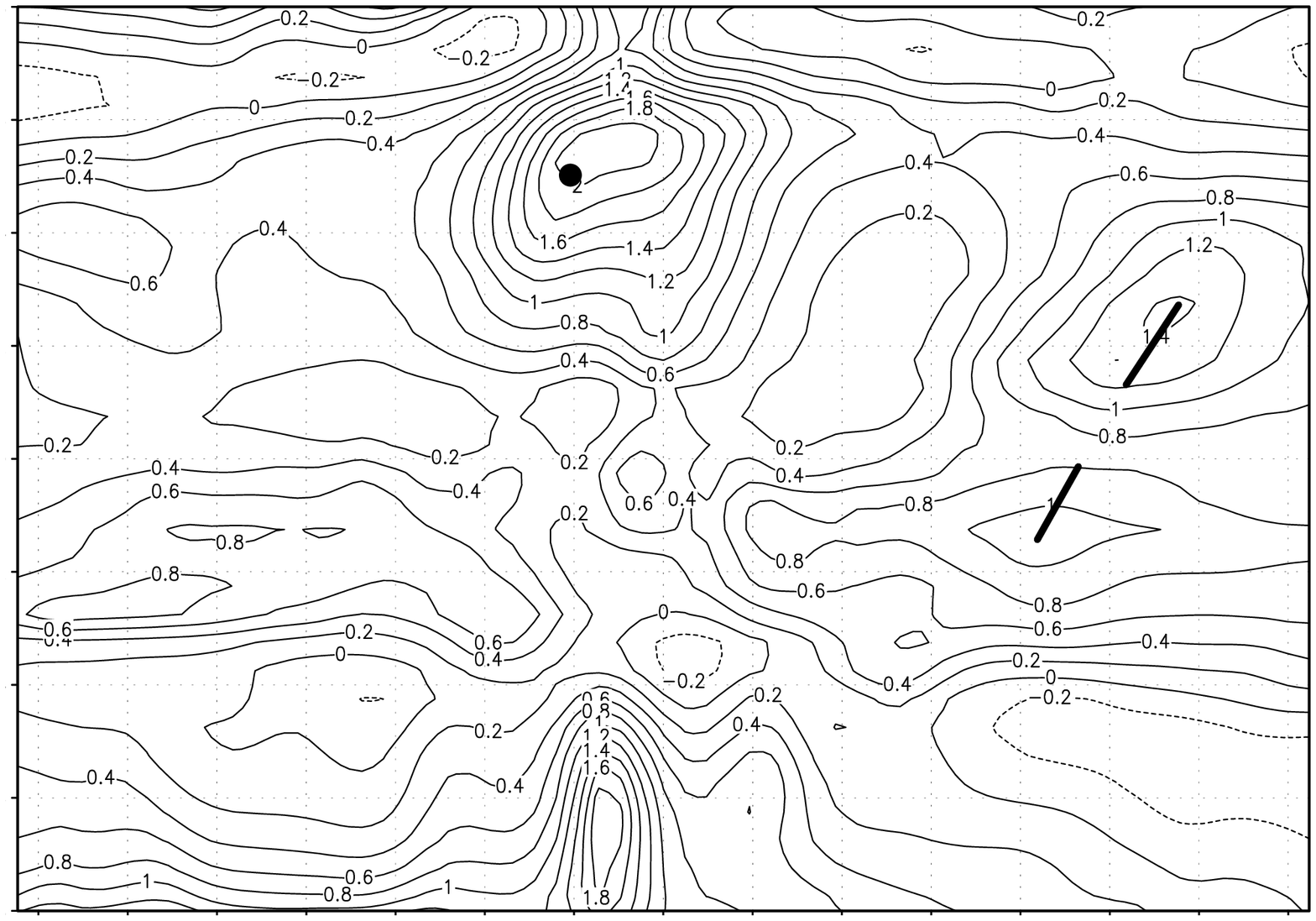,angle=-90,width=\figwidth}\\
\unitlength\figwidth
\begin{picture}(0,0)(-.01,0)
%\put(-.5,0.045){\makebox(0,0)[r]{$-8^\circ$}}
%\put(-.5,0.208){\makebox(0,0)[r]{$-4^\circ$}}
\put(.5,0.045){\makebox(0,0)[l]{8\dg S}}
\put(.5,0.208){\makebox(0,0)[l]{4\dg S}}
\put(.5,0.370){\makebox(0,0)[l]{EQ}}
\put(.5,0.538){\makebox(0,0)[l]{4\dg N}}
\put(.5,0.705){\makebox(0,0)[l]{8\dg N}}
%\put(-.5,0.538){\makebox(0,0)[r]{$+4^\circ$}}
%\put(-.5,0.705){\makebox(0,0)[r]{$+8^\circ$}}
\put(0.0 ,0.0){\makebox(0,0){$0$}}
\put(0.13,0.0){\makebox(0,0){$1$}}
\put(0.27,0.0){\makebox(0,0){$2$}}
\put(0.40,0.0){\makebox(0,0){$3$}}
\put(-.13,0.0){\makebox(0,0){$-1$}}
\put(-.27,0.0){\makebox(0,0){$-2$}}
\put(-.40,0.0){\makebox(0,0){$-3$}}
\put(0.48,-.06){\makebox(0,0)[r]{c [m/s]}}
\end{picture}\\[5pt]
\Large\vphantom{b}b\quad}
}\\[10pt]
\mbox{\small
\shortstack{\psfig{file=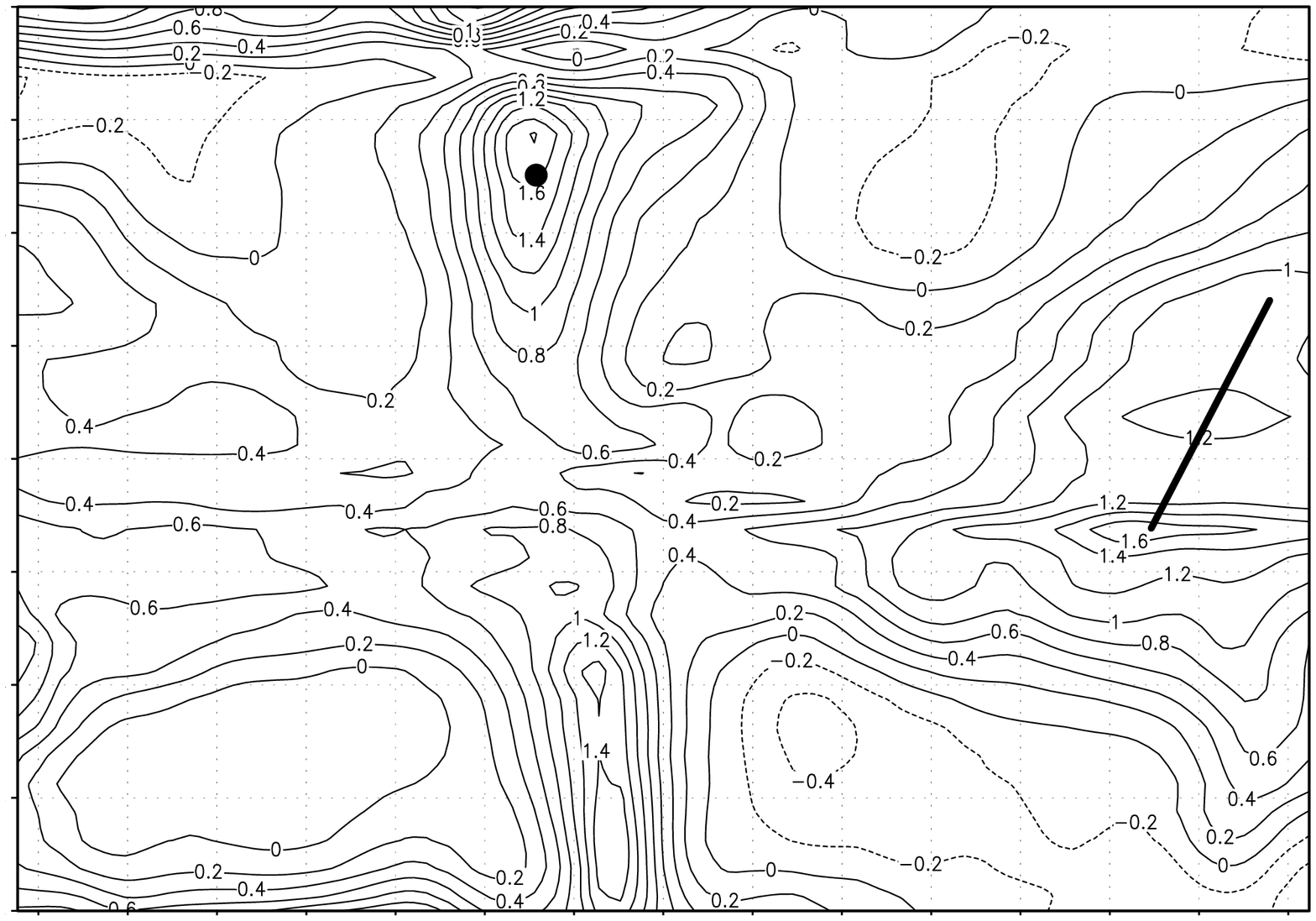,angle=-90,width=\figwidth}\\
\unitlength\figwidth
\begin{picture}(0,0)(-.01,0)
%\put(-.5,0.045){\makebox(0,0)[r]{$-8^\circ$}}
%\put(-.5,0.208){\makebox(0,0)[r]{$-4^\circ$}}
\put(-.5,0.045){\makebox(0,0)[r]{8\dg S}}
\put(-.5,0.208){\makebox(0,0)[r]{4\dg S}}
\put(-.5,0.370){\makebox(0,0)[r]{EQ}}
\put(-.5,0.538){\makebox(0,0)[r]{4\dg N}}
\put(-.5,0.705){\makebox(0,0)[r]{8\dg N}}
%\put(-.5,0.538){\makebox(0,0)[r]{$+4^\circ$}}
%\put(-.5,0.705){\makebox(0,0)[r]{$+8^\circ$}}
\put(0.0 ,0.0){\makebox(0,0){$0$}}
\put(0.13,0.0){\makebox(0,0){$1$}}
\put(0.27,0.0){\makebox(0,0){$2$}}
\put(0.40,0.0){\makebox(0,0){$3$}}
\put(-.13,0.0){\makebox(0,0){$-1$}}
\put(-.27,0.0){\makebox(0,0){$-2$}}
\put(-.40,0.0){\makebox(0,0){$-3$}}
\put(0.48,-.06){\makebox(0,0)[r]{c [m/s]}}
\end{picture}\\[5pt]
\Large\vphantom{d}c\quad}
\shortstack{\psfig{file=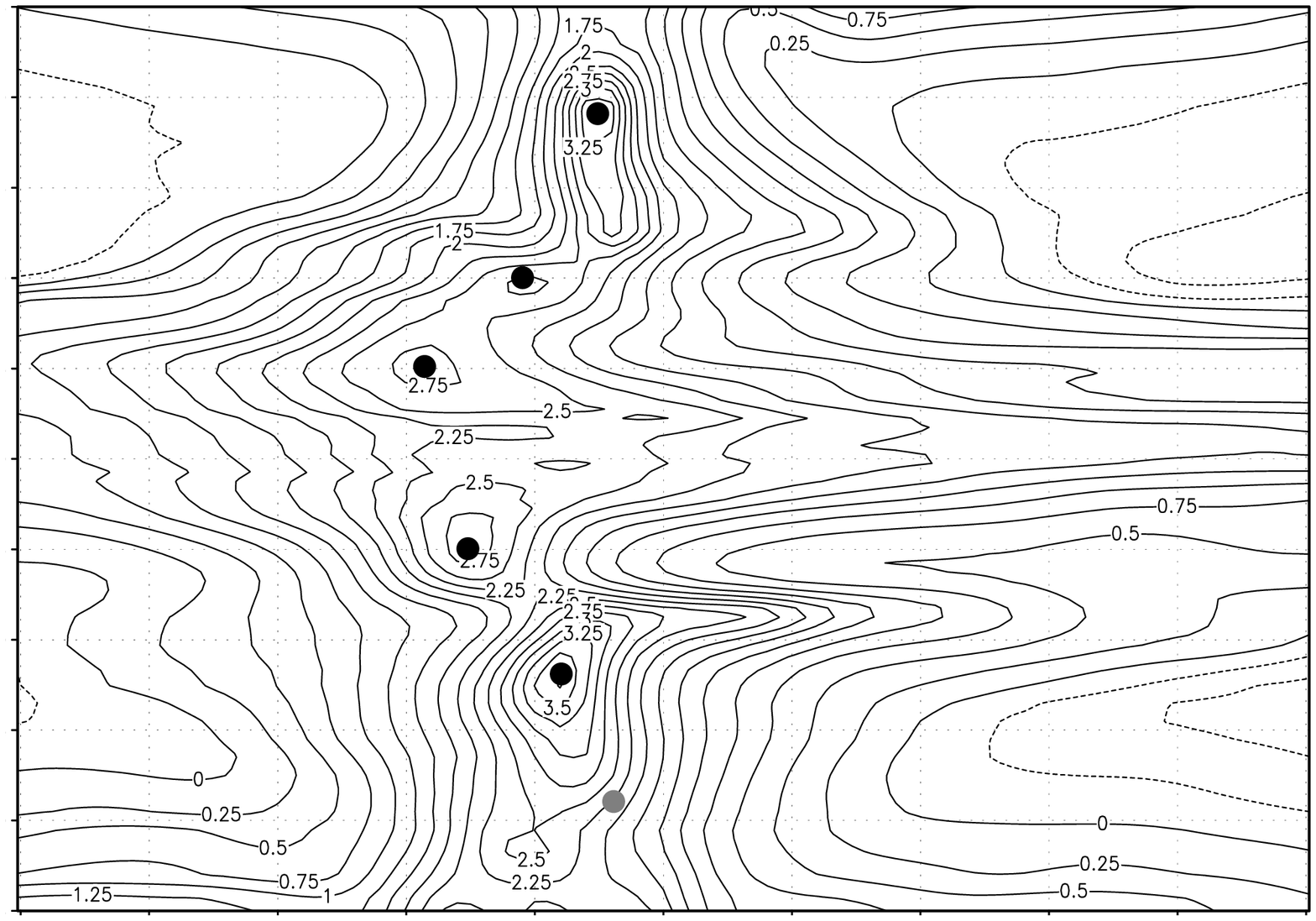,angle=-90,width=\figwidth}\\
\unitlength\figwidth
\begin{picture}(0,0)(-.01,0)
%\put(0.5,0.045){\makebox(0,0)[l]{$-25^\circ$}}
%\put(0.5,0.177){\makebox(0,0)[l]{$-10^\circ$}}
\put(0.5,0.045){\makebox(0,0)[l]{25\dg S}}
\put(0.5,0.242){\makebox(0,0)[l]{10\dg S}}
\put(0.5,0.370){\makebox(0,0)[l]{EQ}}
\put(0.5,0.498){\makebox(0,0)[l]{10\dg N}}
\put(0.5,0.705){\makebox(0,0)[l]{25\dg N}}
%\put(0.5,0.567){\makebox(0,0)[l]{$+10^\circ$}}
%\put(0.5,0.705){\makebox(0,0)[l]{$+25^\circ$}}
\put(0.0 ,0.0){\makebox(0,0){$0$}}
\put(0.19,0.0){\makebox(0,0){$0.6$}}
\put(0.48,0.0){\makebox(0,0){$1.5$}}
\put(-.19,0.0){\makebox(0,0){$-0.6$}}
\put(-.48,0.0){\makebox(0,0){$-1.5$}}
\put(0.48,-.06){\makebox(0,0)[r]{c [m/s]}}
\end{picture}\\[5pt]
\Large d\quad}
}
\end{center}
\caption{The 1-month lag correlations $p$ as a function of latitude
and speed in m/s of $\partial N_3/\partial\zeta$ in the last month of
the Dec 1987 run (a), the Oct 1987 run (b), the Dec 1988 run (c), and
the 2-month lag correlation in January--April 1987 (d).  The contours
refer to $-\log(1-p)$, the black dots and lines refer to the values
mentioned in the text.}
\label{fig:dz87corr}
\end{figure}

To check the interpretation of Kelvin and Rossby waves,
we have measured the speeds of the structures visible in the adjoint run
and compared the results to wave speeds calculated from the density 
structure of the model.

The speed was estimated by
taking 1- and 2-month lag correlations for each latitude.  From
Fig.~\ref{fig:dz87corr} one can read off the following speeds for the
waves.  In the last month of the run leading to the NINO3 index at the
end of December 1987, the equatorial Kelvin wave has a speed that
varies between $1.8\:\mathrm{ms^{-1}}$ at 1\dg S{},
$2.0\:\mathrm{ms^{-1}}$ on the equator and $2.4\:\mathrm{ms^{-1}}$ 
at 2\dg N (Fig.~\ref{fig:dz87corr}a).  In the experiment leading to
October 1987 we find speeds of 2.1 to $2.9\:\mathrm{ms^{-1}}$
(Fig.~\ref{fig:dz87corr}b), and at the end of 1988 they are around
$3\:\mathrm{ms^{-1}}$ (Fig.~\ref{fig:dz87corr}c). 

The Rossby waves during these times in the Eastern Pacific have speeds
at 5\dg N of $0.5\:\mathrm{ms^{-1}}$, $0.5\:\mathrm{ms^{-1}}$ and
$0.7\:\mathrm{ms^{-1}}$, this is somewhat slower 
than 1/3 the speed of the Kelvin waves due to the shallower
thermocline.  The speeds of reflected adjoint Rossby waves in the
central Pacific have been read off Fig.~\ref{fig:dz87corr}d.  The
northern branches of the Rossby waves move with
$-0.55\:\mathrm{ms^{-1}}$ (5\dg N), $-0.34\:\mathrm{ms^{-1}}$ (10\dg
N) and $-0.15\:\mathrm{ms^{-1}}$ (19\dg N).  The southern lobes 
are slower and further from the equator: $-0.45\:\mathrm{ms^{-1}}$
(5\dg S), $-0.23\:\mathrm{ms^{-1}}$ at 12\dg S and
$-0.12\:\mathrm{ms^{-1}}$ (19\dg S, only visible in 1986).  These
speeds are roughly the same in the other two experiments, except that
the $n=1$ mode is often poorly visible.

The observed adjoint Kelvin waves therefore seem to be a mixture of
the first baroclinic mode, with speeds around $3\:\mathrm{ms^{-1}}$,
and the second baroclinic mode, with a speed of around
$2\:\mathrm{ms^{-1}}$.  This is in agreement with other theoretical
studies \citep[e.g.,][]{GieseResponse}.
% KesslerForcing: 2.4 \pm 0.3 from TOGA-TAO, constant
% CheltonRossby: about 2.7 from TOPEX-POSEIDON, slower west of 140W.
% BoulangerMenkes: 2.9\pm0.9 from TOPEX/Poseidon, quote 2.3\pm0.3 TOGA-TAO
Observations of Kelvin waves give a single mode with a speed of
$2.4\pm0.3\:\mathrm{ms^{-1}}$ from TOGA-TAO subsurface temperatures 
\citep{KesslerForcing} and $2.9\pm0.9\:\mathrm{ms^{-1}}$ from
TOPEX/POSEIDON altimetry data \citep{BoulangerMenkes,CheltonRossby}.  
Also, in these observations the  
shallowing of the thermocline towards the east is compensated by a
stronger density contrast across the thermocline, leading to an almost
constant speed up to 110\dg W{}.  In contrast, in HOPE we observe lower
speeds east of 160\dg W (see Fig.~\ref{fig:dz87st}), as the
thermocline is not sharp enough due to discretization effects.

\begin{figure}[tbp]
\setlength{\figwidth}{0.32\textwidth}
\begin{center}
\shortstack{%
%\Large\vphantom{b}a\\
\psfig{file=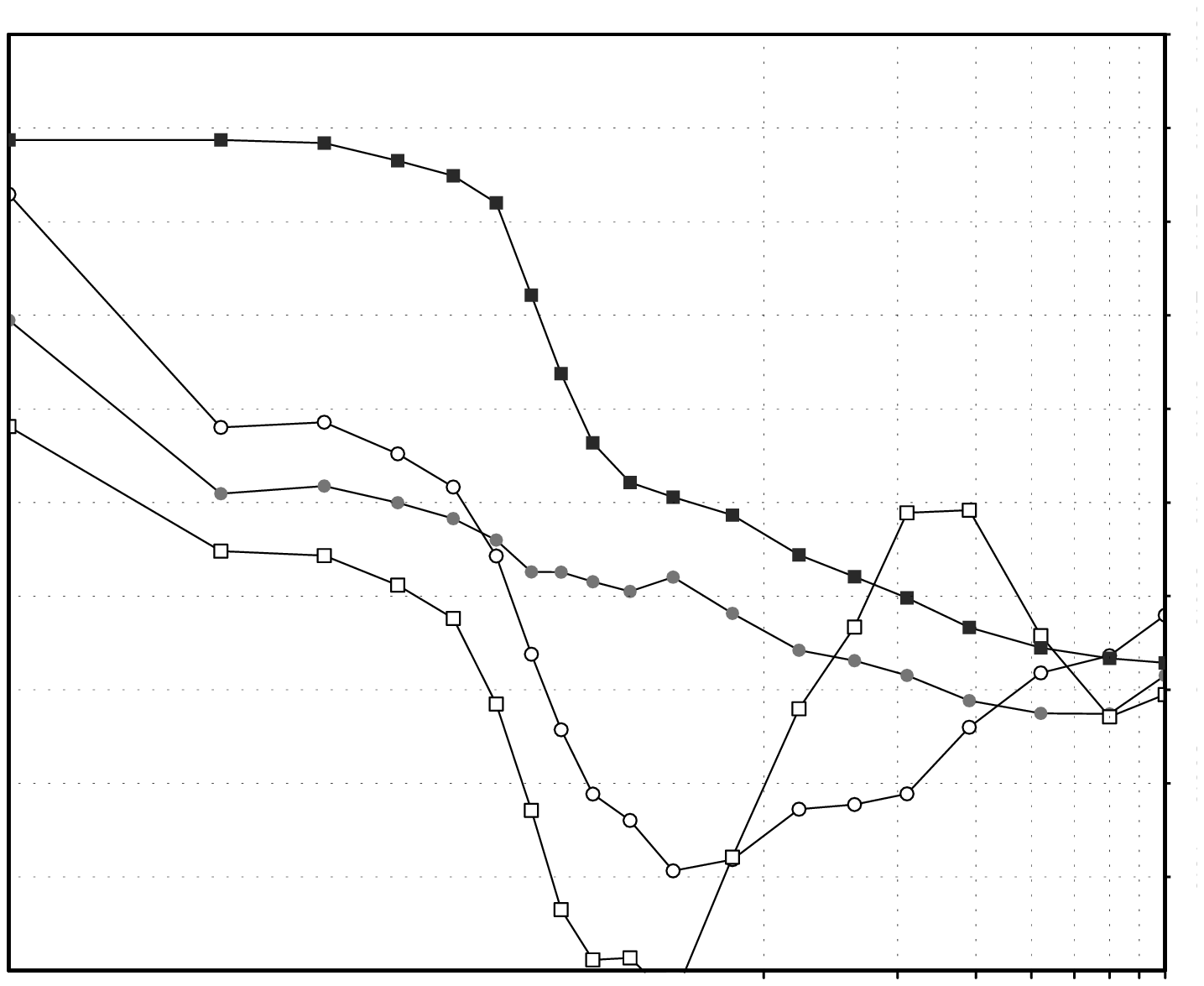,angle=-90,width=\figwidth}\\
\unitlength\figwidth
\begin{picture}(0,0)(0,0)
\put(-0.63,1.14){\makebox(0,0)[tr]{\shortstack{$z$\\\relax[m]}}}
\put(0.30,0.60){\makebox(0,0){\large$T$}}
\put(0.09,1.04){\makebox(0,0){\large1}}
\put(0.25,1.04){\makebox(0,0){\large2}}
\put(-.05,1.04){\makebox(0,0){\large3}}
\put(-0.48,1.14){\makebox(0,0)[r]{10}}
\put(-0.48,0.85){\makebox(0,0)[r]{50}}
\put(-0.48,0.725){\makebox(0,0)[r]{100}}
\put(-0.48,0.60){\makebox(0,0)[r]{200}}
\put(-0.48,0.435){\makebox(0,0)[r]{500}}
\put(-0.48,0.31){\makebox(0,0)[r]{1000}}
\put(-0.48,0.18){\makebox(0,0)[r]{2000}}
\put(-0.48,0.05){\makebox(0,0)[r]{4000}}
\put(-.15,0.){\makebox(0,0){0}}
\end{picture}
\\\Large\vphantom{b}a
}
\shortstack{%
%\Large b\\
\psfig{file=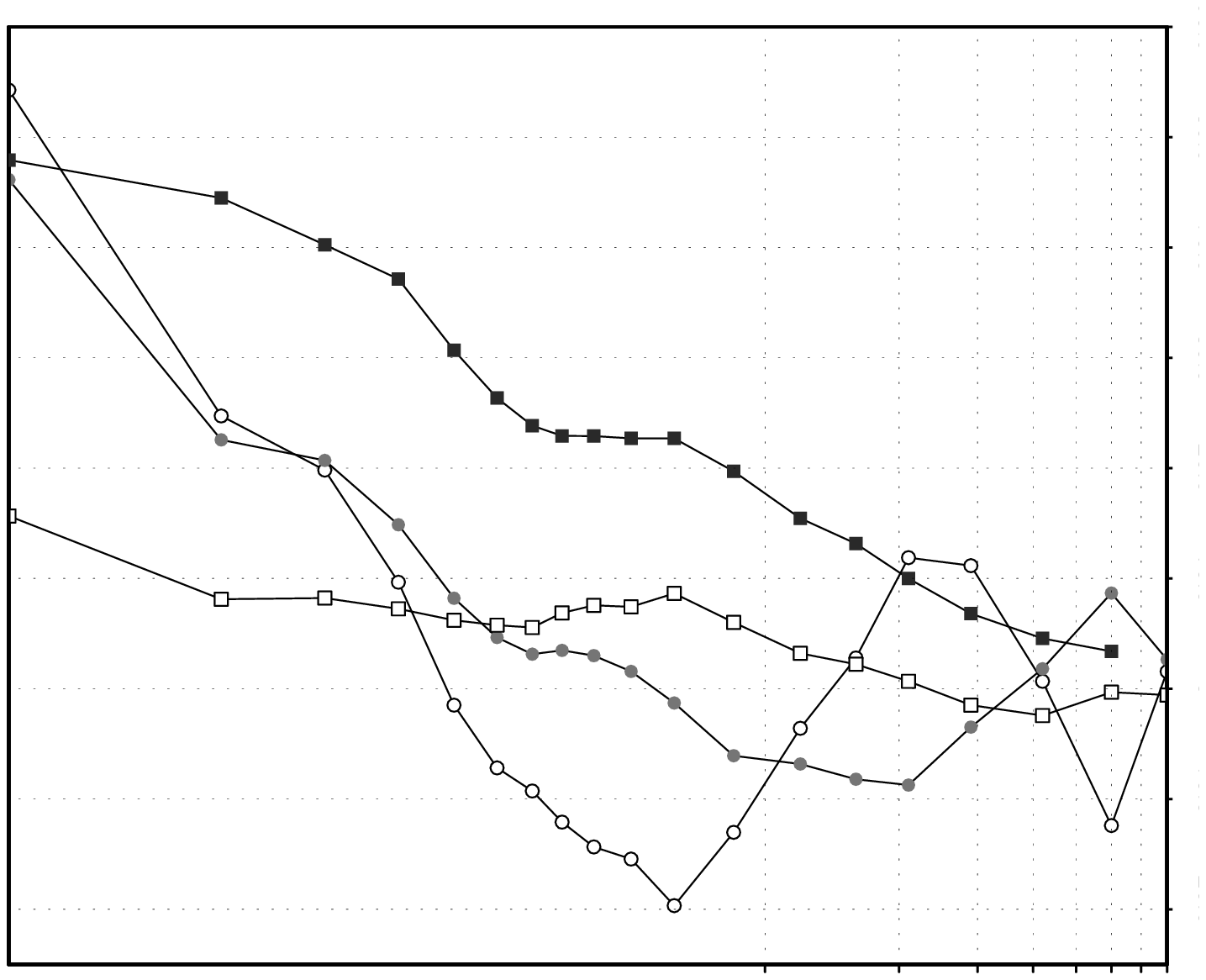,angle=-90,width=\figwidth}\\
\unitlength\figwidth
\begin{picture}(0,0)(0,0)
\put(0.15,0.60){\makebox(0,0){\large$T$}}
\put(-.10,1.04){\makebox(0,0){\large1}}
\put(-.19,0.60){\makebox(0,0){\large2}}
\put(-.24,0.75){\makebox(0,0){\large3}}
\put(-.16,0.){\makebox(0,0){0}}
\end{picture}
\\\Large b
}
\\[10mm]
\shortstack{%
\psfig{file=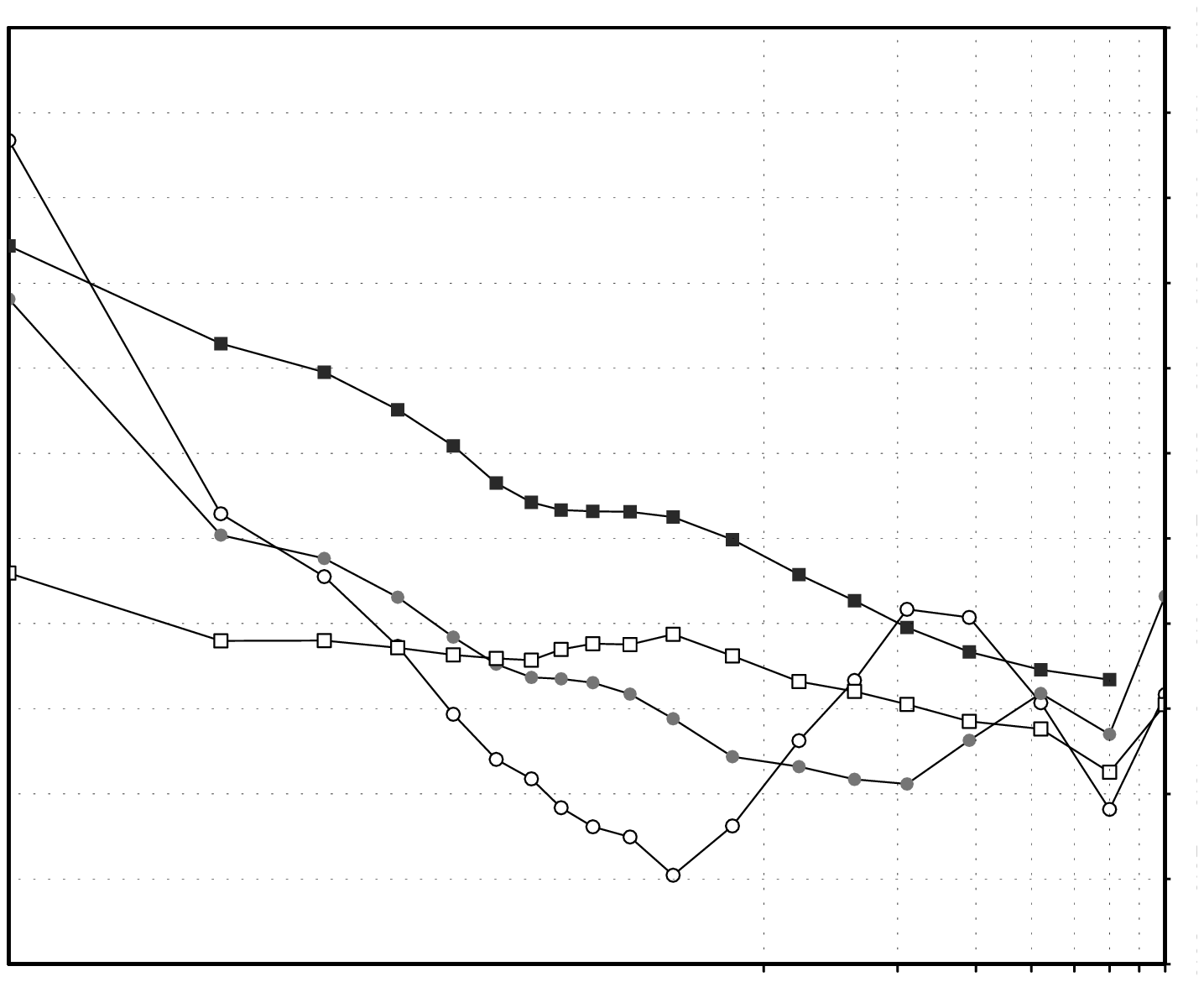,angle=-90,width=\figwidth}\\
\unitlength\figwidth
\begin{picture}(0,0)(0,0)
\put(-0.63,1.14){\makebox(0,0)[tr]{\shortstack{$z$\\\relax[m]}}}
\put(0.23,0.90){\makebox(0,0){\large$T$}}
\put(-.12,1.10){\makebox(0,0){\large1}}
\put(0.06,1.10){\makebox(0,0){\large2}}
\put(-.26,0.75){\makebox(0,0){\large3}}
\put(-0.48,1.14){\makebox(0,0)[r]{10}}
\put(-0.48,0.85){\makebox(0,0)[r]{50}}
\put(-0.48,0.725){\makebox(0,0)[r]{100}}
\put(-0.48,0.60){\makebox(0,0)[r]{200}}
\put(-0.48,0.435){\makebox(0,0)[r]{500}}
\put(-0.48,0.31){\makebox(0,0)[r]{1000}}
\put(-0.48,0.18){\makebox(0,0)[r]{2000}}
\put(-0.48,0.05){\makebox(0,0)[r]{4000}}
\put(-.175,0.){\makebox(0,0){0}}
\end{picture}\\
\Large\vphantom{d}c\quad}
\shortstack{%
\psfig{file=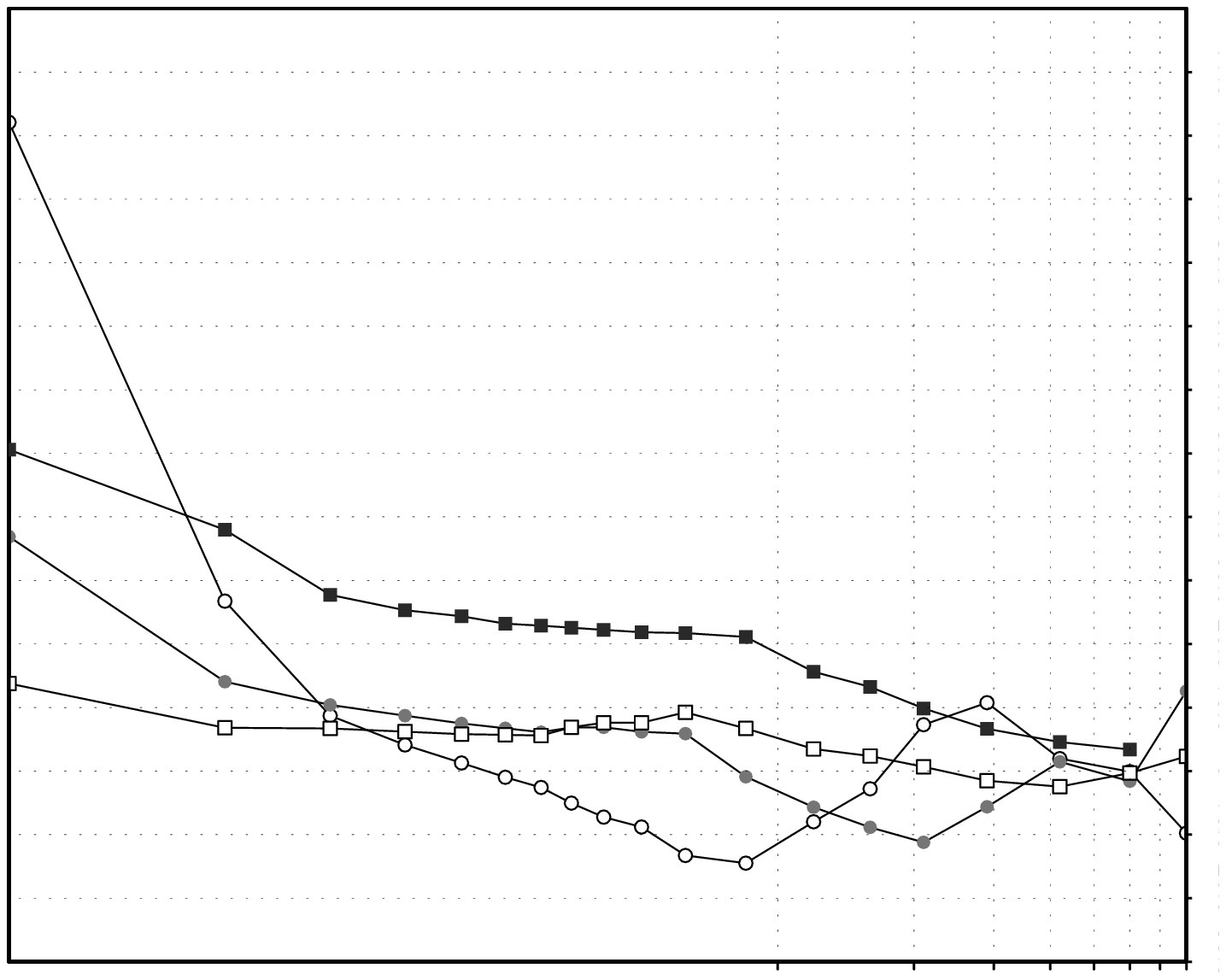,angle=-90,width=\figwidth}\\
\unitlength\figwidth
\begin{picture}(0,0)(0,0)
\put(-.03,0.80){\makebox(0,0){\large$T$}}
\put(-.03,1.10){\makebox(0,0){\large1}}
\put(-.22,1.10){\makebox(0,0){\large2}}
\put(0.35,1.10){\makebox(0,0){\large3}}
\put(-.23,0.){\makebox(0,0){0}}
\end{picture}\\
\Large d\quad}
\end{center}
\caption{Profiles of the first three baroclinic modes and
temperature at the equator, 180\dg\ (a) and 100\dg W (b) in Dec 
1987, and at 100\dg W in Oct 1987 (c) and Dec 1988 (d) in the HOPE
model.}
\label{fig:modes}
\end{figure}

These speeds can also be computed from the density profiles%
\footnote{This computation depends somewhat on the mixing scheme, and
is done for mixing parameters proportional to $1/N^2$, which differs
from the Richardson-number based scheme with a mixed layer of the HOPE
model.  The difference will be mainly in the absence of a mixed layer
in the computed modes.} 
\citep[see, e.g.,][]{GillBook,PhilanderBook}, using a rigid
lid to eliminate the barotropic mode.  The first three modes of the
horizontal velocity are shown in Fig.~\ref{fig:modes}.  The first
baroclinic mode has a zero around 1500 m and a Kelvin wave speed of
\mbox{2.6--$3.3\:\mathrm{ms^{-1}}$}, in agreement with the profiles
discussed 
in \citet{PhilanderBook,GieseResponse}.  This mode seems to be
associated with the rise in salinity around that depth in the Pacific,
which is absent in the Atlantic.  The next modes have their
first zeroes near the thermocline, at around 100--150 m in the
equatorial zone, substantially deeper elsewhere.  The speed fields are
shown in Fig.~\ref{fig:modespeed}.
% Added section comparing with Picaut and Sombardier
These maps can be compared with the ones obtained from much
higher-resolution data in \citet{PicautSombardier1993}.  They find
speeds that are 20--25\% lower than the ones shown here, whereas
\citet{Kessler9193} use a speed measured at one point (150\dg--158\dg
W) that is about 10\% lower than the value we find there.  We assume
the differences are due to the coarse vertical grid we employed, which
is just the 20-level HOPE grid.

With a well-developed thermocline, as in the western and central
Pacific, and in the east during the December 1987 run
(Fig.~\ref{fig:modes}a,b), a Kelvin wave in the first baroclinic mode
mode does not influence the NINO3 index of surface temperatures as
much as one in the second mode.  The second mode changes the
thermocline depth, which most effectively affects the surface
temperature in the upwelling regions (see section
\ref{sec:sealevel}).  Also, the eigenfunction of the first mode has a
smaller amplitude at the surface so that a zonal stress at the surface
does not excite this mode as much as the second one in the derivatives
to wind stress (section \ref{sec:windstress}).  Finally, the 
Richardson-number dependent mixing
scheme, Eqs~(\ref{eq:AV}--\ref{eq:DV}), tends to decouple the
ocean below the thermocline, also preferring the second mode.

However, when the thermocline is absent
in the cold tongue, as our model generates in December 1988
(Fig.~\ref{fig:modes}d), the \emph{first} mode is most effective in
changing the SST in the NINO3 area, although the adjoint Kelvin wave is 
weaker in this case (see Figs~\ref{fig:dz87st}b,e).  The October 1987 run, 
with a weak thermocline, excites adjoint Kelvin waves in both modes.

The speed of the second mode Kelvin wave in the December 1987 run 
(Fig.~\ref{fig:dz87corr}a) is slightly higher than the 1.8 to 
$2.1\:\mathrm{ms^{-1}}$ in the area west of the NINO3 box in 
Fig.~\ref{fig:modespeed}b.  In 1988 the speed corresponds well
with the little over $3\:\mathrm{ms^{-1}}$ read off from
Fig.~\ref{fig:modespeed}a.  The $n=1$ reflected adjoint Rossby waves
correspond reasonably well with 1/3 the zonally averaged speed of the
second mode ($0.57\:\mathrm{ms^{-1}}$ at 5\dg S,
$0.62\:\mathrm{ms^{-1}}$ at 5\dg N); the same holds for $n=2$ (1/5) at
10\dg N ($0.34\:\mathrm{ms^{-1}}$) in all experiments.  Note that this
means that during the reflection off the western boundary energy can
flow from the second into the first baroclinic mode;  the WKB
approximation breaks down when the equivalent depth changes
significantly on scales comparable to the Rossby radius.

The speeds at the other extrema in Fig.~\ref{fig:dz87corr}d,
corresponding to higher-order Rossby modes, are much lower and seem to
resemble more the speed of the third baroclinic mode
($0.21\:\mathrm{ms^{-1}}$ at 11\dg S, $0.15\:\mathrm{ms^{-1}}$ at
19\dg S and $0.17\:\mathrm{ms^{-1}}$ at 15\dg N).  Unfortunately the
derivatives to the horizontal speeds are too noisy to determine their
vertical structure, we can only assume that the second and third modes
can also be mixed up in the shallow water at the western coasts.

\setlength{\figwidth}{\textwidth}
\addtolength{\figwidth}{-2cm}
\begin{figure}[tbp]
\begin{center}
\Large a\raisebox{2ex}%
{\psfig{file=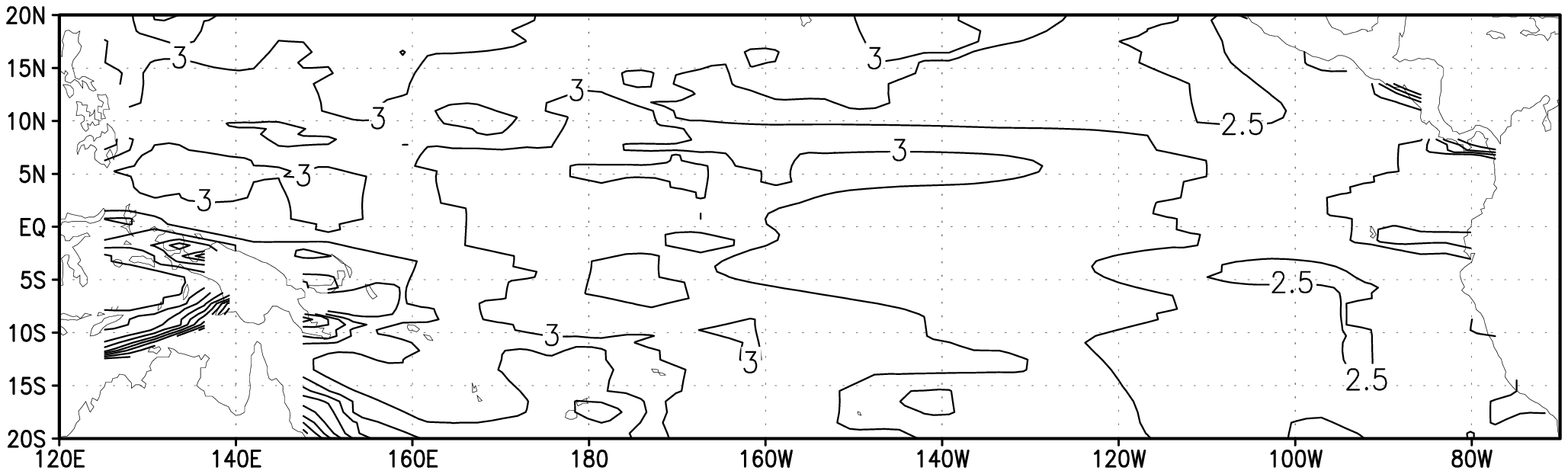,angle=-90,width=\figwidth}}\\
\Large b\raisebox{2ex}%
{\psfig{file=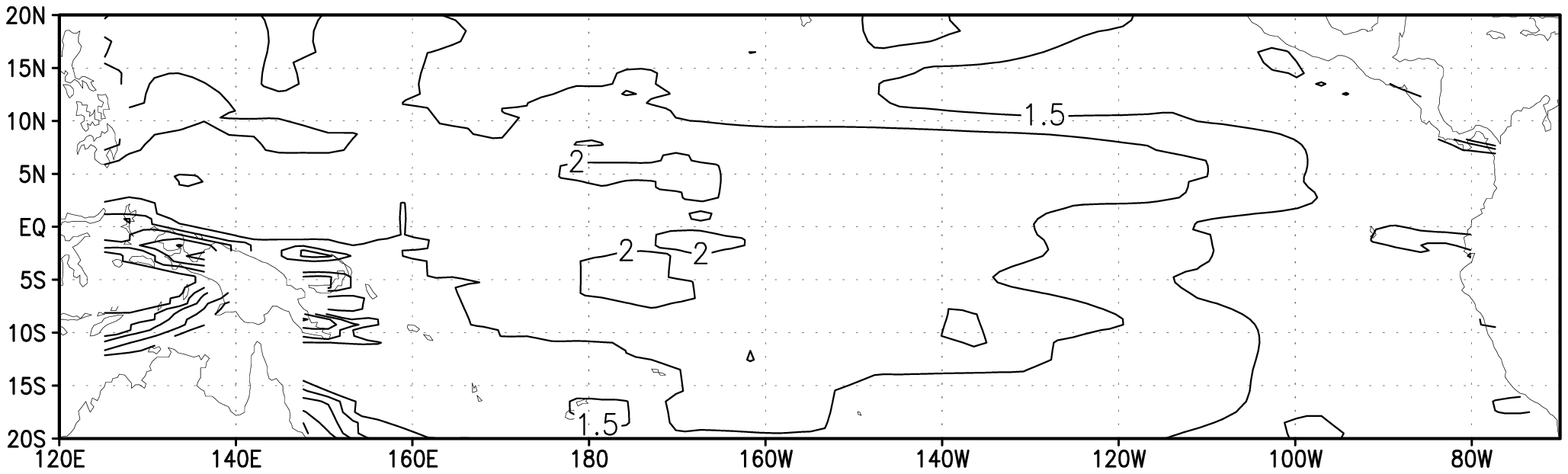,angle=-90,width=\figwidth}}\\
%\Large c\raisebox{2ex}%
%{\psfig{file=speeds3.eps,angle=-90,width=\figwidth}}\\
\end{center}
\caption{The speeds in $\mathrm{ms^{-1}}$ of the first (a) and second (b) 
%and third (c)
baroclinic modes deduced from the density profiles at the end of 1987.}
\label{fig:modespeed}
\end{figure}

%  #] wave speeds: 
%  #[ zonal currents:

\subsection{Zonal currents}
\label{sec:currents}

\setlength{\figwidth}{95mm}
\begin{figure}[htbp]
\begin{center}
\psfig{file=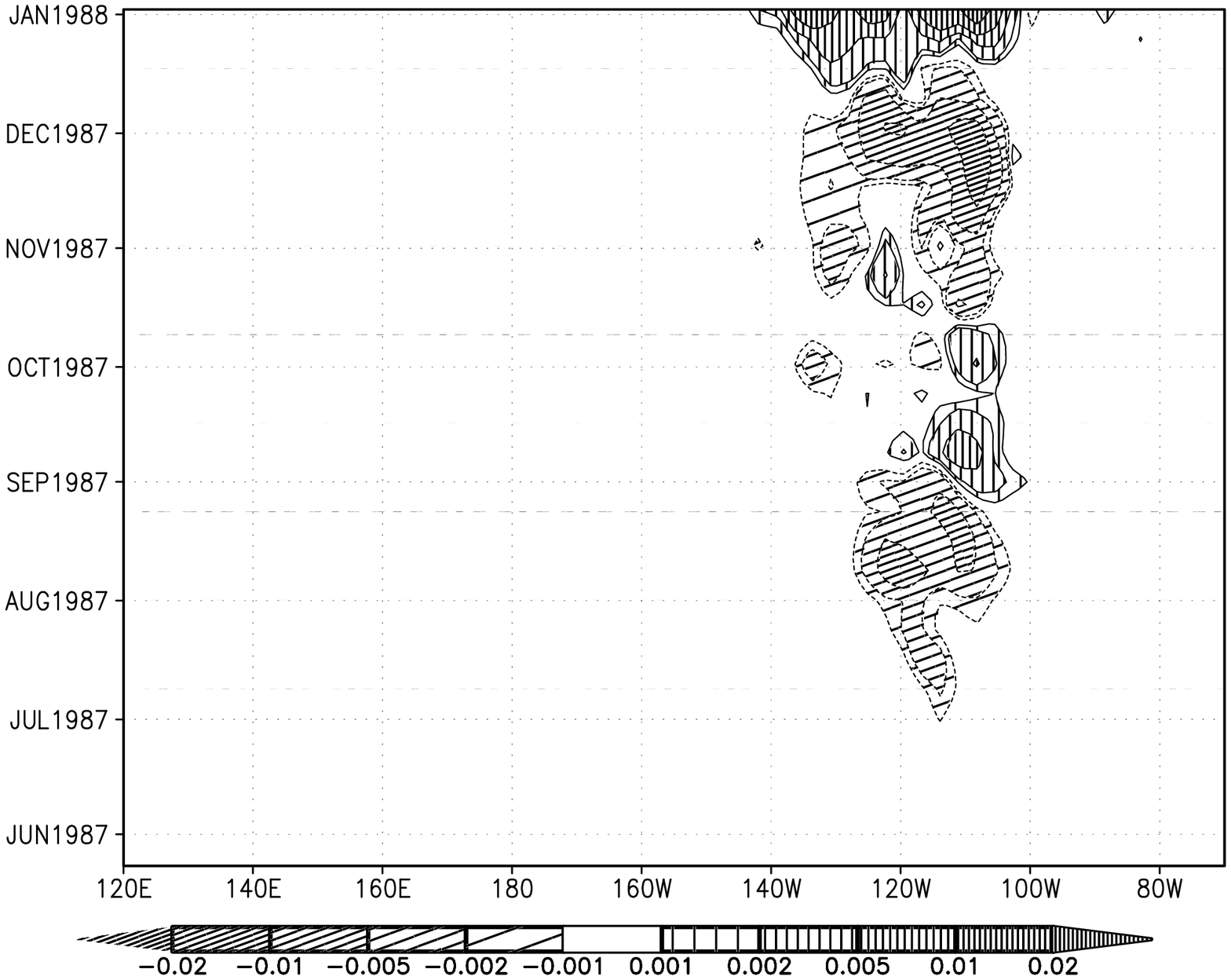,angle=-90,width=\figwidth}\\
\end{center}
\caption{The sensitivity of the NINO3 index to zonal surface currents
$\partial N_3/\partial u_0$ in
$\mathrm{K}/\mathrm{ms^{-1}}/\mathrm{sr}$ averaged over 4\dg S to 4\dg
N.}
\label{fig:du87}
\end{figure}

According to some publications
\citep{PicautDelcroix1995,Picaut1996} zonal currents 
converging at the edge of the warm pool play a major role in ENSO{}.  
As can be seen in Fig.~\ref{fig:du87} this mechanism is not visible in 
the adjoint HOPE model: the NINO3 index only depends significantly on 
surface currents in the measurement region over the last half year.  These 
sensitivities seem to correspond to the surface currents of the adjoint 
Rossby waves visible in Figs~\ref{fig:dz87st}a,c in this area.  The shallow 
thermocline in the cold tongue may increase the effect of changing just the 
surface current.

%  #] zonal currents:
%  #[ wind stress:

%\setlength{\figwidth}{\textwidth}
%\addtolength{\figwidth}{-4.2cm}
\setlength{\figwidth}{95mm}
\begin{figure}[p]
\begin{center}
\Large a \raisebox{2ex}%
{\makebox[0mm][l]%
{\psfig{file=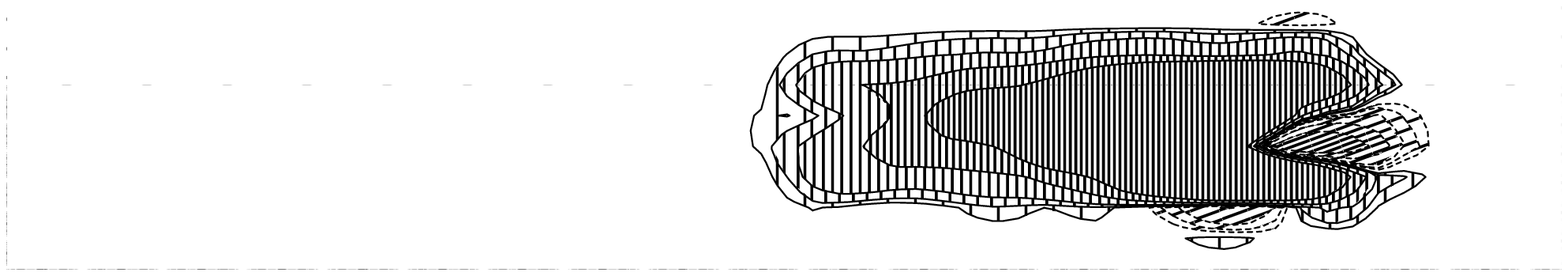,angle=-90,width=\figwidth}}%
\psfig{file=frame.eps,angle=-90,width=\figwidth}}\\
\Large b \raisebox{2ex}%
{\makebox[0mm][l]%
{\psfig{file=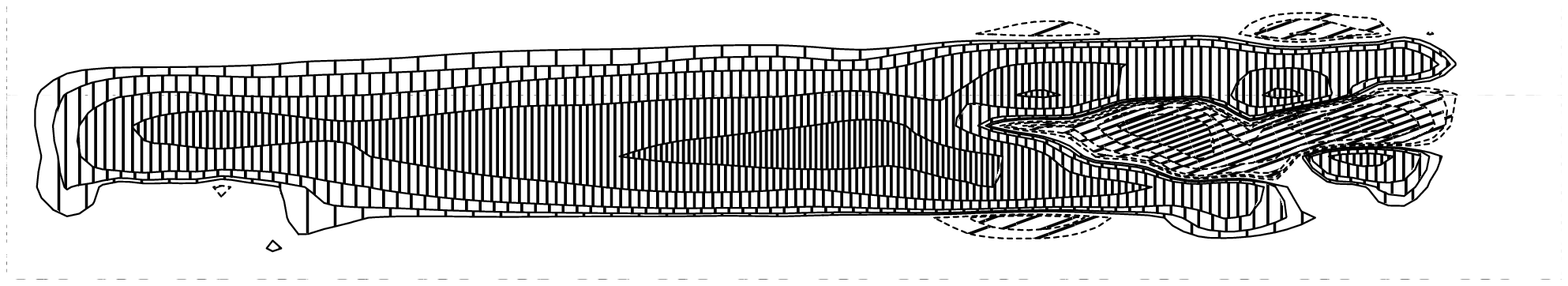,angle=-90,width=\figwidth}}%
\psfig{file=frame.eps,angle=-90,width=\figwidth}}\\
\Large c \raisebox{2ex}%
{\makebox[0mm][l]%
{\psfig{file=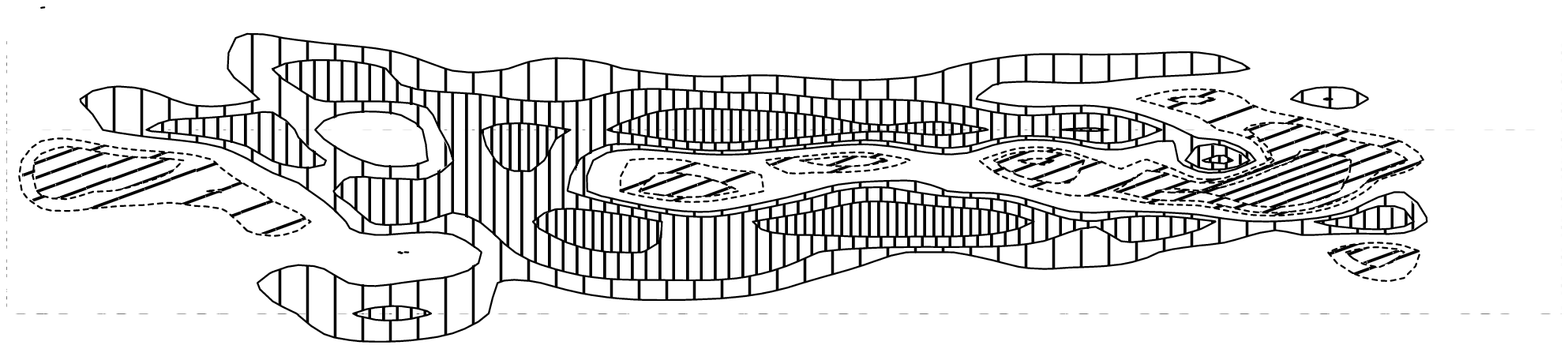,angle=-90,width=\figwidth}}%
\psfig{file=frame.eps,angle=-90,width=\figwidth}}\\
\Large d \raisebox{2ex}%
{\makebox[0mm][l]%
{\psfig{file=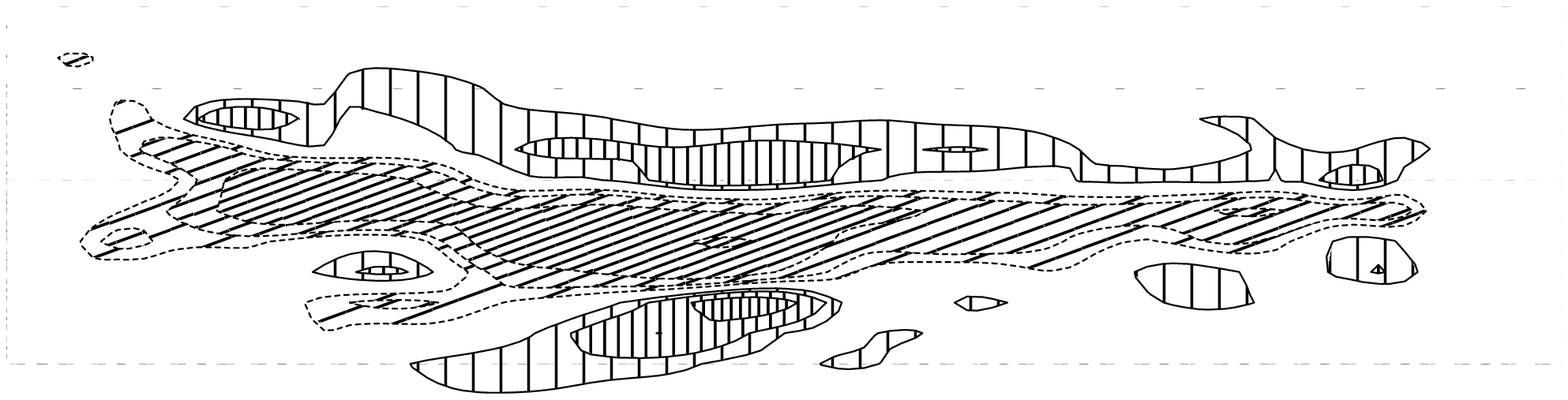,angle=-90,width=\figwidth}}%
\psfig{file=frame.eps,angle=-90,width=\figwidth}}\\
\Large e \raisebox{2ex}%
{\makebox[0mm][l]%
{\psfig{file=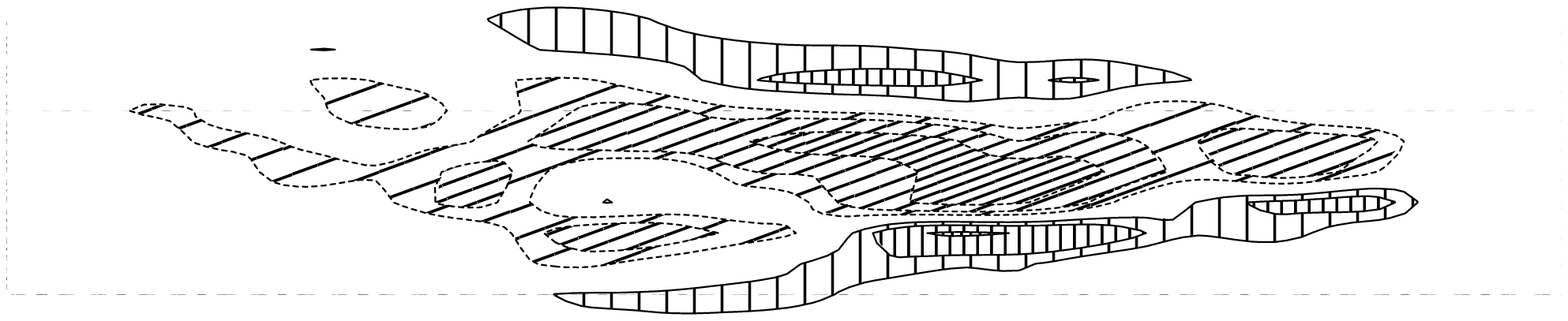,angle=-90,width=\figwidth}}%
\psfig{file=frame.eps,angle=-90,width=\figwidth}}\\
\Large f \raisebox{2ex}%
{\makebox[0mm][l]%
{\psfig{file=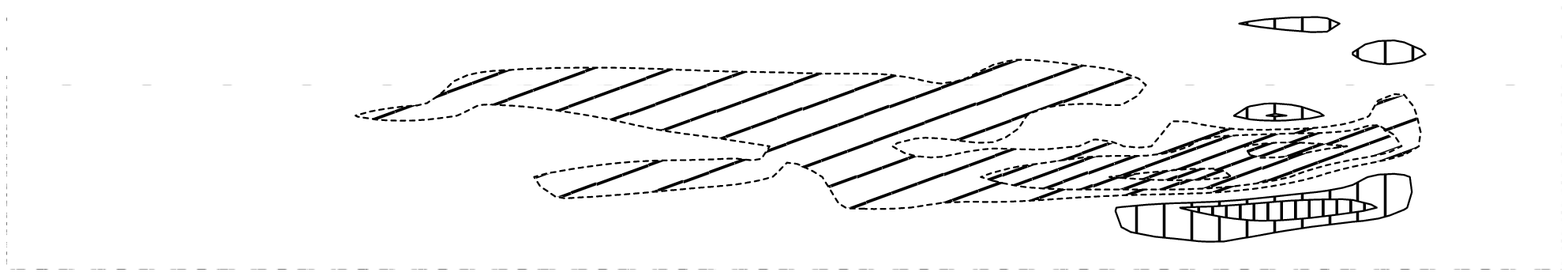,angle=-90,width=\figwidth}}%
\psfig{file=frame.eps,angle=-90,width=\figwidth}}\\
\psfig{file=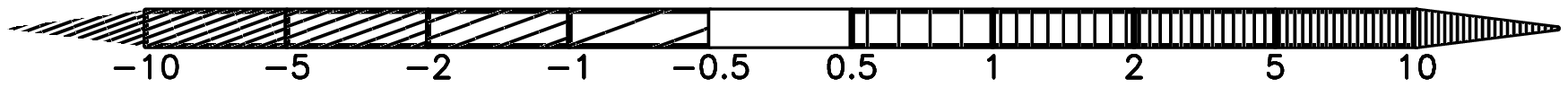,angle=-90,width=\figwidth}
\end{center}
\caption{The sensitivity of the NINO3 index at the end of 1987 to
zonal wind stress, $\partial N_3/\partial\tau_x$ in
$\mathrm{K}/\mathrm{N}\mathrm{m}^{-2}/\mathrm{sr}/\mathrm{day}$, at
the end of 1987 (a), in the first weeks of November (b), July (c),
April (d) and January 1987 (e), and July 1986 (f)}
\label{fig:dtx87}
\end{figure}

\subsection{Wind stress}
\label{sec:windstress}

% Windstress
The derivative to the wind stress is a superposition of the wind
fields associated with the Kelvin and Rossby waves.  In
Fig.~\ref{fig:dtx87} we show the situation at the same times as the
sensitivity to the sea level height in Fig.~\ref{fig:dz87}.  These
fields have the following interpretation: {\em if\/} there would have
been a positive (eastward) anomalous wind field at the position of a
positive derivative, {\em then\/} the NINO3 index would have been
higher at the end of 1987.  To give a quantitative example: suppose
there would have been an extra westerly windstress of
$0.05\:\mathrm{Nm^{-2}}$ at 180\dg--160\dg W, 2.8\dg S--2.8\dg N
(smoothed at the boundaries) during the first eight days of November.
The sensitivity to the zonal windstress in this region is about
$6\:\mathrm{K/Nm^{-2}/sr/day}$, so the effect would be an increase of
the NINO3 index two months later of $\mathrm{0.05\:Nm^{-2}}$$\,\times\,$$
8\:\mathrm{days}$$\,\times\,$$20^\circ\:5.6^\circ\:\pi^2/(180^\circ)^2
\:\mathrm{sr}$$\,\times\,$$6\:\mathrm{K/Nm^{-2}/sr/day = 0.08\:K}$.  Using
the exact dotproduct between 
the sensitivity and the extra stress the prediction is
$0.064\:\mathrm{K}$, a perturbation experiment gave
$0.077\:\mathrm{K}$.  Similarly, in April the sensitivity in this
region is roughly $-2\:\mathrm{K/Nm^{-2}/sr/day}$, so that the same
extra wind stress would have given an increase in the NINO3 index of
$-0.03\:\mathrm{K}$ nine months later.  (Appendix~\ref{sec:kicks}
gives comparisons between the linear sensitivities and full
perturbation experiments.)

During the first week of the adjoint run (Fig.~\ref{fig:dtx87}a), one
sees that a positive wind anomaly (weaker trade winds) reduces
upwelling, and hence heats up the NINO3 region.  At the extreme eastern
end of the region a positive anomaly causes a more divergent wind
field, and hence more upwelling and a lower NINO3 index.

At earlier times the disturbance caused by an anomalous wind field has
to propagate to the eastern Pacific in order to influence this index.
Comparing with the derivatives to the sea level (Fig.~\ref{fig:dz87}), 
one sees the textbook relations \citep[e.g.,][]{GillBook}
that for a Kelvin wave the zonal wind field and sea level have the
same shape, whereas an $n=1$ Rossby wave has an anti-cyclonal
circulation around a positive sea level anomaly.  The $n=2$ wave clearly
develops its antisymmetric wind field (for instance at 160\dg E in 
Fig.~\ref{fig:dtx87}c), whereas the $n=3$ Rossby wave 
is symmetric again.  However, the wind-field sensitivity 
of the higher order modes is much weaker than the sensitivity of 
the $n=1$ mode.
The 1988 experiment, during cold conditions, shows a weaker dependence
on wind stress; discussion of this effect is delayed to section
\ref{sec:delayed}.

The result of these adjoint Kelvin and Rossby waves is that during the 
half year leading up to the measurement the sensitivity to $\tau_x$ at 
the date line is positive, meaning that a weaker easterly trade wind or 
even westerly wind will increase the SST at the NINO3 region.  At 
earlier times, the sensitivity has the opposite sign --- a westerly 
anomaly will cool the NINO3 region at these lead times.

The sensitivity to the meridional wind stress $\tau_y$ is much
smaller, and mainly shows the edges of the structures visible in the
derivative to the sea level, Fig.\ \ref{fig:dz87}.
% maybe mention that there is a numerical instability?

%  #] wind stress:
%  #[ heat flux:

\subsection{Heat flux}
\label{sec:heat}

\begin{figure}[htbp]
\begin{center}
\setlength{\unitlength}{0.16bp}
\begin{picture}(2235,1182)(500,520) % was (3600,2160)(0,0)
\put(0,0){\psfig{file=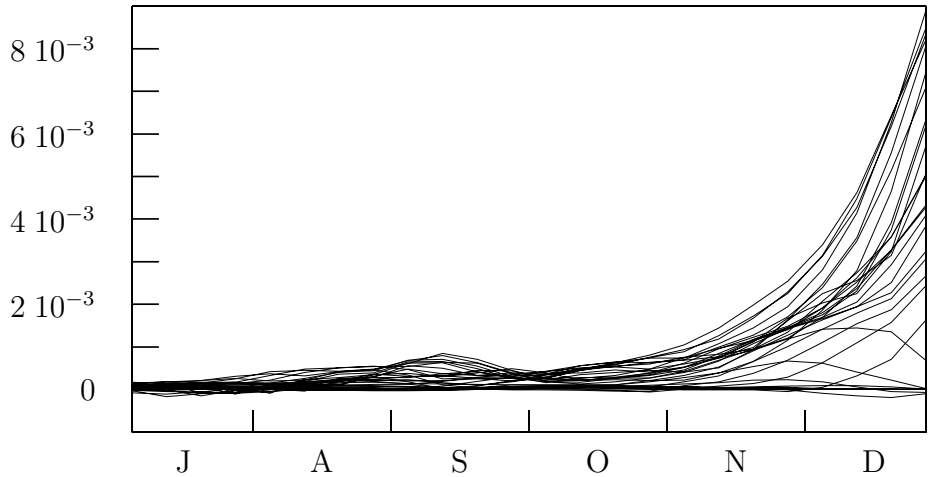}}
%\put(779,1632){\makebox(0,0)[r]{$9\;10^{-3}$}}
\put(779,1531){\makebox(0,0)[r]{$8\;10^{-3}$}}
%\put(779,1431){\makebox(0,0)[r]{$7\;10^{-3}$}}
\put(779,1330){\makebox(0,0)[r]{$6\;10^{-3}$}}
%\put(779,1230){\makebox(0,0)[r]{$5\;10^{-3}$}}
\put(779,1130){\makebox(0,0)[r]{$4\;10^{-3}$}}
%\put(779,1029){\makebox(0,0)[r]{$3\;10^{-3}$}}
\put(779,929){\makebox(0,0)[r]{$2\;10^{-3}$}}
%\put(779,828){\makebox(0,0)[r]{$1\;10^{-3}$}}
\put(779,728){\makebox(0,0)[r]{$0$}}
%\put(779,628){\makebox(0,0)[r]{$-1\;10^{-3}$}}
\put( 986,558){\makebox(0,0){J}}
\put(1149,628){\line(0,1){50}}
\put(1312,558){\makebox(0,0){A}}
\put(1475,628){\line(0,1){50}}
\put(1637,558){\makebox(0,0){S}}
\put(1800,628){\line(0,1){50}}
\put(1962,558){\makebox(0,0){O}}
\put(2125,628){\line(0,1){50}}
\put(2287,558){\makebox(0,0){N}}
\put(2450,628){\line(0,1){50}}
\put(2613,558){\makebox(0,0){D}}
\end{picture}
\end{center}
\caption{The sensitivity of the NINO3 index to the heat flux $\partial 
N_3/\partial Q$ 
in $\mathrm{K/Wm^{-2}/sr}$ for various longitudes in the eastern
Pacific, averaged over 5\dg S to 5\dg N.}
\label{fig:dq87}
\end{figure}

Finally, an obvious influence on the surface temperature in the NINO3
region is the heat flux into this area, or, almost equivalently, the
temperature some time previously due to persistence.  This is
relevant only for the last few months preceding the measurement.  
In Fig.~\ref{fig:dq87} we plot this
derivative as a function of time at the equator.  Except for some
waves that
are excited in August--September the sensitivity drops off
exponentially with a time scale of about one month ($25\pm5$ days).
There is some advection from the north-east, but this does not extend
beyond a few degrees of latitude outside the NINO3 area.

$n=1$ Rossby waves are excited at the point where the thermocline is
very shallow in the HOPE model (around 140\dg W).  We assume that heat 
input at this point lowers the thermocline and thus excites waves.  Only 
the Rossby waves travel in the correct direction to influence the NINO3 
index.  In other circumstances a heat flux does not efficiently excite 
long waves.

%  #] heat flux:

%  #] sensitivity:
%  #[ delayed oscillator:

\section{The delayed oscillator in HOPE}
\label{sec:delayed}

% Introduction
The experiments described in the previous section give the
sensitivities of an uncoupled ocean model driven by prescribed fluxes.
We would like to compare this with the concept of the delayed
oscillator, which is a coupled mechanism.  However, the delayed action
term is due to long waves propagating in the ocean, and is already
visible in the sensitivity to the zonal wind stress changing sign at
the equator in Fig.~\ref{fig:dtx87}.  The necessary feature of an
accompanying atmosphere model is a large-scale and quick adjustment of
$\tau_x$ to zonal SST gradients along the equator.  This can be
represented by the simple statistical atmosphere model described in
\ref{sec:atmosphere}, which generates a zonal wind stress similar to
the forcing fields (see Fig.~\ref{fig:txlat}).

\setlength{\figwidth}{95mm}
\begin{figure}[p]
\begin{center}
\Large a \raisebox{2ex}%
{\makebox[0mm][l]%
{\psfig{file=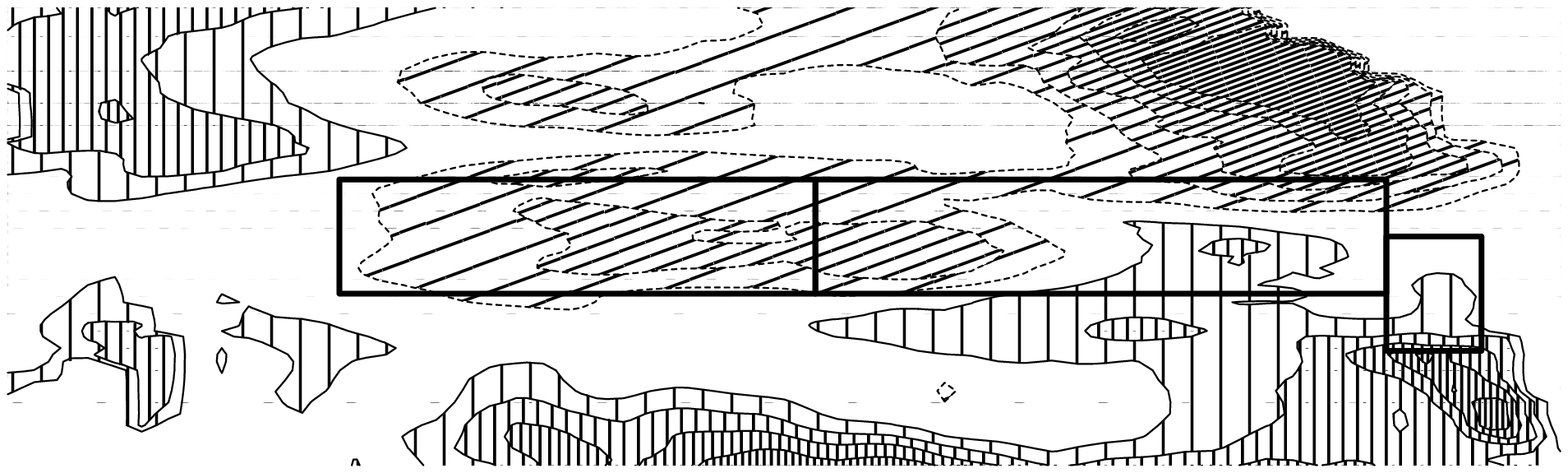,angle=-90,width=\figwidth}}%
\psfig{file=frame.eps,angle=-90,width=\figwidth}}\\
\Large b \raisebox{2ex}%
{\makebox[0mm][l]%
{\psfig{file=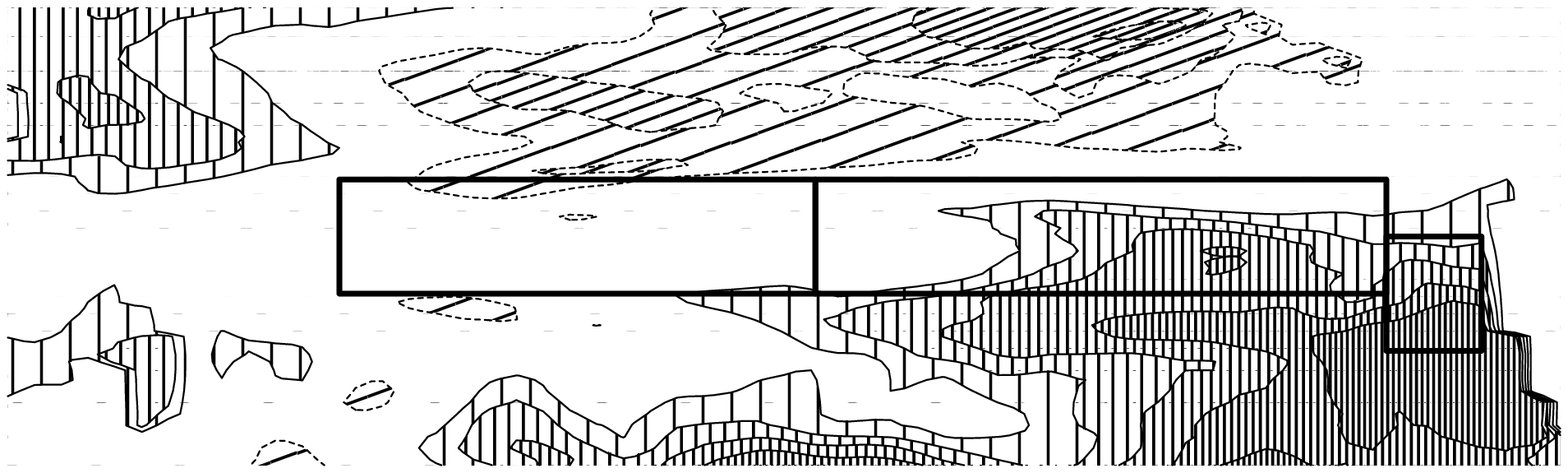,angle=-90,width=\figwidth}}%
\psfig{file=frame.eps,angle=-90,width=\figwidth}}\\
\Large c \raisebox{2ex}%
{\makebox[0mm][l]%
{\psfig{file=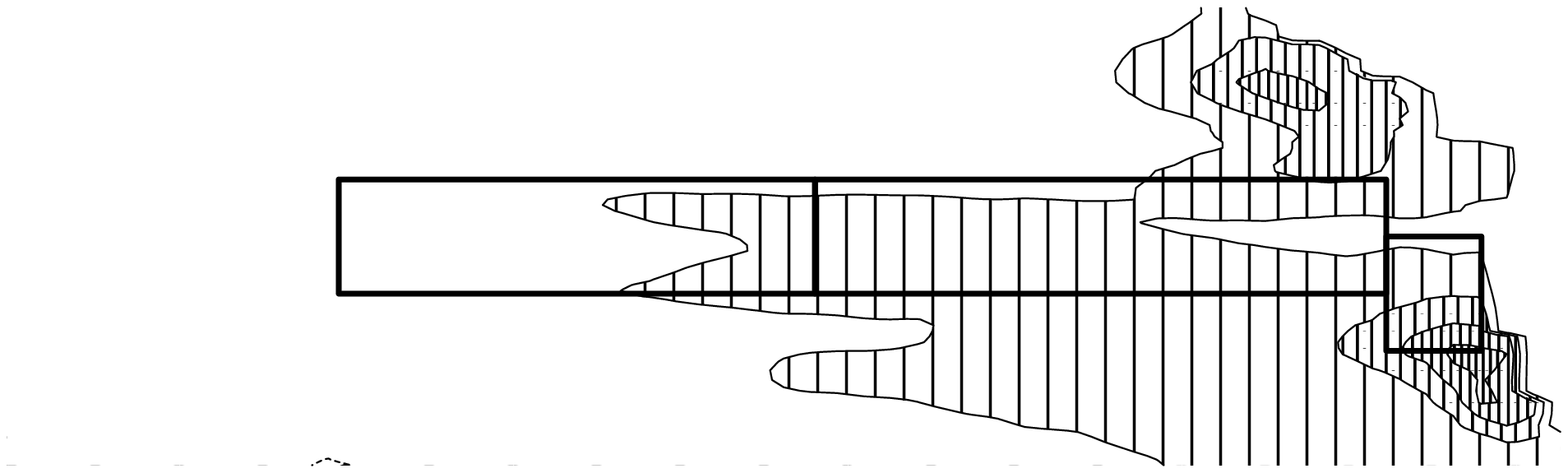,angle=-90,width=\figwidth}}%
\psfig{file=frame.eps,angle=-90,width=\figwidth}}\\
\Large d \raisebox{2ex}%
{\makebox[0mm][l]%
{\psfig{file=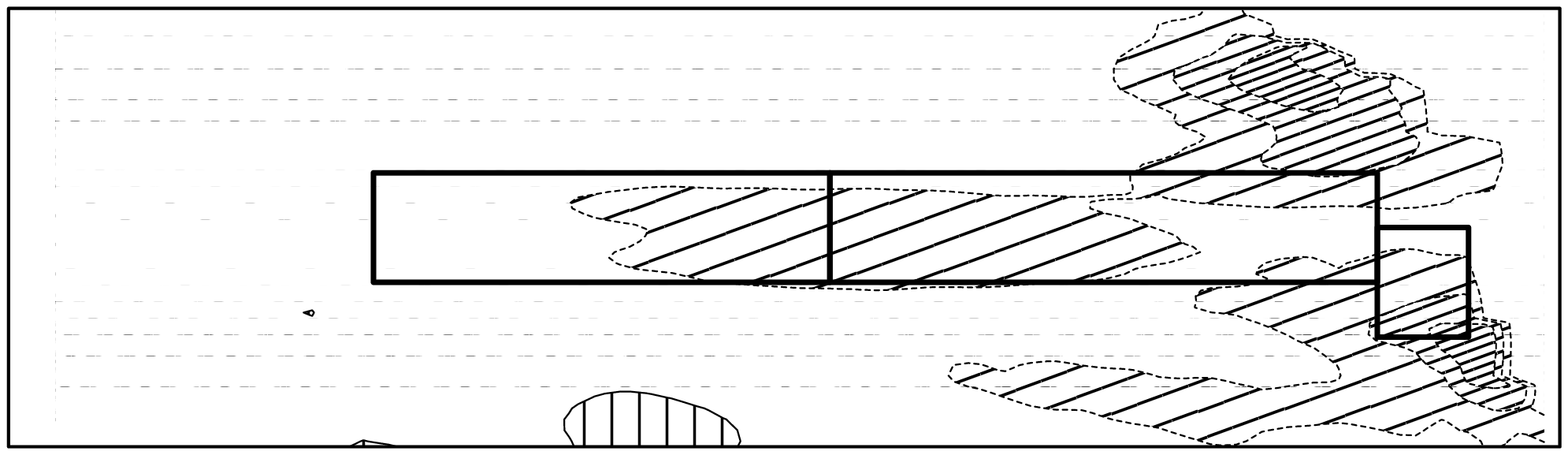,angle=-90,width=\figwidth}}%
\psfig{file=frame.eps,angle=-90,width=\figwidth}}\\
\Large e \raisebox{2ex}%
{\makebox[0mm][l]%
{\psfig{file=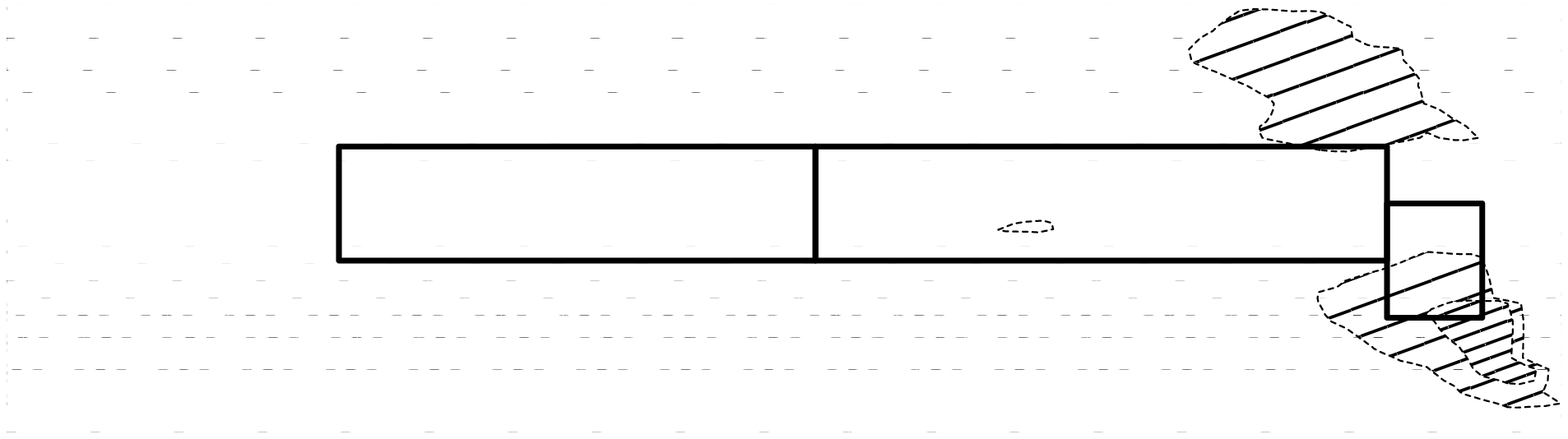,angle=-90,width=\figwidth}}%
\psfig{file=frame.eps,angle=-90,width=\figwidth}}\\
{\psfig{file=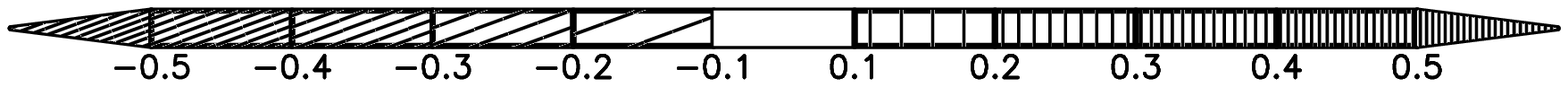,angle=-90,width=\figwidth}}
\end{center}
\caption{The sensitivity of the NINO3 index for forcing the
statistical atmosphere with an extra SST field $\partial N_3/\partial
\mathcal{F}_i \: \partial \mathcal{F}_i/\partial 
T$ in $\mathrm{K/K/sr}$ in November 1987 (a), October (b), June (c),
March (d) and December 1986 (e)}
\label{fig:dto87}
\end{figure}

Using the adjoint of the statistical atmosphere model mentioned in
section \ref{sec:adjointHOPE} we can relate the sensitivities of the
NINO3 or NINO3.4 index $N_n$ to the forcing fields $\mathcal{F}_i$ (the
adjoint fields $\partial N_n/\vec{\tau}_i$ and $\partial N_n/\partial Q_i$
we presented in sections \ref{sec:windstress} and \ref{sec:heat}) to
sensitivities to SST
\begin{equation}
\label{eq:dN}
\delta\!N_n = \!\!\sum_{\mathrm{months}\;i} 
	\frac{\partial N_n}{\partial \mathcal{F}_i} 
		\cdot \delta \mathcal{F}_i \approx
        \sum_i \frac{\partial N_n}{\partial \mathcal{F}_i} 
                \frac{\partial \mathcal{F}_i}{\partial T_{i-1}}
                \cdot \delta T_{i-1}
\;.
\end{equation}
The fields $\partial N_n/\partial \mathcal{F}_i\:\partial
\mathcal{F}_i/\partial T_{i-1}$ are plotted in 
Fig.~\ref{fig:dto87} for the months corresponding to
Fig.~\ref{fig:dtx87} (note the linear scale).  These are again a kind
of forcing fields: as we did not run in coupled mode they are not
related to the actual SST fields and only serve to change the fluxes.
Hence the summation over previous months in Eq.~(\ref{eq:dN}).

The largest influences are along the Peruvian coast and off southern
Mexico.  The latter is a consequence of a mismatch between the version
of HOPE we used and the version to which the statistical atmosphere
was tuned, and should be
discarded.  The runs with as target the NINO3.4 index should be less
sensitive to this effect.  We also find large influences of the 
extra-tropics, probably due to spurious correlations in the
statistical atmosphere model.  The sensitivity to ocean temperatures
in the cold tongue gives the effect of the weakening and
strengthening of the Walker circulation.  Up to May
(Figs~\ref{fig:dto87}a--c) the influence of heating the cold tongue is
positive: the warmer water suppresses the trade winds, exciting 
downwelling Kelvin waves.  At longer lead times the
effect is negative (Figs~\ref{fig:dto87}d,e): the same change in trade
winds also gives rise to downwelling Rossby waves that reflect at the
western boundary as a series of upwelling Kelwin waves.
During the last month (Fig.~\ref{fig:dto87}a) there also is a local
negative damping feedback, and the effect of warmer waters west of
the NINO3 region is to enhance the trade winds and decrease the index.

This picture can be reduced to a one-parameter dependence of the
NINO3 index on itself at earlier times.   Temperature
changes in the eastern Pacific are highly correlated, and also the
sensitivities to these changes in the eastern Pacific are due to the
ENSO pattern.  We expand Eq.~(\ref{eq:dN}) in EOFs $e_n$ and keep only
the first one $e_1$ (shown in Fig.~\ref{fig:EOFs}a),
\begin{equation}
\delta\!N_n = \sum_i %{\mathrm{months}\;i} 
	\sum_n \frac{\partial N_n}{\partial \mathcal{F}_i} 
                \frac{\partial \mathcal{F}_i}{\partial T_{i-1}} 
		\cdot e_n
                \; e_n \cdot \delta T_{i-1}
%\nonumber\\
     \approx C \sum_i %{\mathrm{months}\;i} 
	 \frac{\partial N_n}{\partial \mathcal{F}_i} 
                \frac{\partial \mathcal{F}_i}{\partial T_{i-1}} 
		\cdot e_1
                \; e_1 \cdot \delta T_{i-1}
\;.
\label{eq:infe1}
\end{equation}
This truncation underestimates the variations by a factor
$C\approx1.5$.
Next we realize that all indicators of sea temperature in the
eastern Pacific are highly correlated, and approximate
\begin{eqnarray}
	e_1 \cdot \delta T_i & \approx & C^{(1)} \; \delta\!N_i
\\
	e_1 \cdot \frac{\partial}{\partial T_i} & \approx & 
		C^{(2)}
		\sum_{\shortstack{$\scriptstyle\mathrm{index}$\\
                                  $\scriptstyle\mathrm{region}$}}
        \frac{\partial}{\partial T_i}
\;.
\end{eqnarray}
Over the 42-year simulation period discussed in section~\ref{sec:enso}
we find that $\delta T_i \cdot e_1$ has a correlation of $r=0.95$ with
the NINO3 index, and $r=0.96$ with the NINO3.4 index; the
corresponding constants are $C^{(1)}_3 = 0.57$ and $C^{(1)}_{3.4} =
0.51$.  The derivatives were fitted over the last 12 months of the six
experiments.  Except for the final two months, when the adjoint fields
are still close to the index region, the agreement is very
good, as the adjoint fields are mainly confined to the tropical
region.  The proportionality constants are $C^{(2)}_{3} = 2.8$,
$C^{(2)}_{3.4} = 4.2$.

\begin{figure}[htbp]
%  #[ sensitivity picture:
\begin{center}
\setlength{\unitlength}{0.11bp}
\begin{picture}(3400,2060)(200,100)
%\put(200,100){\circle{20}}
%\put(3600,2160){\circle{20}}
\put(0,0){\psfig{file=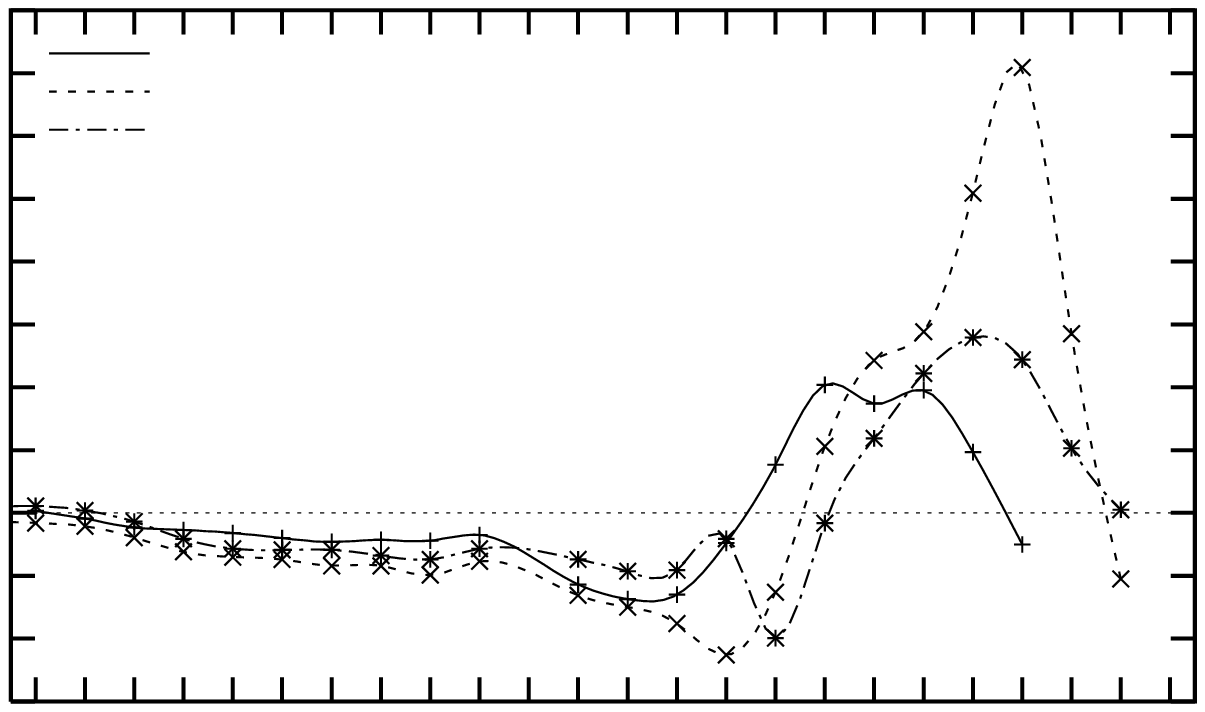}}
\put(863,1747){\makebox(0,0)[l]{\footnotesize NINO3 Dec 1988}}
\put(863,1847){\makebox(0,0)[l]{\footnotesize NINO3 Dec 1987}}
\put(863,1947){\makebox(0,0)[l]{\footnotesize NINO3 Oct 1987}}
\put(3485,150){\makebox(0,0){D}}
\put(3356,150){\makebox(0,0){N}}
\put(3227,150){\makebox(0,0){O}}
\put(3098,150){\makebox(0,0){S}}
\put(2969,150){\makebox(0,0){A}}
\put(2840,150){\makebox(0,0){J}}
\put(2710,150){\makebox(0,0){J}}
\put(2581,150){\makebox(0,0){M}}
\put(2452,150){\makebox(0,0){A}}
\put(2323,150){\makebox(0,0){M}}
\put(2194,150){\makebox(0,0){F}}
\put(2065,150){\makebox(0,0){J}}
\put(1935,150){\makebox(0,0){D}}
\put(1806,150){\makebox(0,0){N}}
\put(1677,150){\makebox(0,0){O}}
\put(1548,150){\makebox(0,0){S}}
\put(1419,150){\makebox(0,0){A}}
\put(1290,150){\makebox(0,0){J}}
\put(1160,150){\makebox(0,0){J}}
\put(1031,150){\makebox(0,0){M}}
\put(902,150){\makebox(0,0){A}}
\put(773,150){\makebox(0,0){M}}
\put(644,150){\makebox(0,0){F}}
\put(515,150){\makebox(0,0){J}}
\put(400,2060){\makebox(0,0)[r]{$0.08$}}
%\put(400,1895){\makebox(0,0)[r]{$0.07$}}
\put(400,1731){\makebox(0,0)[r]{$0.06$}}
%\put(400,1566){\makebox(0,0)[r]{$0.05$}}
\put(400,1402){\makebox(0,0)[r]{$0.04$}}
%\put(400,1237){\makebox(0,0)[r]{$0.03$}}
\put(400,1073){\makebox(0,0)[r]{$0.02$}}
%\put(400,908){\makebox(0,0)[r]{$0.01$}}
\put(400,744){\makebox(0,0)[r]{$0$}}
%\put(400,579){\makebox(0,0)[r]{$-0.01$}}
\put(400,415){\makebox(0,0)[r]{$-0.02$}}
%\put(400,250){\makebox(0,0)[r]{$-0.03$}}
\end{picture}
\begin{picture}(3400,2060)(200,100)
%\put(200,100){\circle{20}}
%\put(3600,2160){\circle{20}}
\put(0,0){\psfig{file=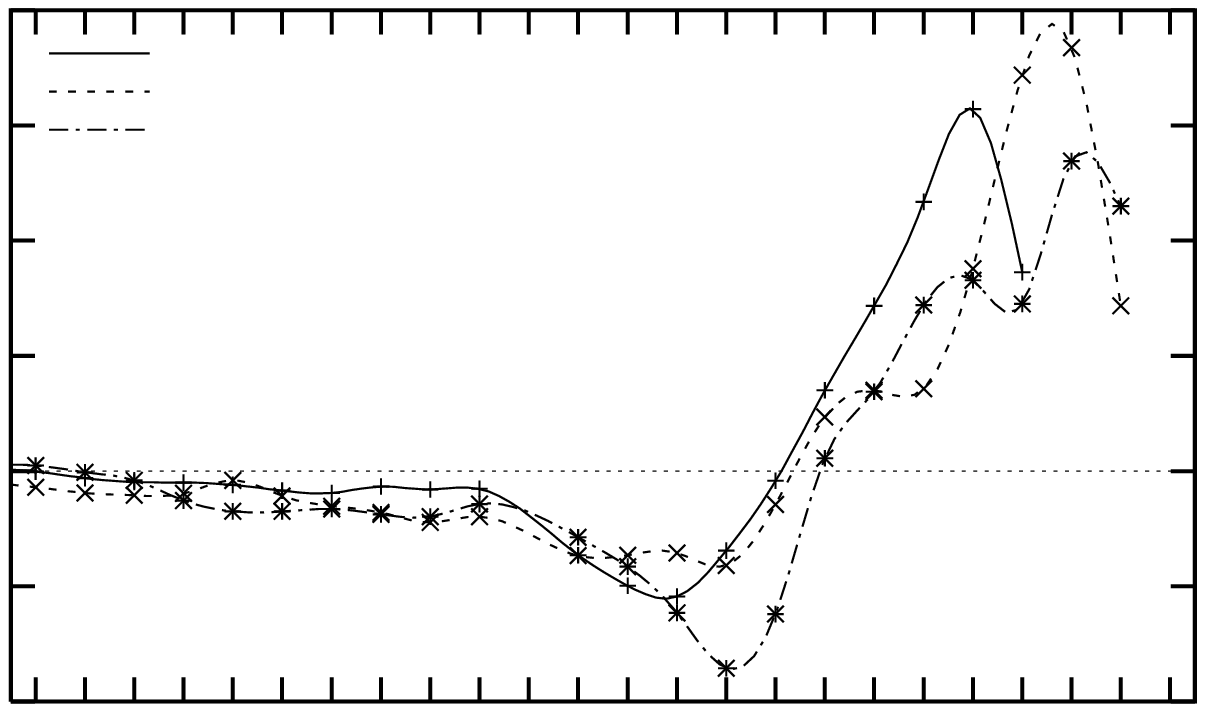}}
\put(863,1747){\makebox(0,0)[l]{\footnotesize NINO3.4 Dec 1988}}
\put(863,1847){\makebox(0,0)[l]{\footnotesize NINO3.4 Dec 1987}}
\put(863,1947){\makebox(0,0)[l]{\footnotesize NINO3.4 Oct 1987}}
\put(3485,150){\makebox(0,0){D}}
\put(3356,150){\makebox(0,0){N}}
\put(3227,150){\makebox(0,0){O}}
\put(3098,150){\makebox(0,0){S}}
\put(2969,150){\makebox(0,0){A}}
\put(2840,150){\makebox(0,0){J}}
\put(2710,150){\makebox(0,0){J}}
\put(2581,150){\makebox(0,0){M}}
\put(2452,150){\makebox(0,0){A}}
\put(2323,150){\makebox(0,0){M}}
\put(2194,150){\makebox(0,0){F}}
\put(2065,150){\makebox(0,0){J}}
\put(1935,150){\makebox(0,0){D}}
\put(1806,150){\makebox(0,0){N}}
\put(1677,150){\makebox(0,0){O}}
\put(1548,150){\makebox(0,0){S}}
\put(1419,150){\makebox(0,0){A}}
\put(1290,150){\makebox(0,0){J}}
\put(1160,150){\makebox(0,0){J}}
\put(1031,150){\makebox(0,0){M}}
\put(902,150){\makebox(0,0){A}}
\put(773,150){\makebox(0,0){M}}
\put(644,150){\makebox(0,0){F}}
\put(515,150){\makebox(0,0){J}}
\put(400,2060){\makebox(0,0)[r]{$0.08$}}
\put(400,1758){\makebox(0,0)[r]{$0.06$}}
\put(400,1457){\makebox(0,0)[r]{$0.04$}}
\put(400,1155){\makebox(0,0)[r]{$0.02$}}
\put(400,853){\makebox(0,0)[r]{$0$}}
\put(400,552){\makebox(0,0)[r]{$-0.02$}}
\put(400,250){\makebox(0,0)[r]{$-0.04$}}
\end{picture}
\end{center}
%  #] sensitivity picture:
\caption{The sensitivity of the NINO3 and NINO3.4 indices 
to the same index at the end of 1987 and 1988 through the statistical
atmosphere in K/K/month.  Including the sensitivity to correlated SST
changes would roughly result in a factor three increase.}
\label{fig:osc}
\end{figure}

This finally gives the dependence of the index to itself some time
earlier
\begin{equation}
    \delta\!N_n \approx C C^{(1)} C^{(2)} \sum_{i=1}^{n-1} 
	 \frac{\partial N_n}{\partial N_i} \delta\!N_i
\;.
\label{eq:delayeddisc2}
\end{equation}
The factor $C C^{(1)} C^{(2)}$ is
about 2--3 for the NINO3 index, 3--4 for the NINO3.4 index.
The sensitivities ${\partial N_n}/{\partial N_{i}}$ are shown in
Fig.~\ref{fig:osc} for the six experiments we have performed (to the
NINO3 and NINO3.4 indices in October 87, December 87 and December 88).
The interpretation of Eq.~(\ref{eq:delayeddisc2}) is not entirely
straightforward, as a change $\delta\!N_i$ not only causes a change
$\delta\!N_n$ directly, but also through intermediate steps $\delta\!
N_j$ with $i<j<n$, which are not explicitly mentioned in this form.
This way the coupled response to a perturbation at 6 months' lead time
will not only consist of the negative feedback indicated in
Fig.~\ref{fig:osc}, but also of the square of the 3-month positive
response, and contributions from all other partitions of six.  The
effect is to move the negative feedback to longer times, in better
agreement with the NINO3, NINO3.4 autocorrelation functions.  In
principle we could have computed this coupled response function from
Fig.~\ref{fig:osc}, but the large uncertainty in the scale make this a
futile exercise.

On the other hand, Fig.~\ref{fig:osc} and Eq.~(\ref{eq:delayeddisc2})
can be compared directly to a discretized form of the delayed
oscillator equation Eq.~(\ref{eq:delayed}) 
\begin{eqnarray}
N_n & = & (1+a\Delta t) N_{n-1} - b\Delta t \, N_{n-1}^3 
	- c\Delta t \, N_{n-\delta} 
\;,
\label{eq:delayeddisc}
\\
\delta\!N_n & = & \Bigl( (1+a\Delta t) - 3b\Delta t N_{n-1}^2 \Bigr)
	\delta\!N_{n-1} - c\Delta t \: \delta\!N_{n-\delta}
\;,
\end{eqnarray}
with $\Delta t$ one month.
As expected from this equation, the `immediate' positive feedback
differs very much 
between a situation where the index is falling to zero (Dec 1987) and
a situation in which the index is strongly positive (Oct 1987) or
negative (Dec 1988).  This is the effect of the non-linear effects
that are parametrized by the $N^3$ term in
Eqs~(\ref{eq:delayed},\ref{eq:delayeddisc}).
In cold conditions, the thermocline has disappeared in our model,
so that a Kelvin wave in the second baroclinic mode, varying the
thermocline depth, is not very effective at changing SST.  During 
a warm epsiode the thermocline is so deep that the effect of such
a wave is also smaller.  The effect of a Kelvin wave in the first 
baroclinic mode is not sensitive to the thermocline depth.
The effect of these nonlinearities in the positive feedback 
is larger for the NINO3 index than for the NINO3.4 index, 
as the thermocline is shallower in the NINO3 region and 
hence variations of its depth have a larger effect.

The approximation in the delayed oscillator equation 
that the Bjerknes feedback is instantaneous does not
seem to be a very good one, especially for the NINO3 index.  Part of
the delay is due to the artificial 1-month delay of the statistical
atmosphere, but another 2--3 months are due to the time it takes for a
Kelvin wave generated at the date line to reach the signal region;
before this time local negative effects play a role.  This effect is
also smaller for the NINO3.4 index.

The negative feedback consists of an overlapping train of $n=1$ Rossby
waves that are reflected off different parts of the coast, instead of
an idealized peak.  Combined with the lower effective group velocity
shown in Fig.~\ref{fig:dz87st} we obtain negative influences from 6 to
about 18 months prior to the measurement.  Higher order Rossby waves
do not contribute at these timescales.

Due to the normalization uncertainties it is difficult to gauge the
extend to which the summed positive feedback exceeds one, and compare
this quantitatively with the negative feedback terms, \emph{i.e.}, to
determine the parameters in Eq.~(\ref{eq:delayed}) or
Eq.~(\ref{eq:delayeddisc}) in our model.
However, it is clear from the figure that the negative terms do not
depend on the initial state in the same non-linear way as the positive
terms.  The negative feedback is much stronger than expected for the
December 1988 experiments.  We suspect that this is due to the
non-linear interactions between the different baroclinic modes that we
observed in section \ref{sec:speeds}.  The weak positive response was
due to the small amplitude at the surface of the first baroclinic mode
adjoint Kelvin wave that is excited.  On the other hand, the adjoint
Rossby waves that determine the negative feedback propagate in the
usual second baroclinic mode in the thermocline, which has a larger
amplitude at the surface for the same energy content and is thus more
easily excited by a wind anomaly.  At the reflection in shallower
waters at the western boundary some energy seems to be transferred down
from Rossby waves in the second baroclinic mode to Kelvin waves in the
deeper first mode.

%  #] delayed oscillator:
%  #[ conclusions:

\section{Summary and conclusions}
\label{sec:conclusions}

We studied the causes of variations of the sea surface temperature in
the eastern equatorial Pacific with an adjoint OGCM and statistical
atmosphere.  The adjoint gives us first-order estimates of
the sensitivity of the NINO3 and NINO3.4 indices of this temperature
to forcing fields and state fields some time earlier.  We generated
sensitivity fields to these indices at the end of October 1987,
December 1987 and 
December 1988; respectively warm, transition and cold phases of ENSO.
In these experiments we used prescribed fluxes to drive the ocean model
HOPE.

The sensitivities either damp out exponentially (heat flux, previous
SST, zonal surface current) with a time scale of about one month, or
propagate as equatorial Kelvin and
Rossby waves (wind stress, sea level).  Of course, without a full
atmosphere model we cannot observe small-scale coupled modes.

The temperature of the NINO3 region can be affected by a
Rossby wave coming from the eastern coast.  In our model some of this 
influence is reflected back --- an equatorial Kelvin wave
impinging on the eastern coast can generate such a
Rossby wave, albeit not very efficiently.  
Another way the temperature can be changed is by the
arrival of a Kelvin wave from the west; these are clearly visible in
the sensitivities to the wind stress and sea level.  This Kelvin wave
can be generated by Rossby waves reflecting off the coast of New
Guinea and the Philippines, either directly or with a few months'
delay at the western boundary area.  These delays, combined with the
eastern reflections, give rise to an effective group velocity of the
sensitivities that is lower than the speed of the individual Kelvin
and Rossby waves.  Higher order adjoint Rossby waves are
observed, but do not have a strong enough wind stress sensitivity 
field to be able to influence the eastern Pacific.
These waves sum to a sensitivity to the thermocline depth in the
warm pool off the equator, but there is no sensitivity to the 
zonally-averaged thermocline depth along the equator at long
timescales.

The speed of the adjoint Kelvin waves depends very much on the state of the
ocean in our model at the measurement region.  Note that we do not
simulate Kelvin waves as they occur in the ocean, but as they
influence the NINO3 index.  Normally these propagate in the second
baroclinic mode (with a zero at the thermocline), with $c\approx
2\:\mathrm{ms^{-1}}$ in the western and central Pacific.  However, in
the cold tongue during the cold phase (La Ni\~{n}a), the thermocline is
so shallow that the first baroclinic mode is excited, with a speed
$c\approx 3\:\mathrm{ms^{-1}}$ in the central Pacific.  These
different modes mix at the reflection in shallow waters at the west
coast; the reflected adjoint Rossby waves have speeds compatible with
the second baroclinic modes (the third for higher order Rossby waves).
These results should be compared to an observed speed of 
$2.4\pm0.3\:\mathrm{ms^{-1}}$, independent of longitude, from the
TOGA-TAO array, which is ascribed to a mix of the first and second
modes.

We used an adjoint statistical atmosphere model to convert the
sensitivities into fluxes (mainly zonal wind stress) into sensitivities to
the SST{}.  Next we approximated these fields by sensitivities to
a scalar index: the sensitivity to the index region.  This underestimates
the size of the influence by a factor of roughly four, but does not
change the shape as a function of time very much, due to the high
correlations of the neglected terms with the included ones.

The resulting response function (Fig.~\ref{fig:osc}) shows the essence of 
the delayed oscillator mechanism.  Going back from the measurement time,
the first few months exhibit the positive feedback of the change in
trade winds reinforcing the SST anomaly.  At about half a year before
the target date the feedback changes sign as Rossby waves reflect off
the western coast as Kelvin waves with opposite polarity.  Both these
feedback loops are spread out in time.  Kelvin waves are excited
most easily in the central Pacific and take about two months to
travel to the signal region, and the reflections at the western side
occur at different longitudes so that the influence of Rossby waves is
spread out over half a year as well.

We also observe that the positive feedback is much stronger when the
system is in transition (Dec 1987), than when it is in a warm (Oct
1987) or cold (Dec 1988) phase, pointing to a role of non-linear terms
in restraining the delayed oscillator.  We do not observe this effect
in the delayed response.

%new: say something what we learned from this study.
In this study we have shown that the adjoint HOPE model gives a good
approximation of the sensitivity of the full non-linear model more
than a year back in time.  To our surprise most features of the
sensitivity fields in this primitive equation OGCM can be described by
linear long-wave dynamics, modified by the background state.  Zonal
advection and equatorial thermocline recharge mechanisms are not
visible.  The western-boundary reflections give rise to a feedback
structure similar to the one postulated in the delayed-oscillator
mechanism.  Unfortunately we cannot measure whether the strength of
the feedbacks is enough to sustain oscillations, \emph{i.e.}, whether
other mechanisms, such as stochastic forcing, are needed to excite
ENSO episodes.

% - compare with other theories/studies.
Singular vector analysis of simpler but coupled models can be compared
to the results presented here if the evolved perturbation pattern is
similar to an ENSO pattern and the difference between our uncoupled
response function and a coupled one are (qualitatively) accounted
for (cf.\ the discussion of Eq.~(\ref{eq:delayeddisc2})).  The
3-month singular vectors of \cite{Moore97a,Moore97b} are dominated
by dynamical processes in the atmosphere, and do not correspond to our
oceanic sensitivities, although their one-month results could be
compared to our sensitivity to wind fields at two months' lead time
(the difference is due to the more westerly location of their signal
pattern); both indicate a high sensitivity to westerly wind bursts.
In \citet{Xue97a} the final patterns have a typical ENSO signature
and should be comparable to our sensitivities at 6 months' lead
time.  They find a predominance of local forcing due to a
reduction of upwelling.  This mode seems to correspond to an iteration
of our local forcing in Fig.~\ref{fig:dto87}a.  A new norm
\citep{Xue97b} adds to this a wind field comparable with our
remote-forcing result (Fig.~\ref{fig:dtx87}d), which is dominated by
variability of the thermocline depth.  This signal includes
a reflection at the western boundary.

\paragraph{Acknowledgements} Geert Jan van Oldenborgh was supported by
the Geosciences Foundation (GOA) of the Netherlands Organization of
Scientific Research (NWO).  We would like to thank Arie Kattenberg,
Gerbrand Komen, Femke Vossepoel and Jean-Philippe Boulanger for many
helpful comments on the manuscript.

%  #] conclusions:
%  #[ technical:

\appendix
\section{Technical details of the adjoint model}
\label{technical}

Most of the adjoint code was generated by automatic differentiation of
the Fortran code of the HOPE and statistical atmosphere models by the
(T)AMC \citep{AMC,TAMC}.  The exceptions were the horizontal and
vertical diffusion and viscosity computations in HOPE.  These involve
the inversion of a tridiagonal matrix, the adjoint can be written
much simpler than the automatically-generated algorithm. We have not
checked the code for vectorization as all experiments reported here
were performed on a scalar CPU (MIPS R10000).

The adjoint model $M^\dagger$ depends both on the adjoint
fields $\partial/\partial x$ and the fields of the reference trajectory $x$ 
for non-linear functions $\mathcal{M}${}.  The latter fields are saved
from the forward run of the model to be used in the backward run of
the adjoint model in a three-layer scheme. To 
compute the adjoint of one day (12 ocean time steps), the model is run
forward over this period while storing fields needed in non-linear 
transformations on a
direct-access file.  Saving these in single precision, and eliminating
the multiple storage of the slowly-varying temperature and salinity
fields per time step one obtains a 250 MB storage file (21 3D fields of
2*97*61*20 numbers per time step times 12 steps per day).  These daily
runs are restarted from restart files generated by a one-month run.
Each (modified) restart file takes 8 MB (8 3D fields).  The monthly
restart files are generated by a top-level run.  In all the storage
requirements are about 500 MB disk space.  The run-time size of the
program is 130 MB.

Apart from end effects (the last month can be skipped on the highest
level) the running time for an adjoint run is three forward runs
(total, month and day) and one backward run.  The latter includes many
forward recomputations, and takes about 2--3 times longer than a
forward run.  The total cost of a full adjoint run is therefore 5--6
times the cost of a forward run in CPU time.  To this one should add
the cost of writing and reading 250 MB temporary storage per simulated
day.

% Checks
We verified whether finite perturbations of the initial state fields
did in fact give results in agreement with the adjoint model, which is
only strictly valid for infinitesimal perturbations.  For small random
perturbations (1--6 wavelengths in $x$ and $y$ and 0--1 in $z$
tapering off with depth) of all initial state variables except
temperature and salinity we obtain good agreement with the derivative
up to more than 6 months.  We did not include variations of the
temperature and salinity fields as these give rise to different
convection patterns, mainly in the mixed layer at high latitudes.  The
derivative to the most important input variable, $\tau_x$, is validated 
more systematically in Appendix \ref{sec:kicks}.

%  #] technical:
%  #[ kicks:

\section{Validation}
\label{sec:kicks}

Although the linear dynamics we found in
section~\ref{sec:sensitivity} looks plausible, 
it remains to be verified that non-linear effects do not in fact
dominate in the model (let alone in nature).  For this purpose we
compared the influence of zonal wind fields on the NINO3 index at the
end of 1987 with the predictions of the adjoint model, cf.\
Eq.~(\ref{eq:limit})
\begin{equation}
	\delta N_3 \approx \frac{\partial N_3}{\partial\tau_x}
	         \delta\tau_x
\;.
\end{equation}
All deviations are relative to a forward run with prescribed
fluxes\footnote{Due to an oversight, relaxed to climatology rather
than observed SST; this should not influence the findings.}.  For the
extra wind field $\delta\tau_x$ we chose a longitude with a large
influence, the size was two grid points at latitude $\pm
1.4^\circ$ with a zonal extent of $20^\circ$; the zonal edges are
smoothed over $\pm5^\circ$.  The duration of the storm was matched
to the sampling of the derivative fields, 8 days, and its intensity
was $\pm0.01$ and $\pm 0.05\:\mathrm{Nm^{-2}}$, resulting in
predicted deviations of the NINO3 index of roughly $0.05\,\mathrm{K}$
(1 month) to $0.01\,\mathrm{K}$ (1 year) for the higher wind stress.

\begin{figure}[tbp]
\begin{center}
\setlength{\unitlength}{0.14bp}
%\begin{picture}(3600,2160)(0,0)
\begin{picture}(2840,1620)(320,230)
%\put(3160,1850){\circle{20}}
%\put(320,230){\circle{20}}
\put(0,0){\makebox{\psfig{file=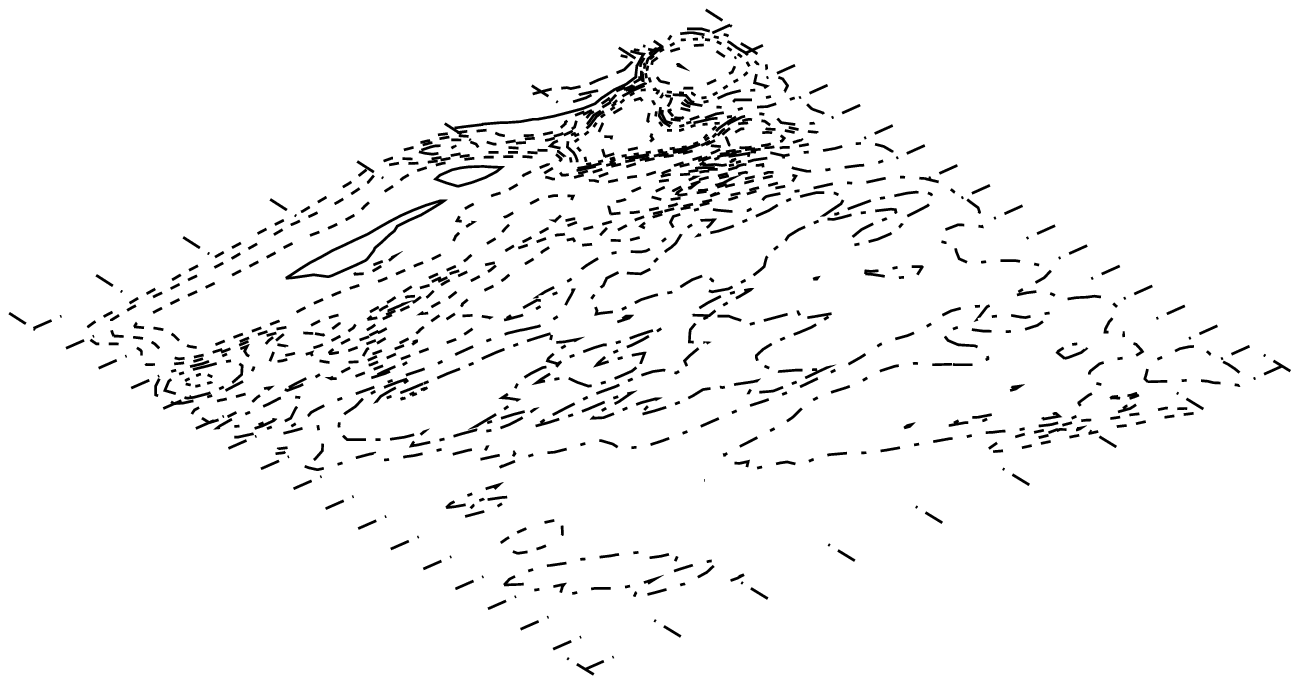}}}
\put(0,0){\makebox{\psfig{file=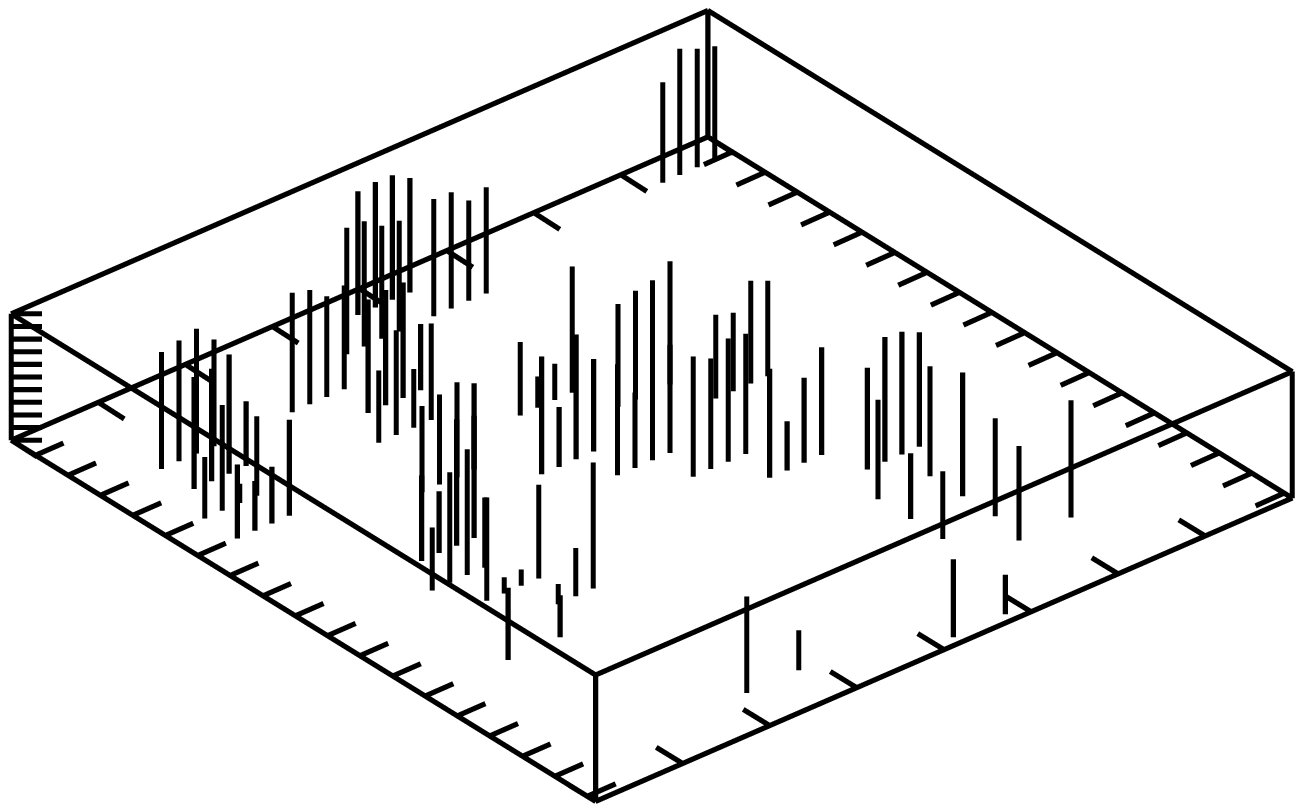}}}
% added by hand
\put( 847,720){\makebox(0,0)[tr]{1987}}
\put(1448,348){\makebox(0,0)[tr]{1986}}
\put(2528,541){\makebox(0,0)[tl]{lon}}
\put(426,1369){\makebox(0,0)[rb]{\large$q$}}
% shifted 70 to the right and 10 up
\put(1622,286){\makebox(0,0)[c]{J}}
\put(1555,327){\makebox(0,0)[c]{A}}
\put(1488,368){\makebox(0,0)[c]{S}}
\put(1421,410){\makebox(0,0)[c]{O}}
\put(1354,451){\makebox(0,0)[c]{N}}
\put(1288,492){\makebox(0,0)[c]{D}}
\put(1221,533){\makebox(0,0)[c]{J}}
\put(1154,575){\makebox(0,0)[c]{F}}
\put(1087,616){\makebox(0,0)[c]{M}}
\put(1021,657){\makebox(0,0)[c]{A}}
\put( 954,698){\makebox(0,0)[c]{M}}
\put( 887,740){\makebox(0,0)[c]{J}}
\put( 820,781){\makebox(0,0)[c]{J}}
\put( 753,822){\makebox(0,0)[c]{A}}
\put( 687,864){\makebox(0,0)[c]{S}}
\put( 620,905){\makebox(0,0)[c]{O}}
\put( 553,946){\makebox(0,0)[c]{N}}
\put( 486,987){\makebox(0,0)[c]{D}}
%
%\put(3224,873){\makebox(0,0){280}}
% shifted 50 to the left, 10 up
\put(2995,805){\makebox(0,0){260}}
\put(2816,727){\makebox(0,0){240}}
\put(2637,649){\makebox(0,0){220}}
\put(2458,571){\makebox(0,0){200}}
\put(2278,493){\makebox(0,0){180}}
\put(2099,415){\makebox(0,0){160}}
\put(1920,337){\makebox(0,0){140}}
\put(456,1319){\makebox(0,0)[r]{1}}
%\put(356,1293){\makebox(0,0)[r]{0.9}}
%\put(356,1267){\makebox(0,0)[r]{0.8}}
%\put(356,1241){\makebox(0,0)[r]{0.7}}
%\put(356,1215){\makebox(0,0)[r]{0.6}}
\put(456,1189){\makebox(0,0)[r]{0.5}}
%\put(356,1163){\makebox(0,0)[r]{0.4}}
%\put(356,1137){\makebox(0,0)[r]{0.3}}
%\put(356,1111){\makebox(0,0)[r]{0.2}}
%\put(356,1085){\makebox(0,0)[r]{0.1}}
\put(456,1059){\makebox(0,0)[r]{0}}
\end{picture}
\end{center}
\caption{The quality of the adjoint model 1986--1987 run compared with
perturbed forward runs.  On the bottom plane a space-time diagram of
the sensitivity to the zonal wind field, $\partial
N_3/\partial\tau_x$, along the equator is sketched (dashed contours).
The groups of four bars represent the quality of the adjoint field,
checked with perturbed wind stresses of $\delta\tau_x = -0.05$,
$-0.01$, $+0.01$, $+0.05\:\mathrm{N}\mathrm{m}^{-2}$ respectively.
The position of the bars indicates the longitude and date of the
perturbation.  The height gives the quality, defined as $q =
1-2/\pi\,\arctan |1- \delta N_3 / (\delta\tau_x \partial N_3 /
\partial\tau_x) |$.  It is 1 for perfect agreement, 0.6 when the ratio
is 0.5 or 1.5, and it tends to zero for very bad agreement}.
\label{fig:kicks}
\end{figure}

In Fig.~\ref{fig:kicks} we present the results of these experiments.
In the initial adjoint Kelvin wave (Sep--Dec) the ratios almost always
are within the range 0.8--1.3.  Smaller wind fields
give better results, as the error is dominated by non-linear effects.
The exceptions are in September, when the reflected Rossby
waves influence the picture.  The previous two months (Jul--Aug), when
the reflections are building up and the perturbations are near the
western coast, the adjoint performs poorly.  For higher wind fields
the ration lies between 0.2 and 2.  These perturbations send waves of
both signs across the Pacific, so the final perturbation of the NINO3
index is a sum of positive and negative patches that is very
sensitive to details.

The situation improves when the adjoint Rossby waves are clearly
established (Jan---Jun).  When the signal is not too weak the
prediction for the stronger winds is accurate to within a factor
0.4--1.6 (it tends to be a bit too low).  The error is dominated
by noise, so that the lower wind fields give worse results.  In the
previous year we only
investigated the stronger wind fields.  One sees the same pattern:
the well-defined first Rossby wave gives reasonable agreement
(although the derivative again underestimates the real effect by a
factor 1.5 on average).  However, the effect of a wind field outside
the strongest wave is predicted poorly by the adjoint.

\ifx\stupidformat\undefined
\begin{table}[htbp]
\else
\begin{table}[b]
\fi
\begin{center}
\begin{tabular}{|rcr|rrr|}
\hline
\multicolumn{1}{|c}{Date}   
         & \multicolumn{1}{c}{longitude} 
                          & \multicolumn{1}{c|}{$\delta\tau_x$} 
                                   & $\delta$NINO3 & adjoint & ratio \\
         &                & \multicolumn{1}{c|}{$[\mathrm{Nm^{-2}}]$} 
                                   & $[\mathrm{K}]$ & $[\mathrm{K}]$ & \\
\hline
Oct 1987 & 160\dg--180\dg & $ 0.2$ & $ 0.295$ & $ 0.199$ & $1.48$ \\
         &                & $-0.2$ & $-0.173$ & $-0.199$ & $0.86$ \\
Jun 1987 & 160\dg--180\dg & $ 0.2$ & $-0.041$ & $-0.069$ & $0.60$ \\
         &                & $-0.2$ & $ 0.062$ & $ 0.069$ & $0.90$ \\
Jan 1987 & 225\dg--245\dg & $ 0.2$ & $-0.048$ & $-0.048$ & $1.00$ \\
         &                & $-0.2$ & $ 0.060$ & $ 0.048$ & $1.25$ \\
Aug 1986 & 160\dg--180\dg & $ 0.2$ & $ 0.000$ & $-0.019$ &$-0.02$ \\
         &                & $-0.2$ & $ 0.012$ & $ 0.019$ & $0.64$ \\
\hline
\end{tabular}
\end{center}
\caption{The effect of 8-day stronger anomalous wind 
bursts over 20\dg$\times$4\dg\ along the equator.}
\label{tab:kicks}
\end{table}

We also studied the effect of stronger anomalous wind stress fields,
$\delta\tau_x = 0.2\:\mathrm{Nm^{-2}}$, 
which is the typical strength of a westerly wind 
burst \citep{GieseResponse}.  The results are presented in 
Table~\ref{tab:kicks}.  Non-linear effects worsen the agreement at short 
lead times compared to more gentle perturbations, but at longer lead times 
the results are comparable.

In all, it seems we can trust the sensitivities of the adjoint model
most of the time to within a factor 2, even to one year back.  This
does not hold when the expected deviation is small, or when the
perturbation is near the western coasts so that reflected and direct
waves interfere.

%  #] kicks:
%  #[ postamble:

%\bibliographystyle{mwr}
%\bibliography{ocean}

\ifx\stupidformat\undefined\else
\newpage
\listoffigures
\listoftables
\end{spacing}
\fi
\end{document}